\documentclass[preview,12pt]{elsarticle}

\usepackage{epsfig} 
\usepackage{amsmath}
\usepackage{bm}
\usepackage{graphicx}
\usepackage{booktabs,caption}
\usepackage[flushleft]{threeparttable}
\usepackage[linesnumbered,ruled,vlined]{algorithm2e}
\usepackage{float}
\usepackage[utf8]{inputenc}
\usepackage{wrapfig}
\usepackage{amsfonts}
\usepackage{lscape}
\usepackage[table]{xcolor}
\usepackage{ulem}
\usepackage{lineno,hyperref}
\usepackage{multirow}
\usepackage{gensymb}
\usepackage{color}
\usepackage{float}
\usepackage[table]{xcolor}
\usepackage{multirow}
\usepackage{latexsym}
\usepackage{lscape}
\usepackage{amsthm}
\newtheorem*{remark}{Remark}
\modulolinenumbers[5]

\journal{Journal of Computational Physics}

\begin{document}

\begin{frontmatter}

\title{HeartSimSage: Attention-Enhanced Graph Neural Networks for Accelerating Cardiac Mechanics Modeling}

\author[mymainaddress2]{Lei Shi}
\author[mymainaddress]{Yurui Chen}
\author[mymainaddress]{Vijay Vedula\corref{mycorrespondingauthor}}
\cortext[mycorrespondingauthor]{Corresponding author, 500 W 120th street, MC 4703 Mudd 220, New York, NY, 10027}
\ead{vv2316@columbia.edu; TEL(212) 853-2547}

\address[mymainaddress2]{Department of Mechanical Engineering, Kennesaw State University, Marietta, GA, 30060, USA}
\address[mymainaddress]{Department of Mechanical Engineering, Columbia University, New York, NY, 10027, USA}

 

\begin{abstract}
Finite element analysis (FEA) is a powerful tool that forms the cornerstone of modeling cardiac biomechanics. However, FEA is computationally expensive for creating digital twins, which typically involves performing tens or hundreds of FEA simulations to estimate tissue parameters, limiting its clinical application. We have developed an attention-enhanced graph neural network (GNN)-based FEA emulator, \textit{HeartSimSage}, to rapidly predict passive biventricular myocardial displacements from patient-specific geometries, chamber pressures, and material properties. Designed to overcome the limitations of current emulators, \textit{HeartSimSage} can effectively handle diverse three-dimensional (3D) biventricular geometries, mesh topologies, fiber directions, structurally based constitutive models, and physiological boundary conditions. It also supports flexible mesh structures, allowing variable node count, ordering, and element connectivity. To optimize information propagation, we developed a neighboring connection strategy inspired by Graph Sample and Aggregate (GraphSAGE) that prioritizes local node interactions while maintaining mid-to-long-range dependencies. Additionally, we integrated Laplace-Dirichlet solutions to enhance spatial encoding and employed subset-based training to improve computational efficiency. Incorporating the attention mechanism allows \textit{HeartSimSage} to adaptively weigh neighbor contributions and filter out irrelevant information flow, enhancing prediction accuracy. As a result, \textit{HeartSimSage} achieves a computational speedup of approximately 13000X on a GPU and 190X on a CPU compared to traditional FEA while maintaining a nominal averaged error of $0.13\% \pm 0.12\%$ in predicting biventricular myocardial displacements. We validated our model on a published left ventricle dataset and analyzed the model’s sensitivity to hyperparameters, neighboring connection strategies, and the attention mechanism.
\end{abstract}



\end{frontmatter}


\newpage 

\section{Introduction}

Computational modeling is poised to become a quintessential tool for studying cardiovascular disease, biomechanics and mechanobiology, and clinical applications~\cite{trayanova2011electromechanical, campbell2011multi, marsden2015multiscale, palit2015computational, mittal2016computational, sotiropoulos2016fluid, vedula2017method, finsberg2018efficient, lee2018spatial, pfaller2019importance, yoshida2021computational, schwarz2023beyond, green2024myocardial}. By integrating patient imaging data with advanced numerical simulations, these \textbf{digital twins} provide valuable insights into the mechanisms of cardiac disease and function, enable biomarker identification, assessment of structural and functional abnormalities, and prediction of patient-specific outcomes~\cite{corral2020digital, qureshi2023imaging, qureshi2022mechanistic, bazzi2022experimental, crozier2016image, marx2022robust, balaban2017high, gonzalo2024multiphysics}. Beyond diagnostics, computational modeling plays a vital role in treatment planning, surgical optimization, and medical device design by simulating interventions before actual clinical application~\cite{mangion2018advances, boyle2019computationally, niederer2021scaling}. These advancements contribute to \textbf{precision medicine}, reducing the reliance on trial-and-error or population-based approaches and facilitating personalized treatment strategies~\cite{corral2020digital, stimm2022personalization, arevalo2016arrhythmia}.

Current cardiac biomechanics modeling workflows, particularly simulating the mechanics of the myocardium (heart tissue), rely primarily on finite element analysis (FEA) among other numerical solvers~\cite{finsberg2018efficient, marx2022robust, gillette2022personalized, brown2024modular, shi2024optimization, jiang2024highly, feng2024whole}. These methods require solving complex partial differential equations over high-resolution, patient-specific geometries, often demanding hours to days of computation on high-performance computing clusters~\cite{finsberg2018efficient, marx2022robust, shi2024optimization, augustin2016anatomically, bucelli2023mathematical}. This computational bottleneck limits their feasibility for clinical translation and performing large-scale studies. There is an imminent need for more scalable and efficient modeling techniques, such as data-driven approaches, which can significantly reduce computational costs while maintaining accuracy and robustness~\cite{dalton2022emulation, salvador2023fast, salvador2024digital}.

Machine Learning (ML) has emerged as a promising alternative for accelerating cardiovascular biomechanics modeling, offering data-driven surrogates for traditional FEA. Various ML techniques, including classic probabilistic models (e.g., Gaussian/Bayesian models)~\cite{yun2014statistical, tepole2022data} and neural networks~\cite{quarteroni2025combining, peirlinck2024universal, sahli2020physics}, have been explored. While Gaussian processes and Bayesian methods allow quantifying the effects of uncertain inputs~\cite{melis2017bayesian, ranftl2023bayesian}, they are computationally expensive and often underperform with the high-dimensional, nonlinear, and anisotropic nature of soft tissue mechanical characteristics~\cite{khatibisepehr2013design, liu2023bayesian, sena2021bayesian}. In contrast, neural network-based models are robust, computationally efficient, and well-suited for developing direct emulators~\cite{dalton2022emulation, dalton2023physics} and reduced-order models~\cite{regazzoni2022machine, cicci2024efficient, liang2023synergistic}.

Several neural network architectures have been applied to cardiovascular modeling, including convolutional neural networks (CNNs), diffusion models, physics-informed neural networks (PINNs), and graph neural networks (GNNs). CNNs have demonstrated success in medical image processing applications~\cite{maher2019accelerating, su2020generating, ronneberger2015u, liu2018feature, chen2016deep, sveinsson2024seqseg} but are fundamentally limited by their inability to handle irregular meshes and account for spatial relationships present in finite element models. Diffusion models have shown the potential to generate high-fidelity synthetic data~\cite{du2024conditional, kadry2024diffusion, guo2024diffusion} but are not inherently designed to enforce physical constraints, such as the mechanics of soft tissues. Physics-Informed Neural Networks (PINNs) incorporate physical constraints into neural network training~\cite{kissas2020machine, buoso2021personalising, caforio2024physics, arzani2021uncovering, perdikaris2024probabilistic, raissi2024physics}. However, their ability to handle large deformations, complex geometries, heterogeneous material properties, and intricate boundary conditions remains limited, and they are often challenging to train with constrained accuracy~\cite{sahli2020physics, garay2024physics}. While PINNs have been explored for biomechanics applications~\cite{hu2024physics, haghighat2020deep, haghighat2021physics, aghaee2025pinning}, these challenges hinder their applicability to patient-specific cardiac mechanics modeling.

GNNs offer a natural and inherently well-suited framework for cardiac mechanics modeling, as they generalize CNNs to non-Euclidean domains, such as the unstructured grids typically employed in cardiac mechanics modeling using FEA~\cite{pak2023patient, saleh2024physics, salehi2022physgnn, dalton2023physics}. GNNs effectively capture spatial and topological relationships within cardiac structures, enabling efficient and accurate surrogate modeling of high-fidelity FEA simulations. By leveraging learned representations of cardiac geometry and mechanical behavior, GNN-based approaches substantially reduce computational costs while preserving predictive accuracy. This enhanced efficiency enables scalable, patient-specific modeling with the potential to integrate additional clinical variables, such as patient demographics, genetic factors, or patient clinical history, to uncover complex relationships across diverse patient populations.

Despite significant progress, many existing neural network-based models remain constrained to simplified tasks and lack the versatility necessary for comprehensive cardiac biomechanics simulations. A common limitation among these models is their reliance on fixed geometries or fixed mesh topologies, with static node connections and predetermined structures~\cite{dalton2022emulation, dalton2023physics}, restricting their ability to adapt to heart models constructed from patient images. Further development is required to fully integrate with personalized modeling pipelines and capture realistic cardiac mechanics. Current neural network technologies lack the ability to simultaneously (1) accommodate diverse three-dimensional (3D) cardiac geometries across patient groups, ensuring adaptability to anatomical and pathological variations, (2) integrate heterogeneous tissue properties (e.g., spatially varying fiber directions, constitutive model parameters), structurally-based constitutive models, and boundary conditions vital to reproduce physiologically relevant myocardial deformations, and (3) support dynamic node topologies, where the number of nodes, the nodal ordering, and connectivity can change flexibly. These limitations pose challenges for employing these data-driven models to create personalized models of cardiac biomechanical function in health and disease.

\begin{figure}[h] 
 \begin{center}
 \includegraphics[width=0.75\textwidth]{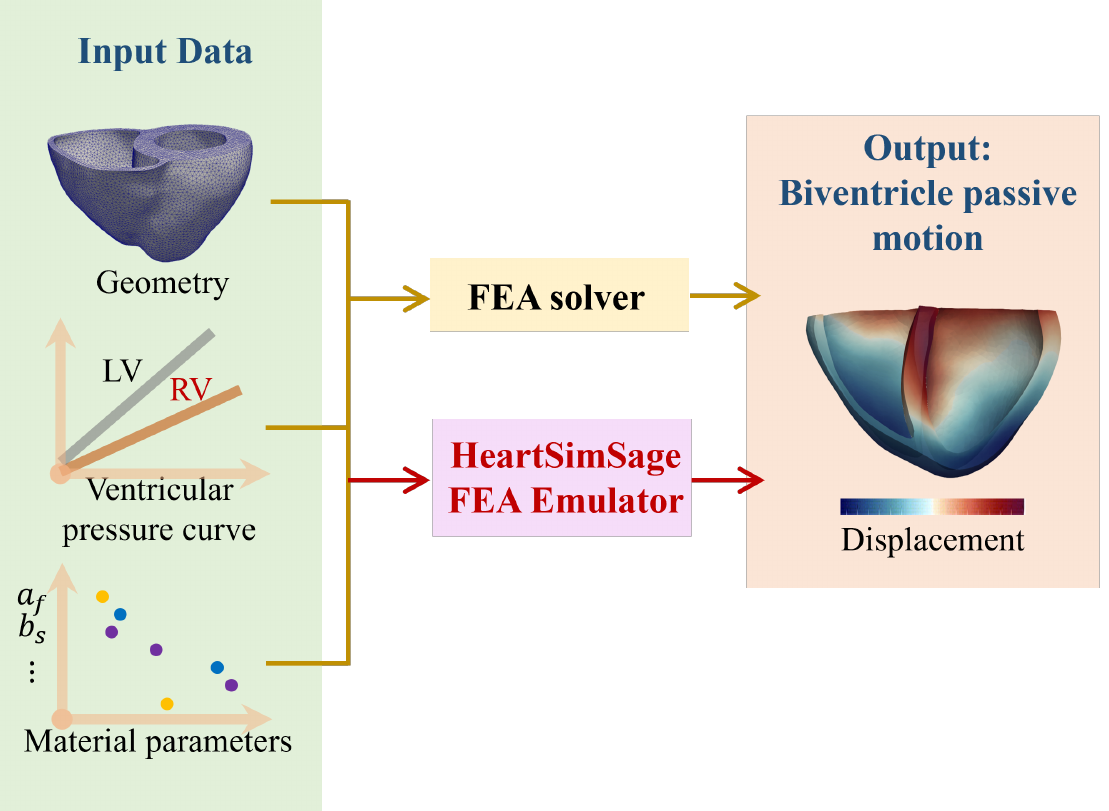}
 \end{center}
 \caption{Illustration of \textit{HeartSimSage}, a biventricular finite element analysis (FEA) emulator to predict passive deformations from an image-based geometry, chamber pressures, and material properties including local fiber directions.}
 \label{fig_whole_structure}
\end{figure}

To address these challenges, we developed an FEA emulator, \textit{HeartSimSage}, with an attention-enhanced GNN architecture designed to predict passive biventricular myocardial deformation under dynamic loading from the reference geometry, material properties, and chamber pressures (Fig.~\ref{fig_whole_structure}). Our approach adapts to intricate 3D cardiac anatomies by incorporating spatially varying material characteristics (e.g., fiber directions), advanced constitutive relations, and clinically relevant boundary representations. Additionally, it accommodates a flexible node topology, allowing dynamic variations in node count, ordering, and connectivity, thereby avoiding the need to retrain the neural network model for changes in the input geometry and mesh configuration frequently encountered in patient-specific modeling workflows.

Within the neural network framework, we incorporated Laplace-Dirichlet solutions encoding to enhance the representation of relative node positions and employed subset-based training to improve computational efficiency. A transformer-type attention mechanism~\cite{vaswani2017attention} refines node and edge interactions, allowing the model to effectively capture tissue heterogeneity and nonlinear deformations while filtering irrelevant information. Additionally, Graph Sample and Aggregate (GraphSAGE)~\cite{liu2020graphsage, oh2019advancing, hamilton2017inductive} improves scalability through inductive learning, prioritizing local node interactions, and preserving long-range dependencies, enabling generalization across patient-specific heart geometries without exhaustive graph traversal. 

By integrating Laplace-Dirichlet solutions encoding, subset training, attention mechanisms, and GraphSAGE, \textit{HeartSimSage} provides a versatile and efficient alternative to traditional physics-based solvers, capturing spatial dependencies in cardiac structures while significantly reducing computational costs without compromising accuracy. Therefore, our \textit{HeartSimSage} framework supports creating robust and efficient cardiac digital twins while laying the groundwork for AI-driven, generalizable solvers applicable to a broad range of mesh-dependent physics-based problems.

The manuscript is organized as follows: Section 2 describes the data acquisition and augmentation processes, the neural network architecture, and the training strategy. Section 3 presents the results, including dataset details, prediction outcomes, and the attention-derived weights. Section 4 investigates the effects of hyperparameters, Laplace-Dirichlet solutions, node connection strategies, and the attention mechanism on model performance. We further assess the model's generalizability using a previously published dataset and discuss its limitations along with potential directions for future research. 

\section{Methods} 

\subsection{Data acquisition}\label{sec:data_acquisition}

\begin{figure}[h] 
 \begin{center}
 \includegraphics[width=1\textwidth]{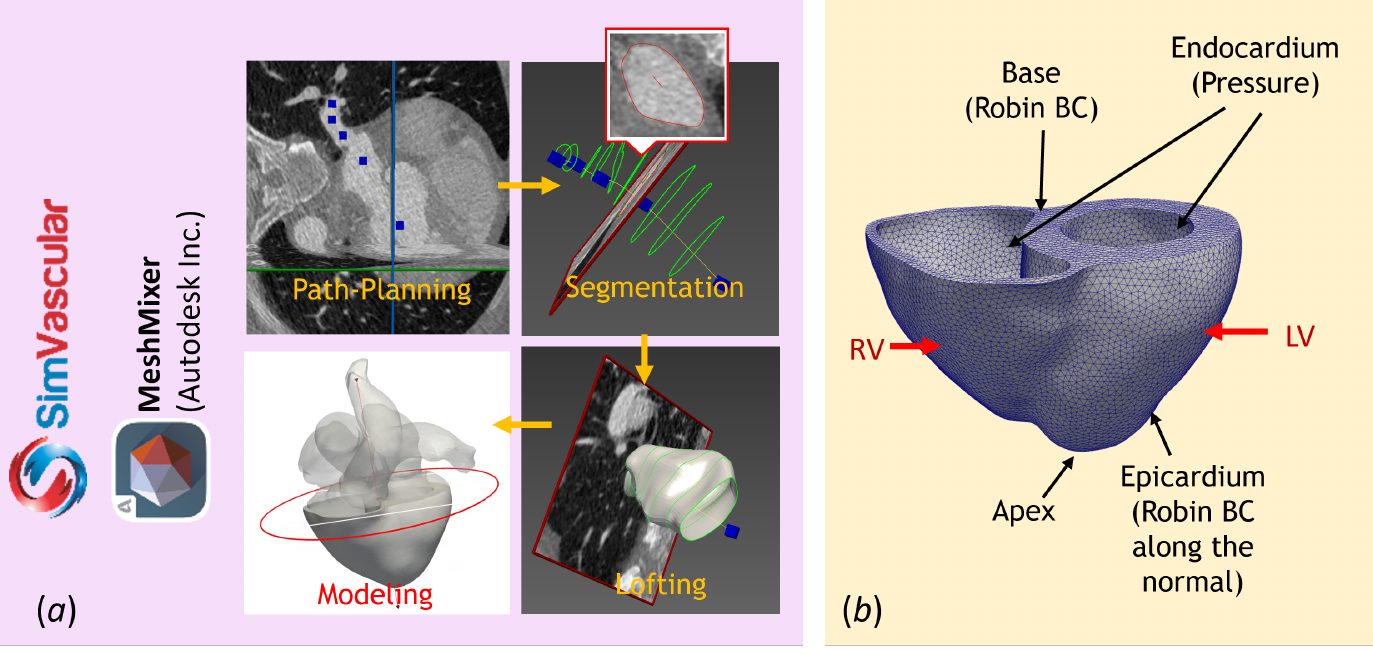}
 \end{center}
 \caption{(a) Workflow for segmenting the biventricular myocardial geometry and creating an FEA-ready mesh from the patient's three-dimensional (3D) image volume (CT/MRI). (b) The biventricular FEA-ready mesh includes labeled surfaces and typically consists of approximately 15K–20K triangular elements for the surfaces and approximately 86K–105K tetrahedral elements for the myocardial volume.}
 \label{fig_mesh}
\end{figure}

We used 150 3D image volumes (95 CT and 55 MR) from the recently published MMWHS dataset~\cite{gao2023bayeseg, zhuang2018multivariate}, the left atrial wall thickness challenge~\cite{karim2018algorithms}, the left atrial segmentation challenge~\cite{tobon2015benchmark}, and the orCalScore challenge~\cite{wolterink2016evaluation}. The selection criteria required the images to clearly delineate the left and right ventricles in the field of view. We then segmented the biventricular myocardium model from the 3D images using our previously established segmentation workflow~\cite{shi2024optimization} with the open-source software SimVascular\footnote{\url{https://simvascular.github.io/}}~\cite{updegrove2017simvascular} and Meshmixer\footnote{Autodesk Inc., \url{https://www.meshmixer.com/}, version: 3.5} (Fig.~\ref{fig_mesh}a). The model creation pipeline involved creating paths and 2D segmentations to loft the endocardial and epicardial surfaces in SimVascular. The surfaces are subsequently smoothed and decimated in Meshmixer to refine the triangulated geometry, followed by a Boolean operation to extract the thick myocardium. Finally, the smoothed model was meshed using TetGen\footnote{\url{https://wias-berlin.de/software/tetgen/}}~\cite{hang2015tetgen} to produce an FEA-ready mesh (Fig.~\ref{fig_mesh}b).

\subsection{Biventricular passive mechanics description}\label{sec:biv_passive_mech}

Here, we briefly formulated the biventricular passive mechanics problem based on our recent works~\cite{shi2024optimization, arostica2025software, shi2025leftatrium}. Additional details are provided in \ref{sec:passive-mechanics-description}. We employed a stabilized finite element formulation to model the biventricular passive mechanics, leveraging its ability to resolve near-incompressible deformations using equal-order basis functions for velocity and pressure degrees of freedom~\cite{liu2018unified}. This method is particularly appealing for complex patient-specific myocardial geometries typically meshed using linear tetrahedral elements, which are otherwise challenging to represent using higher-order elements. In the biventricular myocardium domain $\Omega_{\mathbf{x}}$, the governing equations are expressed as,

\begin{subequations}
\begin{align}
  \frac{d\mathbf{u}}{dt} - \mathbf{v} = & \ \mathbf{0} \quad \mathrm{in} \ \Omega_\mathbf{x} \label{eq_kin}\\
  \beta_{\theta}(p)\frac{dp}{dt} + \nabla_{\mathbf{x}}\cdot\mathbf{v} = & \ 0 \quad \mathrm{in} \ \Omega_\mathbf{x} \label{eq_mass} \\
  \rho(p)\frac{d\mathbf{v}}{dt} + \nabla_\mathbf{x} p - \nabla_\mathbf{x} \cdot \pmb{\sigma}_{dev} = & \ \mathbf{0} \quad \mathrm{in} \ \Omega_\mathbf{x} \label{eq_linMom}
\end{align}
\label{eq_govEqs}
\end{subequations}

\noindent where, $\mathbf{u}$ is the displacement, $\mathbf{v}$ is the velocity, $\pmb{\sigma}_{dev} := J^{-1}\overline{\mathbf{F}} \big(\mathbb{P}:\overline{\mathbf{S}}\big) \overline{\mathbf{F}}^{T} + 2\mu_v \mathrm{dev}[\mathbf{d}]$, $\mu_v$ is the dynamic shear viscosity, $\mathrm{dev}[\mathbf{d}]$ is the deviatoric part of the rate of deformation tensor, $\mathbf{d} := \frac{1}{2}\big(\nabla_\mathbf{x} \mathbf{v} + (\nabla_\mathbf{x} \mathbf{v})^T\big)$, and $\mathbb{P} = \mathbb{I} - \frac{1}{3}\big(\mathbf{C}^{-1} \otimes \mathbf{C}\big) $ is the projection tensor. The first term for $\pmb{\sigma}_{dev}$ represents the isochoric elastic stress, while the second term is the viscous shear stress.

Domain boundary is defined as $\Gamma_\mathbf{x} := \Gamma_{\mathrm{base}}\cup\Gamma_{\mathrm{epi}}\cup\Gamma_{\mathrm{endo-lv}}\cup\Gamma_{\mathrm{endo-rv}}$, as the union of the basal plane ($\Gamma_{\mathrm{base}}$), epicardium ($\Gamma_{\mathrm{epi}}$), and endocardium of the left and the right ventricles ($\Gamma_{\mathrm{endo-lv}}$, $\Gamma_{\mathrm{endo-rv}}$). Physiologically relevant boundary conditions are applied to the myocardial boundaries as,

\begin{equation}
\begin{split}
  \pmb{\sigma}\mathbf{\hat{n}} = & - p_{\mathrm{lv}}\mathbf{\hat{n}} \quad \mathrm{on} \ \Gamma_{\mathrm{endo-lv}} \\
  \pmb{\sigma}\mathbf{\hat{n}} = & - p_{\mathrm{rv}}\mathbf{\hat{n}} \quad \mathrm{on} \ \Gamma_{\mathrm{endo-rv}} \\
  \pmb{\sigma}\mathbf{\hat{n}} + \big(k_\mathrm{epi}\mathbf{u}\cdot\mathbf{\hat{n}_0} + c_\mathrm{epi}\mathbf{v}\cdot\mathbf{\hat{n}_0}\big)\mathbf{\hat{n}_0} = & -p_\mathrm{epi} \mathbf{\hat{n}} \quad \mathrm{on} \ \Gamma_{\mathrm{epi}} \\
  \pmb{\sigma}\mathbf{\hat{n}} + k_\mathrm{base}\mathbf{u} + c_\mathrm{base}\mathbf{v} = & \ \mathbf{0} \quad \mathrm{on} \ \Gamma_{\mathrm{base}}
\end{split}
\label{eq_bcs}
\end{equation}
where $\pmb{\sigma} := \pmb{\sigma}_{dev} - p\mathbf{I}$ is the Cauchy stress. $k_{(\bullet)}$ and $c_{(\bullet)}$ are the stiffness and damping coefficients of the Robin boundaries, respectively, $p_\mathrm{lv}$ and $p_{\mathrm{rv}}$ are the hemodynamic pressures applied to the left and right endocardium, and $p_\mathrm{epi}$ is the thoracic cavity pressure acting on the epicardium, which is typically small and set to zero. Additionally, $\mathbf{\hat{n}_0}$ and $\mathbf{\hat{n}}$ are the unit surface normal vectors in the reference and updated configurations. Robin-type boundaries are employed to model the effects of the pericardium, constraining the motion on the epicardium and the basal plane~\cite{pfaller2019importance, arostica2025software}. The equations (Eqs. (\ref{eq_govEqs}-\ref{eq_bcs})) complete the description of the initial-boundary value problem for the biventricular myocardial passive mechanics.

The myocardium is modeled as a hyperelastic, nearly incompressible material, for which the stress-strain relation is approximated using the orthotropic modified Holzapfel-Ogden (HO) model~\cite{holzapfel2009constitutive, nolan2014robust}. The isochoric Gibbs free energy per unit mass is expressed as follows,

\begin{equation}
\begin{split}
    G_\mathrm{iso}^{\mathrm{HO}}(\overline{\mathbf{C}}) & = \frac{a}{2b}\exp \{b(\overline{I}_1 - 3)\} \\
    & + \sum_{i\in {f,s}} \frac{a_i}{2b_i}\chi(I_{4,i})\bigg(\exp \{b_i (I_{4,i} - 1 )^2\} - 1\bigg) \\
    &  +  \frac{a_{fs}}{2b_{fs}} \bigg(\exp \{b_{fs}I^2_{8fs}\} - 1\bigg) 
\end{split}
\label{eq_const_ho}
\end{equation}
where we incorporated a smooth sigmoid function, $\displaystyle \chi(\eta) := \frac{1}{1+\exp [-k_\chi (\eta - 1)]}$, to approximate the Heaviside function, which prevents numerical instabilities under small compressive strains~\cite{shi2024optimization, arostica2025software}. The set $\{a,\ b,\ a_f,\ b_f,\ a_s,\ b_s,\ a_{fs},\ b_{fs}\}$ defines the parameters for the HO model. Further, $\overline{I}_1$ ($:=\mathrm{tr}(\overline{\mathbf{C}})$) is the isotropic invariant that captures isochoric deformations, $I_{4f}$ ($:= \mathbf{f}_0\cdot \mathbf{C}\mathbf{f}_0$) and $I_{4s}$ ($:= \mathbf{s}_0\cdot \mathbf{C}\mathbf{s}_0$) are the transverse invariants for the fiber and sheet directions, respectively, and $I_{8fs}$ ($:= \mathbf{f}_0\cdot \mathbf{C}\mathbf{s}_0$) is the anisotropic invariant that captures the fiber-sheet interactions. $\mathbf{f}_0$ and $\mathbf{s}_0$ correspond to the local longitudinal fiber and sheet orientations. Please refer to~\ref {sec:passive-mechanics-description} for a detailed description of the passive mechanics formulation, including the kinematic relations, balance laws, and specific forms of Gibbs free energies for the adopted constitutive models.

\subsection{Laplace-Dirichlet solutions}\label{sec:laplace}

We hypothesized that the relative positioning of each node plays a crucial role in the physics of the problem. To capture this relative positional information between the nodes, we solved a series of Laplace-Dirichlet problems on the biventricular domain. We later compared our method against other approaches (Sec.~\ref{sec:cmp_laplace}), such as the one-hot boundary values~\cite{dalton2022emulation}, adopted by other studies. The generic structure of the Laplace-Dirichlet problem on the biventricular domain, $\Omega_\mathbf{x}$, is given as,

\begin{subequations}
\begin{align}
  \nabla^2\Phi = & 0 \quad \mathrm{in} \quad \Omega_\mathbf{x} \label{eq_lap}\\
  \Phi = & 1 \quad \mathrm{on} \quad \Gamma_a \\
  \Phi = & 0 \quad \mathrm{on} \quad \Gamma_b \\ 
  \nabla \Phi \cdot \mathbf{n} = & 0 \quad \mathrm{on} \quad \Gamma_n 
\end{align}
\label{eq_laplace}
\end{subequations}
employing generic partitions of the domain boundary $\Gamma_a, \Gamma_b, \Gamma_n$ with $\Gamma_a\cup \Gamma_b \cup \Gamma_n = \partial \Omega_\mathbf{x} $. The specific Laplace-Dirichlet problems solved are listed in Table~\ref{table_laplace}. 

\begin{table}[H]
\begin{center}
\caption{Laplace-Dirichlet problems are solved to characterize the relative positions of nodes within the biventricular myocardium and evaluate longitudinal and transverse fiber directions across the tissue.}
\label{table_laplace}
\vspace{4pt}
\begin{tabular}{c c c}
\toprule
$\Phi$ & $\Gamma_a$ & $\Gamma_b$ \\
\arrayrulecolor{black!30}\midrule
$\Phi_{AB}$ & $\Gamma_{\mathrm{base}}$ & $\Gamma_{\mathrm{apex}}$ \\
$\Phi_{EP}$ & $\Gamma_{\mathrm{epi}}$ & $\Gamma_{\mathrm{endo-lv}} \cup \Gamma_{\mathrm{endo-rv}}$ \\
$\Phi_{LV}$ & $\Gamma_{\mathrm{endo-lv}}$ & $\Gamma_{\mathrm{endo-rv}} \cup \Gamma_{\mathrm{epi}}$ \\
$\Phi_{RV}$ & $\Gamma_{\mathrm{endo-rv}}$ & $\Gamma_{\mathrm{endo-lv}} \cup \Gamma_{\mathrm{epi}}$ \\
$\Phi_{LV-RV}$ & $\Gamma_{\mathrm{endo-lv}}$ & $\Gamma_{\mathrm{endo-rv}}$ \\
$\Phi_{LV-EP}$ & $\Gamma_{\mathrm{endo-lv}}$ & $\Gamma_{\mathrm{epi}}$ \\
$\Phi_{RV-EP}$ & $\Gamma_{\mathrm{endo-rv}}$ & $\Gamma_{\mathrm{epi}}$ \\
\arrayrulecolor{black}\bottomrule
\end{tabular}
\end{center}
\end{table}

\begin{figure}[H]
 \begin{center}
 \includegraphics[width=1\textwidth]{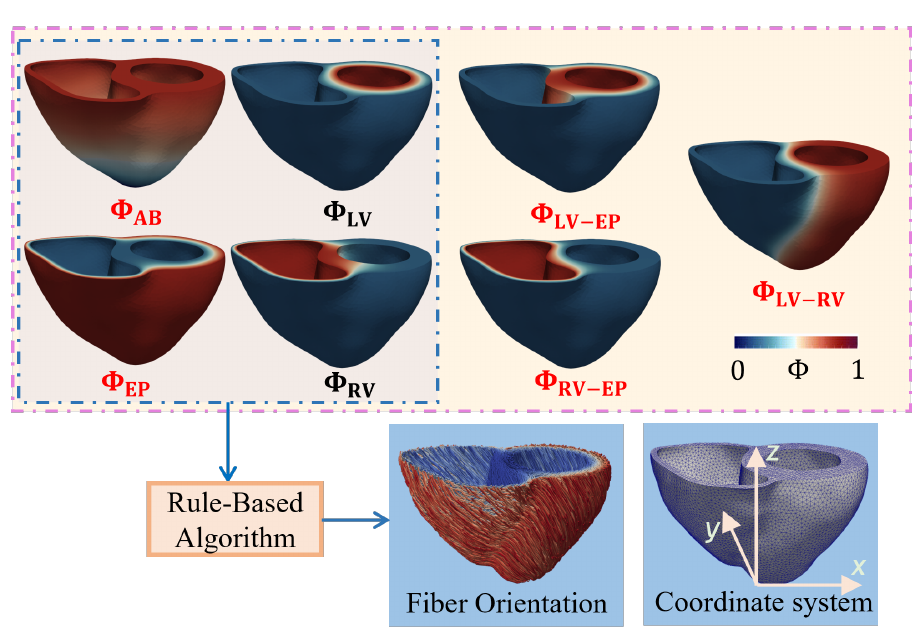}
 \end{center}
 \caption{The solutions to Laplace-Dirichlet problems (defined in Table~\ref{table_laplace}) in a representative biventricular myocardium. Solution fields labeled in bright red ($\Phi_{AB}, \ \Phi_{EP}, \ \Phi_{LV-EP}, \ \Phi_{RV-EP}, \ \Phi_{LV-RV}$) are used as the node feature to characterize relative nodal positions within the myocardium. Solution fields in the bounding box ($\Phi_{AB}, \ \Phi_{EP}, \ \Phi_{LV}, \ \Phi_{RV}$) are used to create longitudinal and transverse fiber orientations across the tissue using a rule-based algorithm~\cite{shi2024optimization, bayer2012novel}. All the biventricular geometries are then reoriented to a common coordinate system with the origin at the ventricular apex, the $x-y$ plane oriented parallel to the base, and the $z$ direction oriented perpendicularly along the apico-basal direction.}
 \label{fig_laplace}
\end{figure}

The solutions to these Laplace-Dirichlet boundary-valued problems are used as node features that characterize the relative positions of nodes within the biventricular myocardium domain (see Fig.~\ref{fig_laplace}). Specifically, $\Phi_{AB}$ characterizes the relative distance of a node along the apicobasal direction, with higher values indicating proximity to the base and lower values indicating proximity to the apex. $\Phi_{EP}$ represents the relative position of a node along the transmural direction, where higher values correspond to locations closer to the epicardium or away from the endocardium. $\Phi_{LV-RV}$ identifies the relative positioning of a node between the two ventricles, with higher values corresponding to the left ventricle and lower values to the right ventricle. Additionally, $\Phi_{LV-RV}$, combined with $\Phi_{LV-EP}$ and $\Phi_{RV-EP}$ helps differentiate nodes located in the septum.

Further, $\Phi_{LV}$ and $\Phi_{RV}$ are also computed and combined with $\Phi_{AB}$ and $\Phi_{EP}$ to determine myocardial longitudinal and transverse fiber orientations using a rule-based method~\cite{shi2024optimization, bayer2012novel}. The longitudinal fibers ($\mathbf{f}_0$) range from $40^{\circ}$ at the endocardium to $-50^{\circ}$ at the epicardium, where the fiber angle is defined as the helical angle relative to the counterclockwise circumferential direction when viewed from the basal plane toward the apex. Likewise, the transverse fiber direction ($\mathbf{s}_0$), or sheet angle, varies from $-65^{\circ}$ at the endocardium to $25^{\circ}$ at the epicardium, with the sheet angle measured relative to the local outward transmural axis, which is perpendicular to the longitudinal fiber direction~\cite{bayer2012novel}. The sheet-normal vector ($\mathbf{n}_0$) is orthonormal to both longitudinal and transverse fiber directions. Our definitions of the fiber $\mathbf{f}_0$, sheet $\mathbf{s}_0$, and sheet-normal $\mathbf{n}_0$ directions align with the $F$, $T$, and $S$ vectors, respectively, as denoted in Bayer et al.~\cite{bayer2012novel}.

After computing the Laplace-Dirichlet solutions for all 150 image-based biventricular geometries employed in the study cohort, we reoriented the meshes into a common coordinate system with its origin centered at the biventricular apex. The positive $z$ direction ($\mathbf{\hat{k}}$) is defined along the normal vector to the base surface representing the apico-basal direction. The positive $x$ direction ($\mathbf{\hat{i}}$) is defined as the vector from the maximum $\Phi_{RV-EP}$ to the maximum $\Phi_{LV-EP}$ on the basal plane, representing the direction passing through the septum from the right ventricle to the left ventricle. The positive $y$ direction ($\mathbf{\hat{j}}$) is set as orthonormal to $\mathbf{\hat{i}}$ and $\mathbf{\hat{k}}$ as, $\mathbf{\hat{j}} = \mathbf{\hat{k}} \times \mathbf{\hat{i}}$ (see Fig.~\ref{fig_laplace}).

\subsection{Data augmentation \& simulation}\label{sec:data_augmentation_sim}

We performed data augmentation on the 150 patient image-based FEA models to increase anatomical variability and improve the generalizability of the downstream learning models. We generated 4,850 meshes by creating pseudo-geometries via kinematic variations and randomly varying material parameters and pressure boundary conditions within reasonably chosen bounds, bringing the total training dataset sample size to 5,000. 

Specifically, we applied random scaling along each coordinate axis (-15\% to 15\%) and random shearing ($-10^{\circ}$ to $10^{\circ}$) in $\mathbf{\hat{i}} - \mathbf{\hat{j}}$ plane on the reoriented biventricle, in addition to elastic deformations~\cite{simard2003best}. For the elastic deformations, 10 control points are placed along each dimension of the enclosing bounding box. At each control point, random displacement perturbations (less than 10\% of the bounding box size) are applied simultaneously. These perturbations served as boundary displacements, and the resulting deformations are computed by solving a linear elastic problem under Dirichlet boundary conditions. The meshes are then warped based on the displacements of the control points using B-spline interpolation. During this process, the base surface was kept flat, with nodes on the basal plane constrained to move together along the z-axis. A set of 40 geometrical parameters $\pmb{\xi}$, characterizing the global shape of the biventricular myocardium, was obtained for each case using a customized Python code. Additional details of each shape parameter are provided in~\ref{sec:shape} and Table~\ref{table_shape-cooeficients}.

For each of the 5000 biventricular geometries created using the above method, a randomly selected set of HO constitutive model parameters, $\pmb{\beta} = (a, b, a_f, b_f, a_s, b_s)$, which characterize the orthotropic myocardial mechanical behavior~\cite{holzapfel2009constitutive}, was chosen from the range $\mathbf{B}$ shown in Table~\ref{table_mater_bounds}. These parameter ranges are defined based on our previous work for simulating passive myocardial mechanics~\cite{shi2024optimization}. The parameters $a$, $a_f$, and $a_s$ are generated on a log scale to prevent bias toward the highest order values~\cite{beauchamp1973corrections}. Note that $a_{fs}$ and $b_{fs}$ are kept constant as they have less influence on the overall myocardial deformation~\cite{shi2024optimization}. Additionally, randomly chosen chamber pressures (pressure feature $\pmb{p} = (p_{\mathrm{lv}}, p_{\mathrm{rv}}) \in \mathbf{P}$, Table~\ref{table_mater_bounds}) within a physiologically relevant range~\cite{msd_heart_pressures, shi2024optimization} are applied to the ventricular endocardial surfaces ($\Gamma_{\mathrm{endo-lv}}, \Gamma_{\mathrm{endo-rv}}$). All other simulation parameters are fixed (Table~\ref{table_fix_params_biv}).

\begin{table}[H]
\begin{center}
\footnotesize
\caption{Constitutive model parameter and pressure ranges used for data augmentation.}
\label{table_mater_bounds}
\vspace{4pt}
\begin{tabular}{c c c c c c c c}
\toprule
$a$ & $b$ & $a_f$  & $b_f$ & $a_s$ & $b_s$ & $p_{\mathrm{lv}}$ & $p_{\mathrm{rv}}$ \\
\arrayrulecolor{black!30}\midrule
$ 10^2-10^4$ & $1-50$ & $10^3-10^6 $ & $1-50$ & $10^2-10^5 $ & $1-50$ & $6-15$ & $(0.3 - 0.8) p_{\mathrm{lv}}$    \\
\arrayrulecolor{black}\bottomrule
\end{tabular}
\end{center}
\begin{tablenotes}
    \centering
    \vspace{-12pt}
    \footnotesize
    \item $a_{(\cdot)}$: $\text{dyn}/\text{cm}^2$; $b_{(\cdot)}$ is dimensionless; $p_{(\cdot)}$: $\text{mmHg}$.
\end{tablenotes}
\end{table}

\begin{table}[ht]
\begin{center}
\footnotesize
\caption{FEA model parameters that are left unchanged for simulating myocardial mechanics for all the 5000 cases created using data augmentation.}
\label{table_fix_params_biv}
\vspace{4pt}
\begin{tabular}{c c c c c c c c c c} 
\toprule
$\rho_0$ & $k_{\chi}$  & $\kappa$ & $\mu_v$ & $k_{\mathrm{base}}$ & $c_{\mathrm{base}}$ & $k_{\mathrm{epi}}$ & $c_{\mathrm{epi}}$    &    $a_{fs}$  & $b_{fs}$   \\
\arrayrulecolor{black!30}\midrule
$1.055$ & 100  & $10^6$  & $10^3$ & $10^4$ & $500$ & $10^4$  & $500$ & $2160$ & $11.436$  \\
\arrayrulecolor{black}\bottomrule
\end{tabular}
\end{center}
\begin{tablenotes}
    \centering
    \vspace{-12pt}
    \footnotesize
    \item $\rho_0$: $\text{g}/\text{cm}^3$; $k_{\chi}$: dimensionless; $\kappa$: $\text{dyn}/\text{cm}^2$; $\mu_v$: $\text{dyn-s}/\text{cm}^{2}$; $k_{(\cdot)}$: $\text{dyn}/\text{cm}^3$; 
    \item $c_{(\cdot)}$: $\text{dyn-s}/\text{cm}^3$; $a_{fs}$: $\text{dyn}/\text{cm}^2$; $b_{fs}$ is dimensionless.
\end{tablenotes} 
\end{table}

We simulated all 5000 cases using our in-house multiphysics finite element solver derived from the open-source solver, \textit{svFSI}~\cite{zhu2022svfsi}, and has been benchmarked for cardiac mechanics simulations~\cite{shi2024optimization, arostica2025software}. The final displacement field for each case was obtained as the ground truth label.

\subsection{Neural network structure }\label{sec:neural_net_structure} 

\begin{figure}[h] 
 \begin{center}
 \includegraphics[width=1\textwidth]{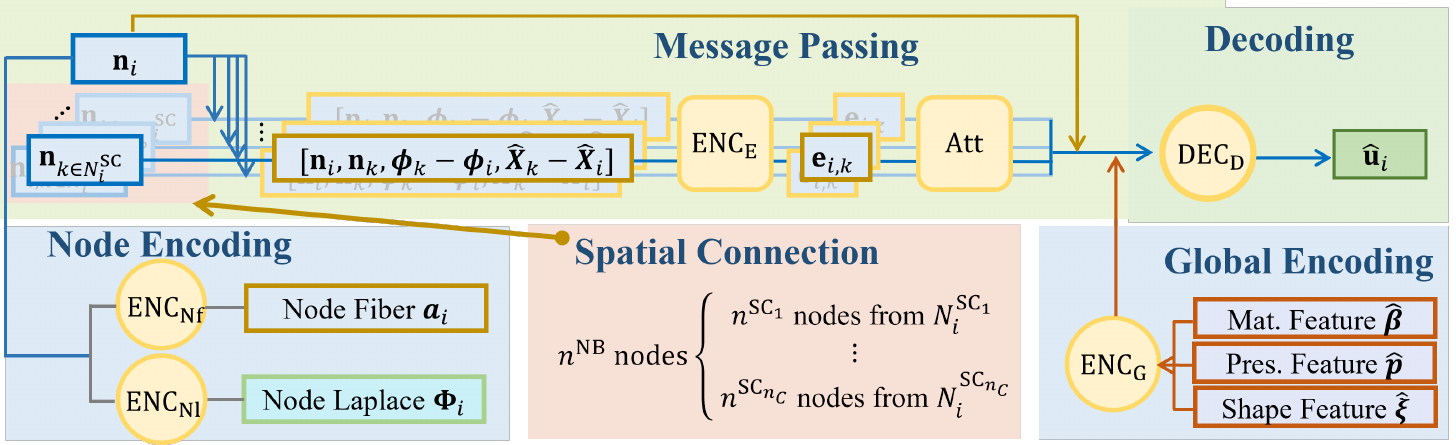}
 \end{center}
 \caption{Schematic representation of the \textit{HeartSimSage} model architecture. The model predicts nodal displacements based on a combination of nodal features—such as fiber orientation and Laplace values—and global features, including material parameters, applied pressures, and shape descriptors. The architecture comprises five key modules: node encoding, spatial connection, message passing, global encoding, and decoding.}
 \label{fig_model_structure}
\end{figure}

The \textit{HeartSimSage} architecture (Fig.~\ref{fig_model_structure}) predicts the displacement $\mathbf{u}_i$ of a node $i$ using its nodal and global features, including self-features $\mathbf{h}_i$, the features of its selected neighboring nodes $\mathbf{h}_k$ ($k \in N^{\mathrm{NB}}_i \subseteq \mathcal{V}$, where $\mathcal{V}$ is the set of all nodes), and the global features $\mathbf{g}$. The node features include local fiber orientation, relative position from boundaries, and coordinates, while the global features comprise material parameters, chamber pressures, and shape parameters (Table~\ref{table_shape-cooeficients},~\ref{sec:shape}). Their relationship is given by:

\begin{equation}
    \mathbf{\hat{u}}_i = \mathbf{HSS}_{\pmb{\theta}}\Big(\mathbf{\hat{h}}_i, \{\mathbf{\hat{h}}_k\}_{k\in N^{\mathrm{NB}}_i}; \mathbf{\hat{g}} \Big)
    \label{equ:hss-node}
\end{equation}
where $\mathbf{HSS}_{\pmb{\theta}}$ denotes the \textit{HeartSimSage} neural network parameterized by $\pmb{\theta}$, $\mathbf{\hat{u}}_i$ represents the normalized displacement, $\mathbf{\hat{h}}_i$ is the normalized feature of node $i$, $\mathbf{\hat{h}}_k$ denotes the normalized features of neighboring nodes $k$ selected within the set of neighbors $N^{\mathrm{NB}}_i$, and $\mathbf{\hat{g}}$ represents the normalized global feature. Normalization in deep learning improves convergence rates, prevents diverging or vanishing gradients, enhances generalization, stabilizes training, and ensures robustness by scaling inputs to a consistent range, enabling efficient and stable learning~\cite{deshai2024deep, huang2023normalization, huang2022normalization, hinton2007learning, lecun2015deep}. In this work, we applied a parametric-specific max-min normalization~\cite{ali2014data} to the original data, as described in the following sections. 

The \textit{HeartSimSage} architecture consists of five key modules: node encoding, spatial connection, message passing, global encoding, and decoding. Below, we provide a detailed explanation of each component.

\subsubsection{Node encoding} 

The node encoding module transforms the features of node $i$ into a node embedding $\mathbf{n}_i$, capturing essential local information such as fiber orientations and relative positioning. This embedding is assumed to play a crucial role in predicting the displacement. Specifically, the fiber feature $\mathbf{a}_i$ and the Laplace solution features $\pmb{\Phi}_i$ for the node $i$ are defined as,

\begin{subequations}
    \begin{align}
        \mathbf{a}_i &=({\mathbf{f}_0}_i, \ {\mathbf{s}_0}_i) \\
        \pmb{\Phi}_i &=(\Phi_{AB}^i, \ \Phi_{EP}^i, \ \Phi_{LV-EP}^i, \ \Phi_{RV-EP}^i, \ \Phi_{LV-RV}^i)
    \end{align}
\end{subequations}
Note that the fiber orientations and Laplace solutions are already scaled within the range $[0, 1]$ and therefore do not require additional normalization. The fiber feature $\mathbf{a}_i$ and the Laplace features $\pmb{\Phi}_i$ are processed through distinct encoding functions (encoders), $\mathrm{ENC}_{\mathrm{Nf}}$ and $\mathrm{ENC}_{\mathrm{Nl}}$, respectively. These encoders, which are essentially multi-layer perceptrons (MLPs)~\cite{goodfellow2016deep, hinton2011transforming, lecun2015deep}, transform the raw input features into their corresponding embeddings, $\mathbf{a}'_i$ and $\pmb{\Phi}'_i$, as,

\begin{subequations}
    \begin{align}
        \mathbf{a}'_i &=\mathrm{ENC}_{\mathrm{Nf}}(\mathbf{a}_i) \\
        \pmb{\Phi}'_i &=\mathrm{ENC}_{\mathrm{Nl}}(\pmb{\Phi}_i) 
    \end{align}
\end{subequations}
where $\mathbf{a}'_i\in \mathbb{R}^{d_{\mathrm{emb}}}$ and $\pmb{\Phi}'_i\in \mathbb{R}^{d_{\mathrm{emb}}}$. $d_{\mathrm{emb}}$ is the dimension of the embeddings set as a constant ($d_{\mathrm{emb}}=128$) in this work unless specified. The two embeddings ($\mathbf{a}'_i, \ \pmb{\Phi}'_i$) are then summed to get the node embedding $\mathbf{n}_i \in \mathbb{R}^{d_{\mathrm{emb}}}$ as,

\begin{equation}
    \mathbf{n}_i = \mathbf{a}'_i + \pmb{\Phi}'_i
\end{equation}

\subsubsection{Spatial connection}\label{sec:spatial_connections} 

The likelihood of a node $i$ exchanging information with node $k$ is higher when their relative distance is smaller, which is particularly relevant for elastodynamic problems~\cite{dalton2022emulation}. At the same time, our experience with neighboring node connection schemes suggests that including short- and mid-to-long-range dependencies minimizes training and validation losses (Sec.~\ref{sec:cmp_node_conn}). As a result, even the most intuitive nearest neighbor type connection strategy underperforms with higher losses (Sec.~\ref{sec:cmp_node_conn}, Fig.~\ref{fig_cmp_node_conn}). 

We, therefore, introduce a partitioning mechanism based on the proximity neighborhood of the node $i$ to enhance computational efficiency and mitigate excessive memory usage due to large node counts. Specifically, we define multiple subsets of candidate neighboring nodes, $N^{\mathrm{SC}_{m}}_i$, based on the relative distances between nodes. We assume that the node $i$ exchanges information only with the nodes defined within the subset, $N^{\mathrm{SC}_{m}}_i$, which is essentially a vector of $n^{\mathrm{SC}_{m}}$ candidate neighboring nodes. We established five such sets, i.e., $m \in \{1,2,3,4,5\}$, as,

\begin{subequations}
    \begin{align}
        N^{\mathrm{SC}_1}_i & \subseteq \{k|k\in \mathcal{V}_{n^{\mathrm{SC}_1}}, \ d(i,k)\leq Q_{0.2\%} \} \\
        N^{\mathrm{SC}_2}_i & \subseteq \{k|k\in \mathcal{V}_{n^{\mathrm{SC}_2}}, \ Q_{0.2\%}\leq d(i,k)\leq Q_{1\%} \} \\
        N^{\mathrm{SC}_3}_i & \subseteq \{k|k\in \mathcal{V}_{n^{\mathrm{SC}_3}}, \ Q_{1\%}\leq d(i,k)\leq Q_{2\%} \} \\
        N^{\mathrm{SC}_4}_i & \subseteq \{k|k\in \mathcal{V}_{n^{\mathrm{SC}_4}}, \ Q_{2\%}\leq d(i,k)\leq Q_{5\%} \} \\
        N^{\mathrm{SC}_5}_i & \subseteq \{k|k\in \mathcal{V}_{n^{\mathrm{SC}_5}}, \ Q_{5\%}\leq d(i,k)\leq Q_{10\%} \} 
    \end{align}
    \label{eq_neiboring_set_partition}
\end{subequations}

\noindent where $Q_{p\%}$ represents the $p$ percentile of all pairwise distances $\{d(i,k) | i,k \in \mathcal{V}_{n^{\mathrm{SC}_i}}, i \neq k\}$. Based on our tuning experience, we set $n^{\mathrm{SC}_1}=20$, $n^{\mathrm{SC}_2}=30$, $n^{\mathrm{SC}_3}=30$, $n^{\mathrm{SC}_4}=10$, and $n^{\mathrm{SC}_5}=10$. The selected set of candidate neighboring nodes for node $i$ is then denoted as $N^{\mathrm{SC}}_i=N^{\mathrm{SC}_1}_i \cup N^{\mathrm{SC}_2}_i \cup N^{\mathrm{SC}_3}_i \cup N^{\mathrm{SC}_4}_i \cup N^{\mathrm{SC}_5}_i$. Finally, $n^{\mathrm{NB}}$ neighboring nodes are selected from $N^{\mathrm{SC}}_i$ to exchange information with each node $i$, and the selected set of neighboring nodes is denoted as $N^{\mathrm{NB}}_i$. 

Our tuning experience suggests $n^{\mathrm{NB}} = 12$ is sufficient for the biventricular myocardial geometries considered in this work. The current partitioning mechanism shares conceptual similarities with neighborhood sampling strategies used in GraphSAGE~\cite{liu2020graphsage, oh2019advancing, hamilton2017inductive} but is specifically designed for elastodynamic problems, where node connectivity is determined by spatial proximity rather than stochastic sampling. We have analyzed various schemes for designing the spatial connection sets in the Discussion section (Sec.~\ref{sec:cmp_node_conn}).

\subsubsection{Message passing}

The node $i$ exchanges information with each neighboring node $k$, selected based on the spatial connection scheme described in Sec.~\ref{sec:spatial_connections}, by concatenating their embeddings, the relative Laplace solution values, and the relative normalized distance as,

\begin{equation}
    \mathbf{\hat{e}}_{i,k} = (\mathbf{n}_i, \mathbf{n}_k, \pmb{\Phi}_k - \pmb{\Phi}_i, \mathbf{\hat{X}}_k - \mathbf{\hat{X}}_i)
\end{equation} 
where $\mathbf{\hat{X}}_i$ is the normalized quantity of the original coordinates $\mathbf{X}_i$. The resulting edge representative vector $\mathbf{\hat{e}}_{i,k}$ is encoded as an edge embedding via the edge encoder $\mathrm{ENC}_{\mathrm{E}}$,

\begin{equation}
    \mathbf{e}_{i,k} = \mathrm{ENC}_{\mathrm{E}}(\mathbf{\hat{e}}_{i,k})
\end{equation}

These edge embeddings $\mathbf{e}_{i,k} \in \mathbb{R}^{d_{\mathrm{emb}}}$ for each $k \in N^{\mathrm{NB}}_i$ are processed using a transformer-style attention layer~\cite{vaswani2017attention},

\begin{equation}
    \mathbf{e}'_{i,k} = \alpha_{i,k}\cdot \mathbf{W}_V \mathbf{e}_{i,k}
    \label{eq:transform_attn}
\end{equation}
where $\mathbf{W}_V$ is the learnable weight matrix for the value embeddings, and the attention weights $\alpha_{i,k}$ are obtained by a softmax function,

\begin{equation}
    \alpha_{i,k} = \frac{\exp(a(\mathbf{e}_{i,k}))}{\sum_{k\in N^{\mathrm{NB}}_i } \exp(a(\mathbf{e}_{i,k}))} 
    \label{eq:attn_wghts}
\end{equation}
where $a(\cdot)$ is the scaled dot-product function, defined as,

\begin{equation}
    a(\mathbf{e}_{i,k}) = \frac{\mathbf{e}^T_{i,k} \mathbf{W}_Q \mathbf{W}_K \mathbf{e}_{i,k}}{\sqrt{d_{\mathrm{emb}}}} 
\end{equation}
where $\mathbf{W}_Q$ is a learnable weight matrix applied to the query embeddings $\mathbf{e}_{i,k}$, $\mathbf{W}_K$ is a learnable weight matrix applied to the key embeddings $\mathbf{e}_{i,k}$, and $d_{\mathrm{emb}}$ is the dimension of the embeddings $\mathbf{e}_{i,k}$.

\begin{remark}
In the attention mechanism, for each node $i$, the query $\mathbf{Q} = \mathbf{W}_Q \mathbf{e}_{i,k}$, key $\mathbf{K} = \mathbf{W}_K \mathbf{e}_{i,k}$, and value $\mathbf{V} = \mathbf{W}_V \mathbf{e}_{i,k}$ are used to compute attention scores and aggregate information from neighboring nodes, allowing node $i$ to focus on the most relevant neighbors based on their importance.
\end{remark}

\begin{remark}
We employ a multi-head attention mechanism with 8 parallel attention layers~\cite{vaswani2017attention}. However, we omit the detailed expressions for the multi-head mechanism to simplify the mathematical exposition.
\end{remark}

The updated edge embeddings $\mathbf{e}'_{i,k} \in \mathbb{R}^{d_{\mathrm{emb}}}$ undergo average pooling to obtain a node embedding with attention weight, and is then added to the original node embedding to get the updated node embedding $\mathbf{n}'_i \in \mathbb{R}^{d_{\mathrm{emb}}}$ as,

\begin{equation}
    \mathbf{n}'_{i} = \frac{1}{n^{\mathrm{NB}}}\sum_{k \in N^{\mathrm{NB}}_i} \mathbf{e}'_{i,k} + \mathbf{n}_i
\end{equation} 

\noindent This final summation step acts similarly to the residual connection in deep residual networks (ResNet)\cite{he2016deep}, where the original input is added back to the transformed output. This mechanism helps preserve essential input information and facilitates more effective gradient propagation during backpropagation, which is especially important as the depth of the neural network increases. Without such connections, deeper networks are more prone to vanishing gradients, leading to slower or stalled training\cite{he2016deep}.

\subsubsection{Global encoding} 

The global features $\mathbf{G}$ include material feature $\pmb{\hat{\beta}}$, pressure feature $\mathbf{\hat{p}}$, and the shape feature $\pmb{\hat{\xi}}$, defined as,

\begin{subequations}
    \begin{align}
        \pmb{\hat{\beta}} &= (\hat{a},\hat{b},\hat{a}_f,\hat{b}_f,\hat{a}_s,\hat{b}_s) \text{, defined in Table~\ref{table_mater_bounds}} \\
        \mathbf{\hat{p}} &= (\hat{p}_{\mathrm{lv}},\hat{p}_{\mathrm{rv}}) \text{, defined in Table~\ref{table_mater_bounds}} \\   
        \pmb{\hat{\xi}} &= \text{normalized shape parameters, defined in Table~\ref{table_shape-cooeficients}}
    \end{align}
\end{subequations}
where quantities with a $\hat{(\cdot)}$ represents normalized values. Since the material parameters are generated on a log scale, we also process the quantities in the log scale. However, to avoid excessive complexity, we have not explicitly shown the log-scale representation here. These normalized global features are encoded into the global embedding $\mathbf{g}' \in \mathbb{R}^{d_{\mathrm{emb}}}$ using a global encoder $\mathrm{ENC}_{\mathrm{G}}$, which can be expressed as:

\begin{equation}
    \mathbf{g}' = \mathrm{ENC}_{\mathrm{G}}((\pmb{\hat{\beta}},\mathbf{\hat{p}}, \pmb{\hat{\xi}}))
\end{equation}

\subsubsection{Decoding}

The updated node embedding $\mathbf{n}'_i$ and the global embedding $\mathbf{g}'$ are concatenated and then decoded into the displacements using the decoders $\mathrm{DEC}_{\mathrm{D}_\alpha}$, which are essentially MLPs, similar to the encoders~\cite{goodfellow2016deep}, where $\alpha = x, y, z$. The decoding process can thus be expressed as,

\begin{subequations}
    \begin{align}
        {\mathbf{\hat{u}}_{x,i}} &= \mathrm{DEC}_{\mathrm{D}_x}((\mathbf{n}'_i, \mathbf{g}')) \\
        {\mathbf{\hat{u}}_{y,i}} &= \mathrm{DEC}_{\mathrm{D}_y}((\mathbf{n}'_i, \mathbf{g}')) \\
        {\mathbf{\hat{u}}_{z,i}} &= \mathrm{DEC}_{\mathrm{D}_z} ((\mathbf{n}'_i, \mathbf{g}')) 
    \end{align}
\end{subequations}
where ${\mathbf{\hat{u}}_{x,i}},{\mathbf{\hat{u}}_{y,i}},{\mathbf{\hat{u}}_{z,i}}$ are the predicted normalized displacement components in $x,y,z$ directions of node $i$.

\subsection{Training scheme}\label{sec:training}

Before training, we applied min-max normalization to the original data~\cite{ali2014data}, including both the input features and the true labels, to preserve the data’s shape. Each quantity was normalized by its minimum and maximum values to ensure that the method remains generalized.
The dataset, consisting of 5000 simulation cases, was split into 4000 training cases, 500 validation cases, and 500 test cases.

The training process begins by setting key hyperparameters, including the learning rate $\eta$, number of epochs $n_{\mathrm{epoch}}$, batch size $n_B$, and neighboring connection number $n^{\mathrm{NB}}$. In this study, we used a fixed learning rate of $1 \times 10^{-4}$, $n_{\mathrm{epoch}} = 3000$, a batch size of $n_B = 20$, and the neighboring connection number $n^{\mathrm{NB}}=12$. The model sensitivities to these hyperparameters are analyzed in Sec.~\ref{sec:cmp_lr}.

The overall training procedure is summarized in Alg.~\ref{alg:training}. During each epoch $j \leq n_{\mathrm{epoch}}$, a subset of $n_S$ nodes per sample is randomly selected to pass through the neural network, as outlined in Eq.~\ref{equ:hss-node}. In this work, we selected $n_S = 300$ nodes per sample per epoch. Importantly, the selected nodes change across epochs, ensuring that different regions of the domain contribute to the training process. This strategy significantly reduces computational costs while preserving accuracy. The impact of varying $n_S$ values is discussed in Sec.~\ref{sec:cmp_downsample}.

During each epoch, the training objective is to optimize the $\mathbf{HSS}_{\pmb{\theta}}$ (Eq.~\ref{equ:hss-node}) neural network parameters $\pmb{\theta}$ by minimizing the L2 norm of the displacement error between the predicted values $\mathbf{\hat{u}}_{\alpha,I_i}$ and the FEA-derived ground truth $\mathbf{\hat{\bar{u}}}_{\alpha,I_i}$ for $\alpha = x, y, z$. The loss function is expressed by

\begin{equation}
    \mathcal{L}_{\mathbf{HSS}}(\pmb{\theta}) = \frac{1}{n_\mathrm{sample}}\frac{1}{n_S}\frac{1}{3}\sum^{n_\mathrm{sample}}_{I}\sum^{n_S}_{i}\sum_{\alpha}^3 ||\mathbf{\hat{u}}_{\alpha,I_i} - {\mathbf{\hat{\bar{u}}}_{\alpha,I_i}}||^2_2 
\end{equation}
where $n_{\mathrm{sample}}$ is the number of samples in training. 

The neural network, data processing, and training scheme were implemented using Python 3.9 and PyTorch 2.0. Various engineering optimizations, such as the TensorFlow Record (TFRecord) format, an efficient binary file format for storing and loading large datasets, and advanced input/output techniques, have been applied to enhance code performance and training efficiency. For further details, readers are encouraged to refer to our code~\footnote{\url{https://github.com/LeiShi3374/phy_gnn/tree/master}}. 

\begin{algorithm}[t]
\caption{Training Scheme for \textit{HeartSimSage} (HSS)}  
\label{alg:training}
\SetAlgoLined
\KwIn{Training dataset $\mathcal{D} = \{(\mathbf{\hat{h}}_{I_i}, \mathbf{\hat{u}}_{\alpha,I_i}, \mathbf{\hat{g}}_I)\}$, \\
      Hyperparameter set: number of epoch $n_{\mathrm{epoch}}$, batch size $n_B$, learning rate $\eta$, neighboring connection number $n^{\mathrm{NB}}$, subset nodes per sample $n_S$}
\KwOut{Trained model parameters $\pmb{\theta}$}

\textbf{Initialize:} Model parameters $\pmb{\theta}$, hyperparameters $n_{\mathrm{epoch}}$, $n_B$, $\eta$, $n^{\mathrm{NB}}$, $n_S$

\For{$j = 1$ \KwTo $n_{\mathrm{epoch}}$}{
    Randomly select $n_S$ nodes from each sample $I$ from the training batch\;
    \For{each selected node $i$ and each sample $I$}{
        Compute the displacements $\displaystyle \mathbf{\hat{u}}_i$ via $\mathbf{HSS}_{\pmb{\theta}}$ (Eq.~\ref{equ:hss-node})\;
    }
    Compute error $\displaystyle \mathcal{L}_{\mathbf{HSS}}(\pmb{\theta}) = \frac{1}{n_\mathrm{sample}}\frac{1}{n_S}\frac{1}{3}\sum^{n_\mathrm{sample}}_{I}\sum^{n_S}_{i}\sum_{\alpha}^3 ||\mathbf{\hat{u}}_{\alpha,I_i} - {\mathbf{\hat{\bar{u}}}_{\alpha,I_i}}||^2_2  $\;
    Update $\pmb{\theta}$ using gradient descent:  
    $\pmb{\theta} \gets \pmb{\theta} - \eta \nabla_{\pmb{\theta}} \mathcal{L}$\;
}
\Return $\pmb{\theta}$
\end{algorithm}

\section{Result}

\subsection{Dataset}

Across the 5,000 biventricular myocardial meshes, the number of nodes across the volumes ranges from 15,864 to 19,873, and the number of tetrahedral elements ranges from 86,596 to 105,395, with notable variation in node ordering and connectivity. The node and element counts have been demonstrated to be sufficiently fine to yield stable results, as shown in our previous work~\cite{shi2024optimization}.

\begin{figure}[h] 
 \begin{center}
 \includegraphics[width=1\textwidth]{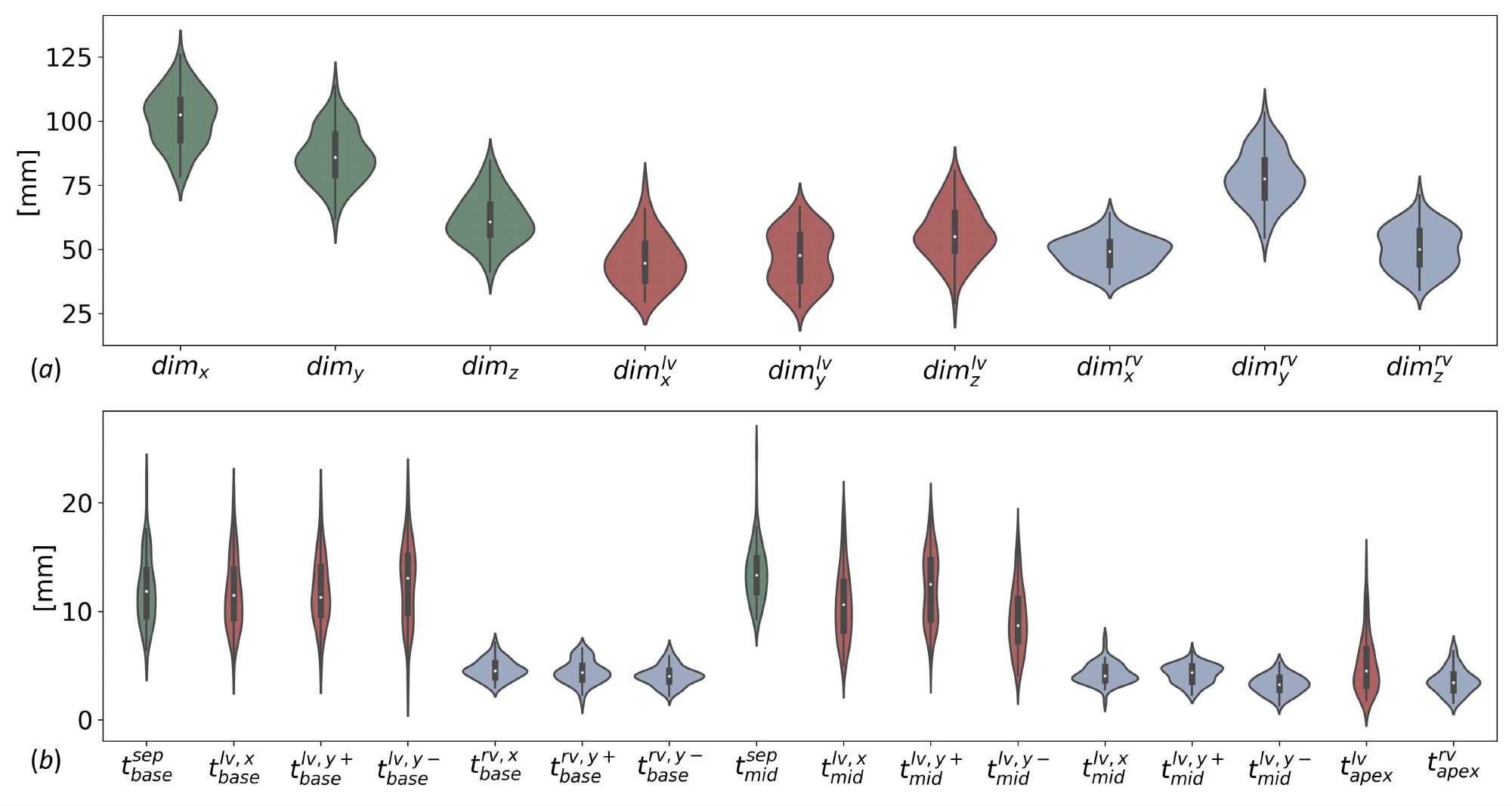}
 \end{center}
 \caption{Distributions of biventricular myocardial dimensions across the entire study cohort, including 150 patients and 4850 additional cases created via data augmentation (Sec.~\ref{sec:data_augmentation_sim}). (a) Dimensions of the entire geometry, the left ventricle (LV), and the right ventricle (RV). (b) Myocardial thickness distributions at various anatomical locations, including ventricular base, mid-wall, and apex for LV, RV, and septum (Table~\ref{table_shape-cooeficients},~\ref{sec:shape}).}
 \label{fig_data_statics}
\end{figure}

Table~\ref{table_shape-summaries} and Fig.~\ref{fig_data_statics} summarize the key dimensions and thicknesses. Most quantities fall within the typical range for a human heart, with some covering both normal and pathological cases. The dataset comprises biventricular heart models derived from CT imaging, augmented to account for anatomical variability. The overall biventricular dimensions, across the entire cohort, span from 78–137 mm ($x$-axis), 61–114 mm ($y$-axis), and 42–95 mm ($z$-axis), consistent with physiological norms~\cite{kou2014echocardiographic, kawel2025reference, cheema1996dimensions}. Ventricular wall thickness measurements show left ventricular (LV) thickness ranging from 3.8–24.0 mm and right ventricular (RV) thickness from 1.3–7.6 mm, with the LV range including both normal and hypertrophied cases~\cite{kawel2025reference, kitzman1988age}. These measurement ranges provide a robust basis for biomechanical modeling and ML applications, ensuring diverse training data while adhering to physiological realism.

\begin{table}[H]
\begin{center}
\footnotesize
\caption{The minimum and maximum ranges of key shape parameters (Table~\ref{table_shape-cooeficients},~\ref{sec:shape}) for the 5000 biventricular myocardial geometries considered in the current study. All quantities are measured in mm.}
\label{table_shape-summaries}
\vspace{4pt}
\begin{tabular}{c c c c c c c c c c}
\toprule
& $t_{\mathrm{base}}^{\mathrm{lv,x}}$ & $t_{\mathrm{base}}^{\mathrm{lv,y+}}$ & $t_{\mathrm{base}}^{\mathrm{lv,y-}}$ & $t_{\mathrm{mid}}^{\mathrm{lv,x}}$ & $t_{\mathrm{mid}}^{\mathrm{lv,y+}}$ & $t_{\mathrm{mid}}^{\mathrm{lv,y-}}$ & $t_{\mathrm{apex}}^{\mathrm{lv}}$ & $t_{\mathrm{base}}^{\mathrm{sep}}$ \\
\arrayrulecolor{black!30}\midrule
max & 20.3 & 20.4 & 20.8 & 19.0 & 24.0 & 16.8 & 14.2 & 21.8 \\
\arrayrulecolor{black!30}\midrule
min & 5.3 & 5.2 & 5.1 & 5.0 & 5.4 & 4.1 & 3.8 & 6.4 \\
\arrayrulecolor{black!80}\midrule
& $t_{\mathrm{base}}^{\mathrm{rv,x}}$ & $t_{\mathrm{base}}^{\mathrm{rv,y+}}$ & $t_{\mathrm{base}}^{\mathrm{rv,y-}}$ & $t_{\mathrm{mid}}^{\mathrm{rv,x}}$ & $t_{\mathrm{mid}}^{\mathrm{rv,y+}}$ & $t_{\mathrm{mid}}^{\mathrm{rv,y-}}$ & $t_{\mathrm{apex}}^{\mathrm{rv}}$ & $t_{\mathrm{mid}}^{\mathrm{sep}}$ \\
\arrayrulecolor{black!30}\midrule
max & 7.1 & 6.6 & 6.5 & 7.6 & 6.3 & 5.3 & 6.7 & 27.1 \\
\arrayrulecolor{black!30}\midrule
min & 3.0 & 1.5 & 2.2 & 1.6 & 2.3 & 1.3 & 1.5 & 9.3\\
\arrayrulecolor{black!80}\midrule
& $\dim_{\mathrm{x}}$ & $\dim_{\mathrm{y}}$ & $\dim_{\mathrm{z}}$ & $\dim^{\mathrm{lv}}_x$ & $\dim^{\mathrm{lv}}_y$ & $\dim^{\mathrm{lv}}_z$ & $\dim^{\mathrm{rv}}_x$ & $\dim^{\mathrm{rv}}_y$ & $\dim^{\mathrm{rv}}_z$  \\
\arrayrulecolor{black!30}\midrule
max & 137 & 114 & 95 & 74 & 78 & 86 & 76 & 103 & 72 \\
\arrayrulecolor{black!30}\midrule
min & 78 & 62 & 42 & 28 & 26 & 28 & 36 & 54 & 34 \\
\arrayrulecolor{black}\bottomrule
\end{tabular}
\end{center}
\end{table}

\subsection{Prediction results}\label{sec:prediction_results} 

\begin{figure}[h] 
 \begin{center}
 \includegraphics[width=1\textwidth]{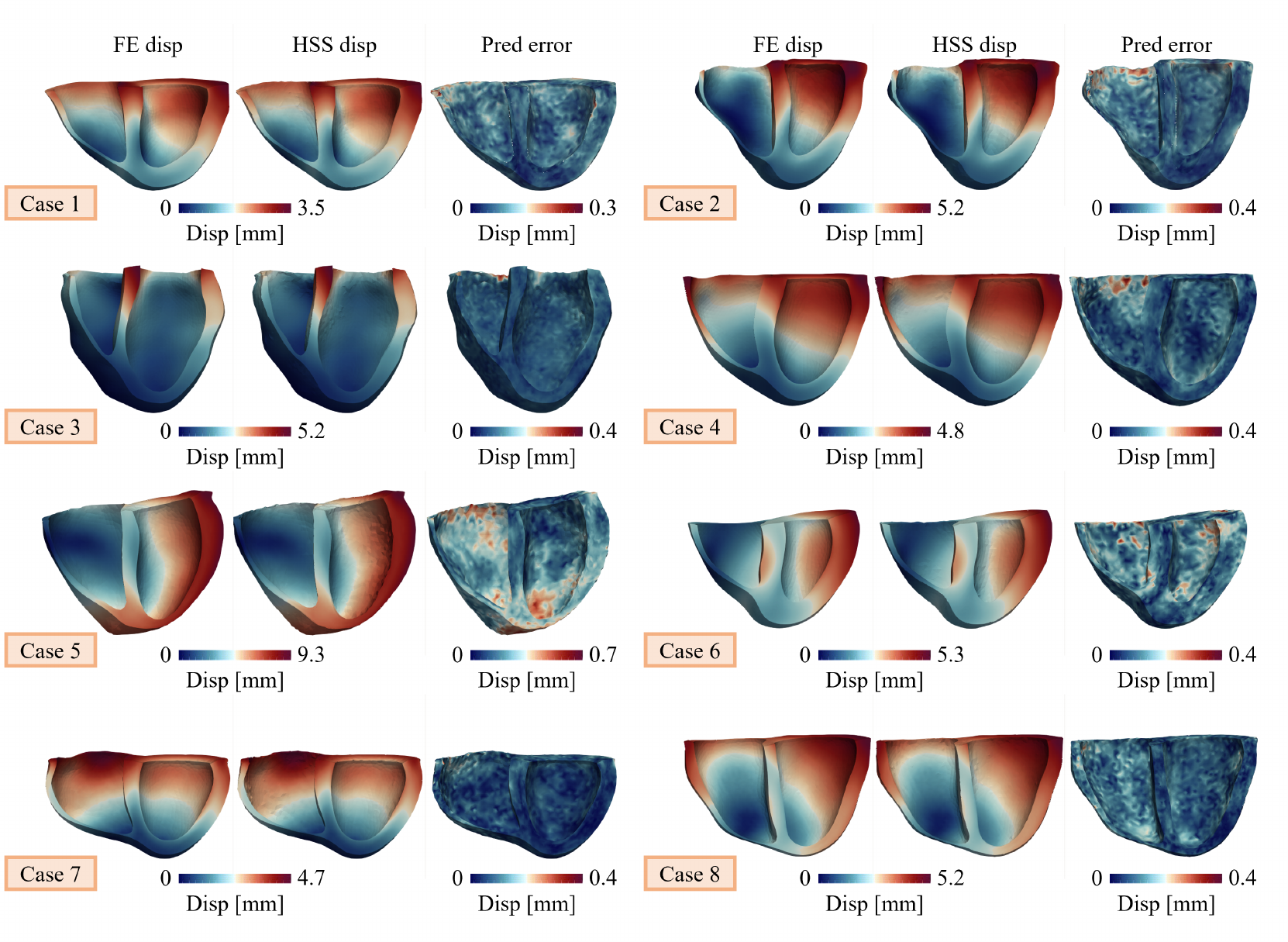}
 \end{center}
 \caption{Comparison between FEA-computed and \textit{HeartSimSage}-predicted biventricular myocardial displacements for passive pressure loading in 9 randomly selected cases. For each case, the FEA-computed displacements (left), the \textit{HeartSimSage}-predicted displacements (center), and the absolute node-wise difference in displacements between the two solutions (right) are shown. The corresponding material parameters and pressure data are provided in Table~\ref{table_mat_params_fig}.}
 \label{fig_data_results}
\end{figure}

In our study, we identified the optimal training hyperparameter set as $n_S = 300$, $n^{\mathrm{NB}} = 12$, $\eta = 0.0001$, and $n_B = 20$. Additionally, we observed that 3000 epochs are sufficient for the training and validation curves to stabilize. The training and validation losses are 0.0067 and 0.0091, respectively, calculated as the average of the last 20 values. The training was conducted on a single Nvidia Titan L40S GPU. The prediction time per case on one L40S GPU is $0.0084 \mathrm{s}$, while on an Intel Core i9-14900 CPU, it is $0.6\mathrm{s}$. In comparison, the FEA calculation for each case takes $112 \pm 35 \mathrm{s}$ on the same CPU, leading to a speedup of $1.3 \times 10^4$ times on the GPU and $1.9 \times 10^2$ times on the CPU. At the same time, \textit{HeartSimSage} performs reasonably well in predicting the biventricular displacements compared to the FEA-based computations (Fig.~\ref{fig_data_results}). A node-wise difference in the FEA-based and the neural network-predicted displacements is below $0.4~mm$ for all cases except for case 5, where the maximum difference is $0.7~mm$ (Fig.~\ref{fig_data_results}).

\begin{figure}[h] 
 \begin{center}
 \includegraphics[width=1\textwidth]{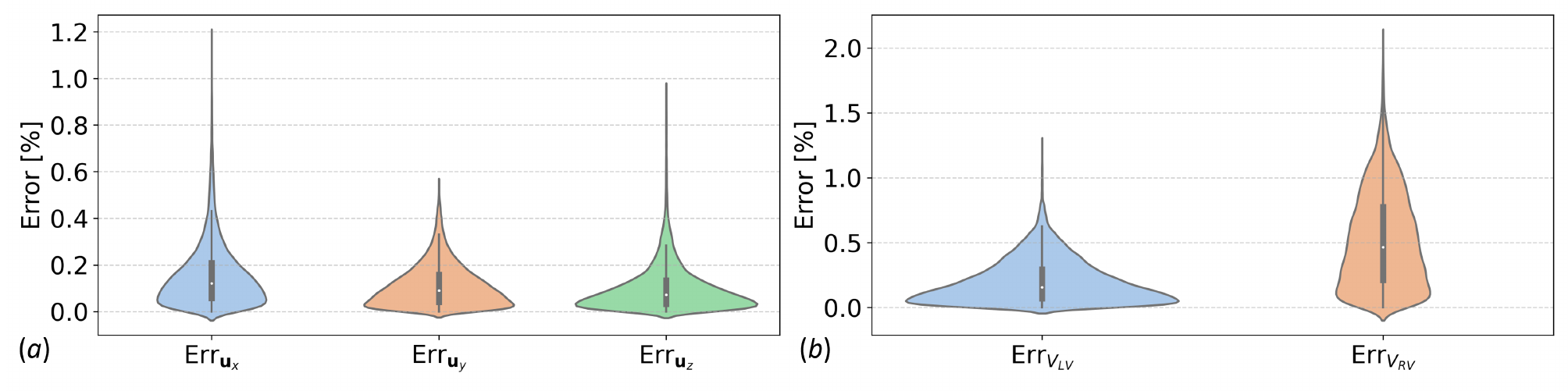}
 \end{center}
 \caption{Normalized prediction error distributions on the test dataset between \textit{HeartSimSage} and FEA for (a) displacement components along the $x$, $y$, and $z$ directions, and (b) the left and right ventricular (LV, RV) chamber volumes.}
 \label{fig_data_error}
\end{figure}

We further quantified the prediction error of \textit{HeartSimSage} by comparing the relative norms of differences in displacement components and LV and RV cavity volumes between the current neural network model and FEA. These errors are expressed as,

\begin{subequations}
    \begin{align}
        \mathrm{Err}_{\mathbf{u}_x} &= \frac{|\mathbf{u}^{HSS}_x-\mathbf{u}^{FEA}_x|}{\mathrm{dim}_x} \\
        \mathrm{Err}_{\mathbf{u}_y} &= \frac{|\mathbf{u}^{HSS}_y-\mathbf{u}^{FEA}_y|}{\mathrm{dim}_y} \\
        \mathrm{Err}_{\mathbf{u}_z} &= \frac{|\mathbf{u}^{HSS}_z-\mathbf{u}^{FEA}_z|}{\mathrm{dim}_z} \\
        \mathrm{Err}_{V_{\mathrm{LV}}} &= \frac{|V^{HSS}_{\mathrm{LV}}-{V}^{FEA}_{\mathrm{LV}}|}{{V}^{FEA}_{\mathrm{LV}}} \\
        \mathrm{Err}_{V_{\mathrm{RV}}} &= \frac{|V^{HSS}_{\mathrm{RV}}-{V}^{FEA}_{\mathrm{RV}}|}{{V}^{FEA}_{\mathrm{RV}}}
    \end{align}
\end{subequations} 
Here, the displacement prediction errors are quantified for all the nodes of the geometry. These displacement errors, along with the prediction errors in chamber volumes, are then computed across the entire study cohort on the test dataset (Table~\ref{table_data_errors}, Fig.~\ref{fig_data_error}). Our results show that the average component-wise displacement errors are at the most $0.15\% \pm 0.14\%$ in the $x$ direction with a maximum error of $1.17\%$, suggesting a highly-skewed distribution (Fig.~\ref{fig_data_error}). The errors in predicting LV and RV volumes are $0.20\%\pm 0.17\%$ and $0.52\%\pm 0.36\%$, respectively, with a maximum error of $1.26\%$ for LV and $2.04\%$ for RV. While the LV volume error exhibits a highly skewed and narrow-band distribution, the RV volume prediction error has a larger spread up to $1.5\%$ before thinning out. This relatively higher error in predicting RV volumes is likely due to its thin wall and fewer data points, making the prediction challenging.

\begin{table}[h]
    \centering
    \caption{Normalized prediction errors between \textit{HeartSimSage} and FEA for the displacement components and chamber volumes over the entire test dataset in the study.}
    \begin{tabular}{c c c c c}
        \toprule
        Error & Mean & STD & Max & Min \\
        \arrayrulecolor{black!30}\midrule
        $\mathrm{Err}_{\mathbf{u}_x}$ & 0.15\% & 0.14\% & 1.17\% & 0\% \\
        $\mathrm{Err}_{\mathbf{u}_y}$ & 0.11\% & 0.10\% & 0.54\% & 0\% \\
        $\mathrm{Err}_{\mathbf{u}_z}$ & 0.10\% & 0.10\% & 0.95\% & 0\% \\
        $\mathrm{Err}_{V_{\mathrm{LV}}}$ & 0.20\% & 0.17\% & 1.26\% & 0\% \\
        $\mathrm{Err}_{V_{\mathrm{RV}}}$ & 0.52\% & 0.36\% & 2.04\% & 0\% \\
        \arrayrulecolor{black}\bottomrule
    \end{tabular}
    \label{table_data_errors}
\end{table}

\subsection{Weight from the attention mechanism} 

\begin{figure}[h] 
 \begin{center}
 \includegraphics[width=0.8\textwidth]{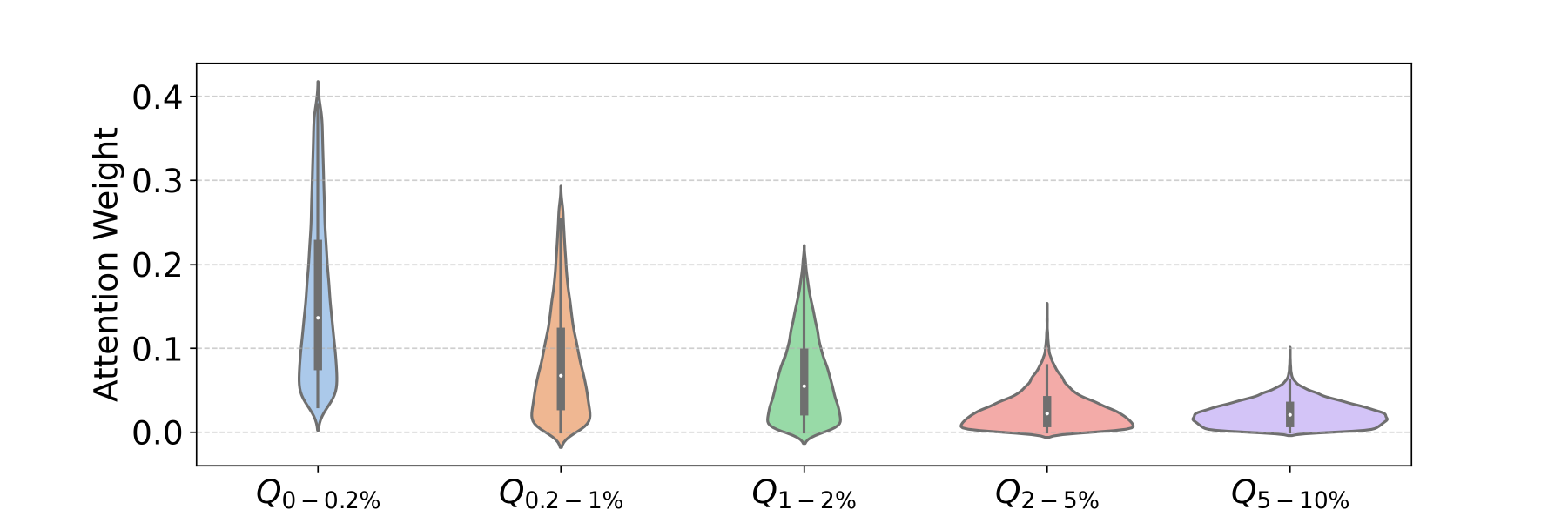}
 \end{center}
 \caption{The attention weight distribution for neighboring nodes in different ranges based on the spatial connection scheme adopted in Sec.~\ref{sec:spatial_connections} measured across all the study cases.}
 \label{fig_attention_weight}
\end{figure}

The attention mechanism for edge embeddings helps determine the influence of the different selected neighboring nodes on node $i$ through the learned attention weights, $\alpha_{i,k}$ (Eqs.~\ref{eq:transform_attn}-~\ref{eq:attn_wghts}). A higher value of $\alpha_{i,k}$ indicates a greater influence of node $k$ on node $i$. To assess this, we analyzed the impact of each neighboring node $k$ on node $i$ across all nodes of all cases and summarized the findings in Table~\ref{table_attention_weight} and Fig.~\ref{fig_attention_weight}. 

Our analysis reveals that the nearest neighboring nodes ($d(i,k) \leq Q_{0.2\%}$) have significantly higher attention weights ($p < 0.05$ from one-way ANOVA) on the displacement prediction of node $i$ compared to all other distances. As the distance between nodes increases, their influence on the displacement prediction decreases, although the farther nodes still contribute to the overall prediction (Sections~\ref{sec:cmp_node_conn},~\ref{sec:cmp_attention}).

\begin{table}[h]
    \centering
    \caption{Statistical summary of the neighborhood nodal attention weight in different neighboring ranges based on the spatial connection scheme adopted in Sec.~\ref{sec:spatial_connections}.}
    \begin{tabular}{c c c c c}
        \toprule
        Neighboring range & Mean & STD & Max & Min \\
        \arrayrulecolor{black!30}\midrule
        $d(i,k)\leq Q_{0.2\%}$ & 0.159 & 0.096 & 0.390 & 0.030 \\
        $Q_{0.2\%}\leq d(i,k)\leq Q_{1\%}$ & 0.082 & 0.063 & 0.275 & 0.000 \\
        $Q_{1\%}\leq d(i,k)\leq Q_{2\%}$ & 0.064 & 0.047 & 0.209 & 0.000 \\
        $Q_{2\%}\leq d(i,k)\leq Q_{5\%}$ & 0.027 & 0.022 & 0.148 & 0.000 \\
        $Q_{5\%}\leq d(i,k)\leq Q_{10\%}$ & 0.022 & 0.014 & 0.097 & 0.000 \\
        \arrayrulecolor{black}\bottomrule
    \end{tabular}
    \label{table_attention_weight}
\end{table}

\section{Discussion}

We developed \textit{HeartSimSage}, an attention-enhanced graph neural network, to predict the biventricular passive mechanics displacement field from the input geometry, material properties, and chamber pressures. Tailored for patient-specific cardiac modeling, the model efficiently handles diverse 3D cardiac geometries, heterogeneous tissue properties, complex constitutive relations, and physiological boundary conditions. It also supports flexible node topologies with variations in node count, ordering, and connectivity.

Our approach leverages smooth Laplace solutions and their gradients from multiple Laplace-Dirichlet boundary value problems to encode each node’s relative position within the biventricular myocardium. To improve efficiency and scalability, we utilized subset-based training and adopted a GraphSAGE-inspired neighbor connection strategy that balances local interactions and long-range dependencies. Additionally, the attention mechanism enhances prediction accuracy by dynamically weighting neighbor contributions and filtering irrelevant information. These advancements establish \textit{HeartSimSage} as a powerful and versatile tool for patient-specific cardiac biomechanics modeling.

This section examines the effect of hyperparameters, Laplace values, node connection strategies, and the attention mechanism on model performance. In particular, we analyzed the sensitivity of each of these variables on the training (Train) and validation (Val) losses (Tables~\ref{table_cmp_lr}--\ref{table_cmp_attention}, Figs.~\ref{fig_cmp_lr}--\ref{fig_cmp_attention}). The losses (Train, Val) are averaged over the last 20 epochs. We also assessed the model’s performance by measuring the speed of training per epoch. The validation losses are compared between different sensitivity variables by computing the relative differences against a baseline case. We also evaluated the model’s effectiveness using a previously published LV mechanics dataset~\cite{dalton2022emulation}. Finally, we discussed the model’s limitations and provided potential directions for future research.

\subsection{Effect of learning rate}\label{sec:cmp_lr}

\begin{figure}[h] 
 \begin{center}
 \includegraphics[width=1\textwidth]{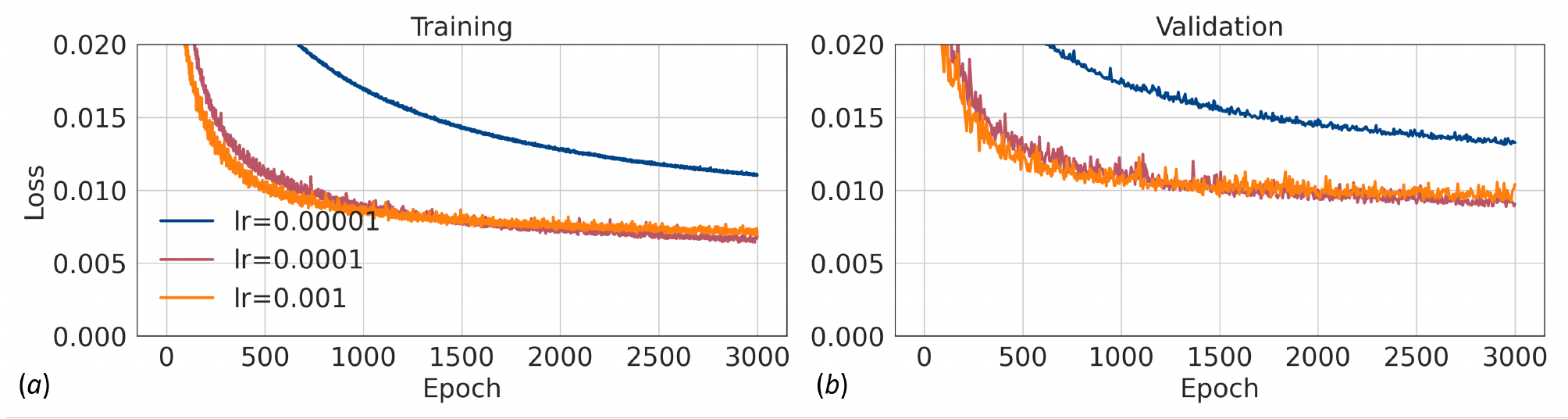}
 \end{center}
 \caption{Training (a) and validation (b) losses for different learning rates. Light red represents the baseline case used in Sec.~\ref{sec:prediction_results}.}
 \label{fig_cmp_lr}
\end{figure}

We evaluated the sensitivity of our neural network model training to the learning rate, $\eta$. Specifically, we examined the effect of the learning rate on the training and validation losses (Table~\ref{table_cmp_lr} and Fig.~\ref{fig_cmp_lr}). We tested three different learning rates for training \textit{HeartSimSage} on the biventricular mechanics dataset: (i) $\eta = 0.001$, (ii) $\eta = 0.0001$, and (iii) $\eta = 0.00001$. The other hyperparameters, node and edge features, and model architecture remained the same as in Sec.~\ref{sec:prediction_results}. We found that $\eta = 0.0001$ provides the best trade-off between convergence speed and stability. While $\eta = 0.001$ resulted in slower initial convergence compared to $\eta = 0.0001$, the case with $\eta = 0.00001$ converged quickly at the start but failed to reach full convergence even after 3000 epochs. This suggests that the update steps are too small for effective optimization. Overall, $\eta = 0.0001$ yielded the most stable and efficient training performance.

\begin{table}[h]
    \centering
    \begin{threeparttable}
    \caption{Effect of learning rate on the model training and validation.}
    \begin{tabular}{c c c c c}
        \toprule
        $\eta$ & Train loss & Val loss & Speed (/epoch) & Comparison \\
        \arrayrulecolor{black!30}\midrule
        0.001 & 0.0111 & 0.0133 & 23s & -46.2\% \\
        \textcolor{red}{0.0001} & 0.0067 & 0.0091 & 23s & /  \\
        0.00001 & 0.0072 & 0.0099 & 23s & -8.8\% \\
        \arrayrulecolor{black}\bottomrule
    \end{tabular}
    \begin{tablenotes}
        \footnotesize
        \item Train (training) and Val (validation) losses represent the average values over the last 20 epochs. Speed refers to the training speed per epoch. The comparison is computed as the difference between a case's validation loss and the baseline value, normalized by the baseline value. A negative value indicates performance worse than the baseline. {\color{red} Red text} indicates the baseline case presented in Sec.~\ref{sec:prediction_results}. The same convention applies to the following tables unless otherwise noted.
    \end{tablenotes}
    \label{table_cmp_lr}
    \end{threeparttable}
\end{table}

\subsection{Effect of batch size}\label{sec:cmp_batch}

\begin{figure}[h] 
 \begin{center}
 \includegraphics[width=1\textwidth]{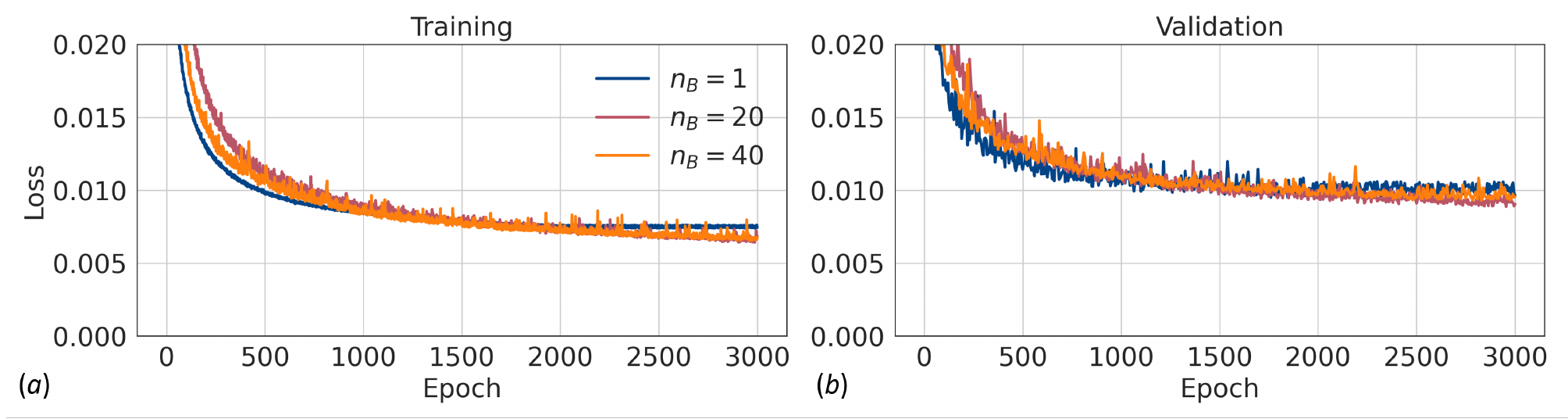}
 \end{center}
 \caption{Training (a) and validation (b) losses for different batch sizes. Light red denotes the baseline case used in Sec.~\ref{sec:prediction_results}.}
 \label{fig_cmp_batch}
\end{figure}

We evaluated the sensitivity of \textit{HeartSimSage} training to batch size $ n_B $ by analyzing its impact on training and validation losses (Table~\ref{table_cmp_batch} and Fig.~\ref{fig_cmp_batch}). Specifically, we compared three batch sizes, (i) $ n_B = 1 $, (ii) $ n_B = 20 $, and (iii) $ n_B = 40 $, while keeping all other hyperparameters, node and edge features, and model structure consistent with Sec.~\ref{sec:prediction_results}. Among these, $n_B = 20$ provided the best trade-off between convergence speed and performance. Training with $n_B = 1$ is notably slow, while $n_B = 40$ also converged at a slower rate than $n_B = 20$. 

\begin{table}[h]
    \centering
    \begin{threeparttable}
    \caption{Effect of batch size on model training and validation.}
    \begin{tabular}{c c c c c}
        \toprule
        $n_B$ & Train loss & Val loss & Speed (/epoch) & Comparison \\
        \arrayrulecolor{black!30}\midrule
        1 & 0.0075 & 0.0105 & 89s & -15.4\% \\
        \textcolor{red}{20} & 0.0067 & 0.0091 & 23s & /  \\
        40 & 0.0067 & 0.0096 & 26s & -5.5\% \\
        \arrayrulecolor{black}\bottomrule
    \end{tabular}
    \begin{tablenotes}
    \footnotesize
    \item See footnotes in Table~\ref{table_cmp_lr} for further details on the adopted convention.
    \end{tablenotes}
    \label{table_cmp_batch}
    \end{threeparttable}
\end{table}

\subsection{Effect of employing Laplace-Dirichlet solutions}\label{sec:cmp_laplace}

\begin{figure}[h] 
 \begin{center}
 \includegraphics[width=1\textwidth]{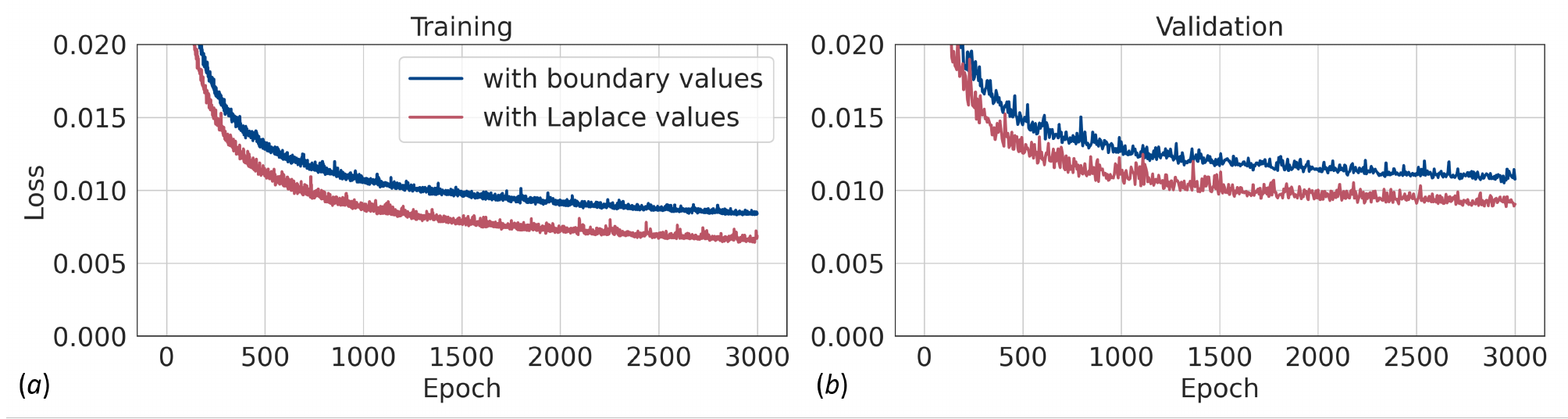}
 \end{center}
 \caption{Comparison of training (a) and validation (b) losses between the case with boundary values and using Laplace solutions as node features. Light red denotes the baseline case used in Sec.~\ref{sec:prediction_results}.}
 \label{fig_cmp_laplace}
\end{figure}

Here, we evaluated the impact of incorporating Laplace solutions in the training of our neural network model. Specifically, we compared the model's performance with and without Laplace values by analyzing their effects on training and validation losses (Table~\ref{table_cmp_laplace} and Fig.\ref{fig_cmp_laplace}). To establish a comparison, we set up an alternative test case where boundary values are represented as one-hot encoded node features, similar to the one previously employed~\cite{dalton2022emulation}. In this setup, each node is assigned a binary value: if the node lies on a particular surface (e.g., base, lv-endo, rv-endo, or epicardium), it receives a value of 1 for the binary value corresponding to that surface; otherwise, it is assigned a value of 0. Consequently, four such one-hot features are used to replace the Laplace values. Apart from this modification, all other hyperparameters, node and edge features, and the overall model architecture remained the same as described in Sec.~\ref{sec:prediction_results}.

Our comparison demonstrates that incorporating Laplace values substantially improves the model performance. The superior results achieved with Laplace values suggest they provide a more effective representation of a node’s relative position relative to the boundaries. This enhanced spatial encoding plays a crucial role in accurately capturing displacement behavior, as it allows the model to better understand the geometric and positional relationships within the cardiac structure.

\begin{table}[h]
    \centering
    \begin{threeparttable}
    \caption{Summary of the Laplace value analysis.}
    \begin{tabular}{c c c c c}
        \toprule
        Boundary/Laplace & Train loss & Val loss & Speed (/epoch) & Comparison \\
        \arrayrulecolor{black!30}\midrule
        \textcolor{red}{Laplace value} & 0.0067 & 0.0091 & 23s & /  \\
        Boundary value & 0.0084 & 0.0110 & 23s & -20.9\% \\
        \arrayrulecolor{black}\bottomrule
    \end{tabular}
    \begin{tablenotes}
    \footnotesize
    \item See footnotes in Table~\ref{table_cmp_lr} for further details on the adopted convention.
    \end{tablenotes}
    \label{table_cmp_laplace}
    \end{threeparttable}
\end{table}

\subsection{Effect of subset node number $n_S$}\label{sec:cmp_downsample}

\begin{figure}[h] 
 \begin{center}
 \includegraphics[width=1\textwidth]{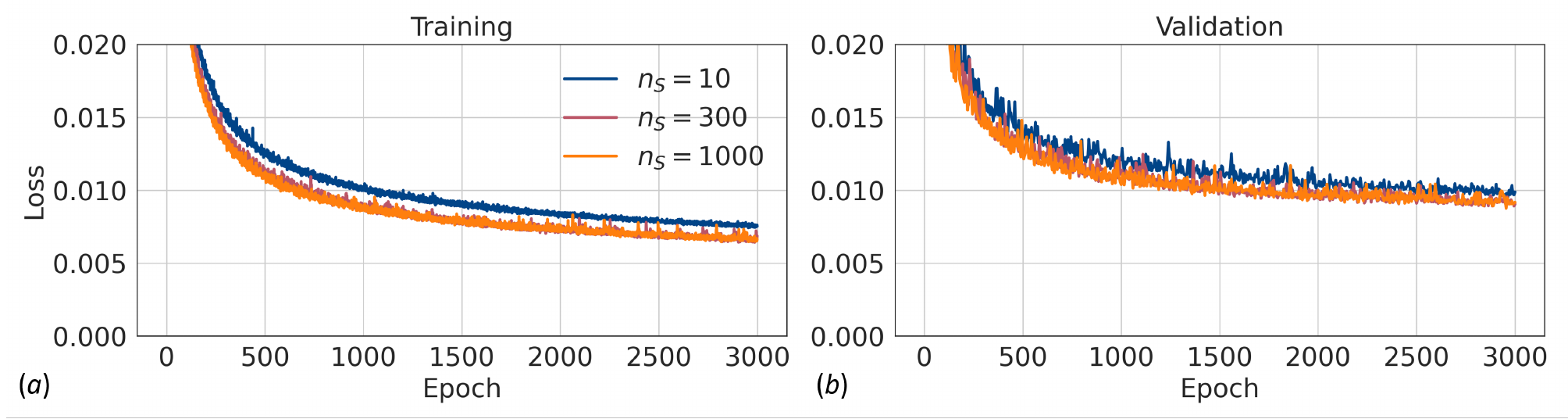}
 \end{center}
 \caption{Training (a) and validation (b) losses for varying subset node numbers, $n_S$. Light red denotes the baseline case which is used in Sec.~\ref{sec:prediction_results}.}
 \label{fig_cmp_downsample}
\end{figure}

We investigated the impact of the subset node number $n_S$ involved in training at each epoch by evaluating its effect on training and validation losses (Table~\ref{table_cmp_downsample} and Fig.~\ref{fig_cmp_downsample}). Three values of $n_S$ are compared: (i) $n_S=10$, (ii) $n_S=300$, and (iii) $n_S=1000$, while keeping all other hyperparameters, node and edge features, and model structure unchanged from Sec.~\ref{sec:prediction_results}. We found that $n_S=300$ and $n_S=1000$ yield nearly identical performance, both significantly outperforming $n_S=10$, suggesting that using too few nodes per epoch leads to suboptimal learning due to insufficient data representation. Additionally, while $n_S=1000$ offers no substantial performance improvement over $n_S=300$, it incurs a much higher computational cost. This indicates that increasing $n_S$ beyond a certain threshold provides diminishing returns, as model convergence is primarily driven by representative sampling rather than sheer data volume per epoch. Our findings suggest that sampling 300 nodes per epoch ($n_S = 300$) achieves a good balance between performance and computational efficiency, making it an optimal choice in this scenario.

\begin{table}[h]
    \centering
    \begin{threeparttable}
    \caption{Effect of subset node number $n_S$ on model training and validation.}
    \begin{tabular}{c c c c c}
        \toprule
        $n_S$ & Train loss & Val loss & Speed (/epoch) & Comparison \\
        \arrayrulecolor{black!30}\midrule
        10 & 0.0076 & 0.0099 & 21s & -8.0\% \\
        \textcolor{red}{300} & 0.0067 & 0.0091 & 23s & /  \\
        1000 & 0.0066 & 0.0091 & 38s & +0\% \\
        \arrayrulecolor{black}\bottomrule
    \end{tabular}
    \begin{tablenotes}
    \footnotesize
    \item See footnotes in Table~\ref{table_cmp_lr} for further details on the adopted convention.
    \end{tablenotes}
    \label{table_cmp_downsample}
    \end{threeparttable}
\end{table}

\subsection{Effect of neighboring node connection strategies}\label{sec:cmp_node_conn}

\begin{figure}[h] 
 \begin{center}
 \includegraphics[width=1\textwidth]{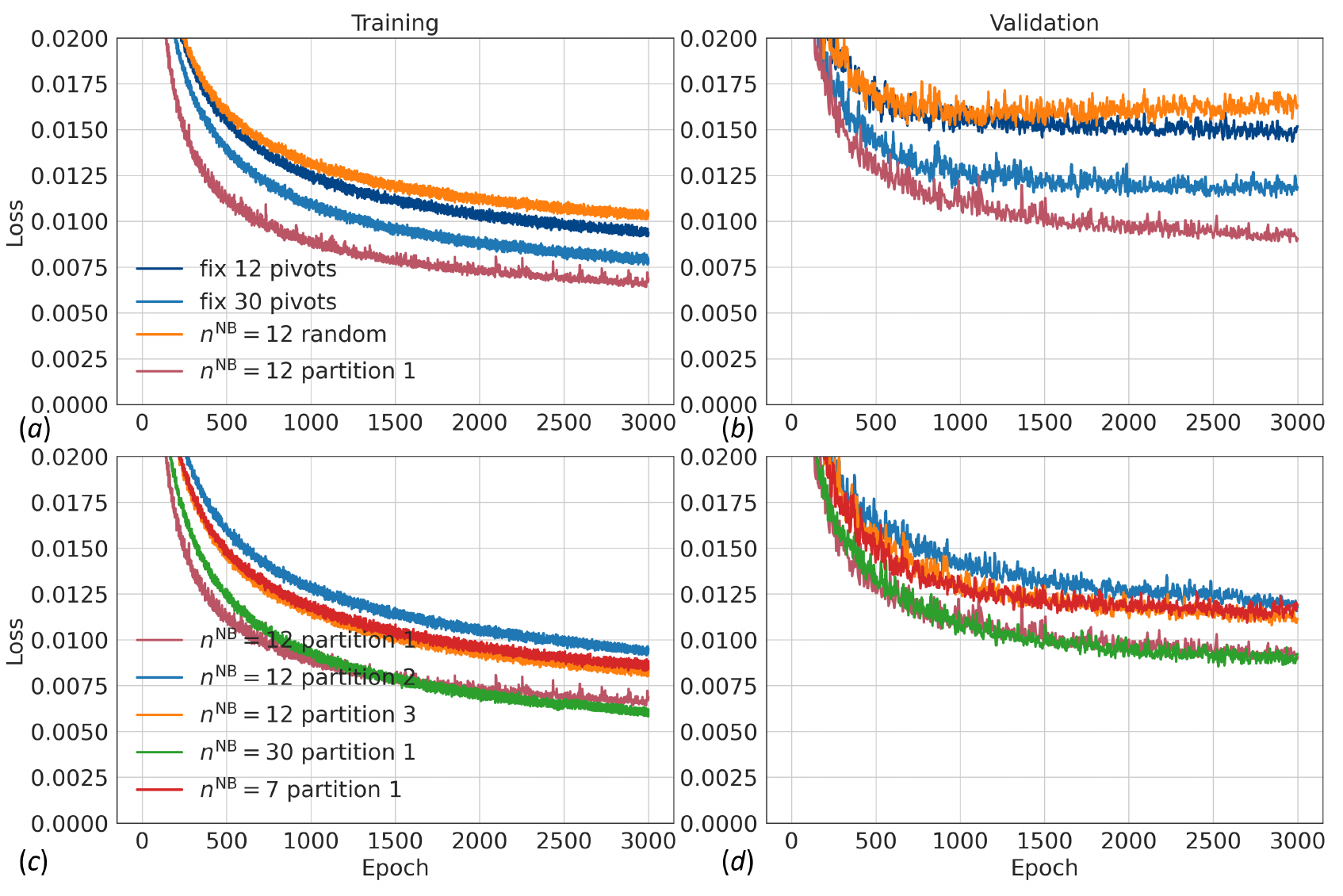}
 \end{center}
 \caption{Training (a,c) and validation (b,d) losses with different neighboring node selection strategies. Light red denotes the baseline case which is used in Sec.~\ref{sec:prediction_results}. Refer to the text for details on the neighboring node connection strategies.}
 \label{fig_cmp_node_conn}
\end{figure}

We investigated the impact of different neighboring connection strategies on the model performance, specifically evaluating their effect on training and validation losses (Table~\ref{table_cmp_node_conn} and Fig.~\ref{fig_cmp_node_conn}). We tested eight different node connection strategies while keeping all other hyperparameters, node and edge features, and model structure consistent with Sec.~\ref{sec:prediction_results}. The eight strategies are as follows: 

\begin{enumerate}

\item Connect each node to 12 fixed pivot nodes, randomly selected from the entire domain, maintaining the same set of neighbors for all samples and across all epochs. 

\item Connect each node to 30 fixed pivot nodes, randomly selected from the entire domain, maintaining the same set of neighbors for all samples and across all epochs. 

\item Connect each node to 12 ($n^{\mathrm{NB}}=12$) randomly selected neighbors per sample, allowing different nodes to be selected at different epochs.

\item Use $n^{\mathrm{NB}}=12$ with the partitioning strategy (denoted as partition 1 and serves as the baseline case), defined in Eq.~\ref{eq_neiboring_set_partition}. 

\item Use $n^{\mathrm{NB}}=12$ with an alternative partitioning strategy (denoted as partition 2), defined as,

\begin{subequations}
    \begin{align}
        N^{\mathrm{SC}_1}_i & \subseteq \{k|k\in \mathcal{V}, d(i,k)\leq Q_{0.2\%} \}
    \end{align}
    \label{eq_neiboring_set_partition_2}
\end{subequations}
with $n^{\mathrm{SC}_1}=20$. 

\item Use $n^{\mathrm{NB}}=12$ with another alternative partitioning strategy (denoted as partition 3), defined as,

\begin{subequations}
    \begin{align}
        N^{\mathrm{SC}_1}_i & \subseteq \{k|k\in \mathcal{V}, d(i,k)\leq Q_{1\%} \} \\
        N^{\mathrm{SC}_2}_i & \subseteq \{k|k\in \mathcal{V}, Q_{0.2\%}\leq d(i,k)\leq Q_{5\%} \} \\
        N^{\mathrm{SC}_3}_i & \subseteq \{k|k\in \mathcal{V}, Q_{1\%}\leq d(i,k)\leq Q_{10\%} \} \\
        N^{\mathrm{SC}_4}_i & \subseteq \{k|k\in \mathcal{V}, Q_{2\%}\leq d(i,k)\leq Q_{20\%} \} \\
        N^{\mathrm{SC}_5}_i & \subseteq \{k|k\in \mathcal{V}, Q_{5\%}\leq d(i,k)\leq Q_{100\%} \} 
    \end{align}
    \label{eq_neiboring_set_partition_3}
\end{subequations}
with $n^{\mathrm{SC}_1}=20,n^{\mathrm{SC}_2}=30,n^{\mathrm{SC}_3}=30,n^{\mathrm{SC}_4}=10,n^{\mathrm{SC}_5}=10$. 

\item Use $n^{\mathrm{NB}}=30$ with the partitioning strategy 1 in Eq.~\ref{eq_neiboring_set_partition}. 
\item Use $n^{\mathrm{NB}}=7$ with the partitioning strategy 1 in Eq.~\ref{eq_neiboring_set_partition}.
\end{enumerate}

\begin{table}[h]
    \centering
    \begin{threeparttable}
    \small
    \caption{Effect of different neighboring node selection strategies on the model training and validation.}
    \begin{tabular}{l c c c c}
        \toprule
        Strategy & Train loss & Val loss & Speed (/epoch) & Comparison \\
        \arrayrulecolor{black!30}\midrule
        fix 12 pivots & 0.0095 & 0.0150 & 17s & -64.8\% \\
        fix 30 pivots & 0.0079 & 0.0121 & 29s & -33.0\% \\
        $n^{\mathrm{NB}}=12$ random & 0.0103 & 0.0165 & 22s & -81.3\% \\
        \textcolor{red}{$n^{\mathrm{NB}}=12$ partition 1} & 0.0067 & 0.0091 & 23s & /  \\
        $n^{\mathrm{NB}}=12$ partition 2 & 0.0095 & 0.0119 & 23s & -30.8\% \\
        $n^{\mathrm{NB}}=12$ partition 3 & 0.0082 & 0.0111 & 23s & -21.9\% \\
        $n^{\mathrm{NB}}=30$ partition 1 & 0.0061 & 0.0090 & 35s & +1.1\% \\
        $n^{\mathrm{NB}}=7$ partition 1 & 0.0086 & 0.0118 & 21s & -29.7\% \\
        \arrayrulecolor{black}\bottomrule
    \end{tabular}
    \begin{tablenotes}
    \footnotesize
    \item See footnotes in Table~\ref{table_cmp_lr} for further details on the adopted convention. Refer to the text for details on the neighboring node connection strategies.
    \end{tablenotes}
    \label{table_cmp_node_conn}
    \end{threeparttable}
\end{table}

Our results suggest that when using the fixed-pivot approach, the greater the number of fixed pivots (strategy 2 vs. strategy 1), the better the model performance (Fig.~\ref{fig_cmp_node_conn}a,b). Moreover, both the fixed pivot connection strategies outperform the entirely random connection approach (strategy 3). However, the random connections based on the baseline partitioning (Eq.~\ref{eq_neiboring_set_partition}, strategy 4) outperform all other strategies compared here (strategies 1-3, Fig.~\ref{fig_cmp_node_conn}a,b). 

These trends can be explained by how different connection strategies impact message passing and representation learning. Increasing the number of fixed pivots (strategy 2) improves performance by providing richer global references and enhancing spatial dependency capture. The fixed-pivot approach (strategies 1, 2) is superior to purely random connections (strategy 3) due to its stability, which helps the model learn consistent spatial relationships. Meanwhile, the partition-based random connection strategy (strategy 4, baseline) performs the best by balancing local and global connections while maintaining flexibility, enabling more effective information propagation across the graph.

Among the partition-based selection strategies (Fig.~\ref{fig_cmp_node_conn}c,d), the approach that prioritizes nearest neighbors while preserving some mid-range connections (strategy 4) outperforms both the strategy that exclusively connects to the nearest neighbors (strategy 5) and the one that includes only mid-range and far-range nodes (strategy 6). Additionally, comparing strategies 4, 7, and 8, we found that increasing $n^{\mathrm{NB}}$ generally enhances model performance but comes at a higher computational cost (Fig.~\ref{fig_cmp_node_conn}c,d, Table~\ref{table_cmp_node_conn}). While $n^{\mathrm{NB}}=30$ performs slightly better than $n^{\mathrm{NB}}=12$, the improvement is only marginal and not worth the additional computational cost. 

Our results indicate that nearer neighbors have the most dominating impact on model performance, while mid-range neighbors provide some benefits, and far-range nodes contribute minimally. This is likely due to stronger local correlations in biomechanical behavior, making distant nodes less relevant. Increasing the number of neighbors improves performance but with diminishing returns, as higher values add computational cost without substantial accuracy gains. Therefore, our baseline partition for the neighboring node connection scheme provides the optimal balance of local and mid-range connections, enhancing efficiency while maintaining accuracy.

\subsection{Effect of the attention mechanism}\label{sec:cmp_attention}

\begin{figure}[h] 
 \begin{center}
 \includegraphics[width=1\textwidth]{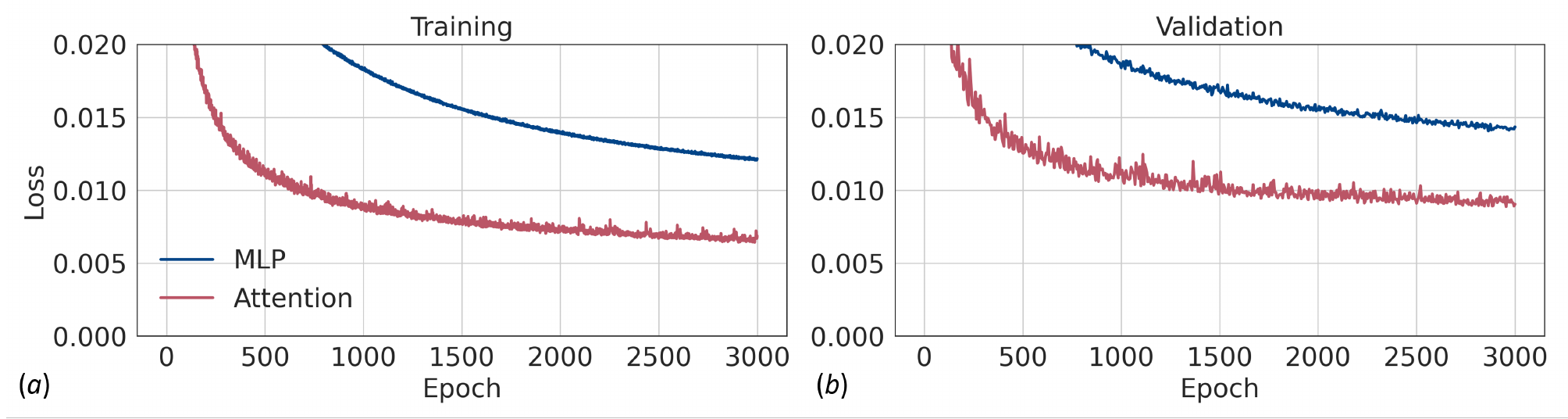}
 \end{center}
 \caption{Comparison of training (a) and validation (b) losses between the case with the attention layer and the same-size MLP layer. Light red denotes the baseline case employed in Sec.~\ref{sec:prediction_results}.}
 \label{fig_cmp_attention}
\end{figure}

We evaluated the impact of the attention mechanism on the training performance of our neural network model by comparing the effects on the training and validation losses (Table~\ref{table_cmp_attention} and Fig.~\ref{fig_cmp_attention}). In the comparison, the attention mechanism is replaced with a similar-sized MLP. The other hyperparameters, node and edge features, and model structure are retained as in Sec.~\ref{sec:prediction_results}. The attention mechanism in our GNN model significantly outperforms the MLP for message passing in predicting nodal displacements.

In the current attention mechanism, for each node $i$, the query $\mathbf{Q} = \mathbf{W}_Q \mathbf{e}_{i,k}$ represents the current node seeking information from its neighboring nodes, while the key $\mathbf{K} = \mathbf{W}_K \mathbf{e}_{i,k}$ corresponds to the neighbor's features, used to compute the attention score. The value $\mathbf{V} = \mathbf{W}_V \mathbf{e}_{i,k}$ holds the actual information from the neighbor, which will be weighted by the attention scores and aggregated to provide the final output for node $i$. This mechanism enables each node to dynamically focus on its most relevant neighbors based on their importance, improving the overall learning process. Unlike an MLP, which treats all neighbors equally, the attention mechanism adaptively assigns importance to the neighboring nodes, allowing the model to better capture local structural dependencies. This enhances the model's context awareness and reduces the influence of less informative nodes. Furthermore, while MLPs apply fixed transformations, the attention mechanism dynamically adjusts weights based on nodal interactions, making it more effective in handling non-uniform influences such as local feature variations. As a result, the attention-based approach improves accuracy and generalization in our FEA emulator.

\begin{table}[h]
    \centering
    \begin{threeparttable}
    \caption{Effect of the attention mechanism on model training and validation.}
    \begin{tabular}{c c c c c}
        \toprule
        Attention/MLP & Train loss & Val loss & Speed (/epoch) & Comparison \\
        \arrayrulecolor{black!30}\midrule
        \textcolor{red}{Attention} & 0.0067 & 0.0091 & 23s & /  \\
        MLP & 0.0121 & 0.0142 & 22s & -56.0\% \\
        \arrayrulecolor{black}\bottomrule
    \end{tabular}
    \begin{tablenotes}
    \footnotesize
    \item See footnotes in Table~\ref{table_cmp_lr} for further details on the adopted convention.
    \end{tablenotes}
    \label{table_cmp_attention}
    \end{threeparttable}
\end{table}

\subsection{Model performance on LV mechanics}\label{sec:cmp_lv}

To evaluate our model’s potential for generalization and versatility, we tested its performance on a published LV mechanics dataset~\cite{dalton2022emulation}.

We will first highlight how our dataset and approach are different in key aspects compared to the seminal work by Dalton et al.~\cite{dalton2022emulation} (Table~\ref{table_lv_data_cmp}). Many of these methodological differences are incorporated into our work to meet our core objective: to develop a GNN emulator that can accommodate variable mesh topology, which is necessary for patient-specific cardiac mechanics modeling. While our current GNN emulator, \textit{HeartSimSage}, accounts for this flexible mesh topology, such as variations in node numbers, orderings, and connectivity, the GNN emulator developed by Dalton et al.~\cite{dalton2022emulation} maintained a fixed node topology across all cases and assumed a constant pressure. 

Further, our GNN model employed Laplace-Dirichlet solutions to represent the relative nodal position, whereas Dalton et al.~\cite{dalton2022emulation} used one-hot boundary values (Sec.~\ref{sec:cmp_laplace}). Additionally, while Dalton et al.~\cite{dalton2022emulation} relied on principal component analysis (PCA)-derived latent space shape coefficients, we use actual geometric measurements to represent shape metrics. The PCA approach can be challenging for our case due to our dynamic node topology, as PCA requires a fixed node number and ordering~\cite{wold1987principal, kherif2020principal, greenacre2022principal}. Dalton et al.~\cite{dalton2022emulation} incorporated an interesting concept of augmented geometry~\cite{mrowca2018flexible,li2018learning}, which introduces multiple layers of virtual nodes to facilitate message passing. While these virtual nodes may enhance message-passing efficiency, they also introduce additional complexity in data processing, and further research is needed to understand the nuances of this approach and how it can be integrated into our framework.

Despite these differences, we trained \textit{HeartSimSage} on this LV mechanics dataset and compared its performance with the GNN model of Dalton et al.~\cite{dalton2022emulation}. It is noted that the original GNN model by Dalton et al. was implemented in JAX. Therefore, to eliminate inter-software differences from overshadowing our interpretation of results, we have implemented their GNN model in PyTorch, \textit{passive-lv-gnn-emul}. We verified it with their JAX version and used it for a direct comparison with \textit{HeartSimSage}. 

\begin{figure}[h] 
 \begin{center}
 \includegraphics[width=1\textwidth]{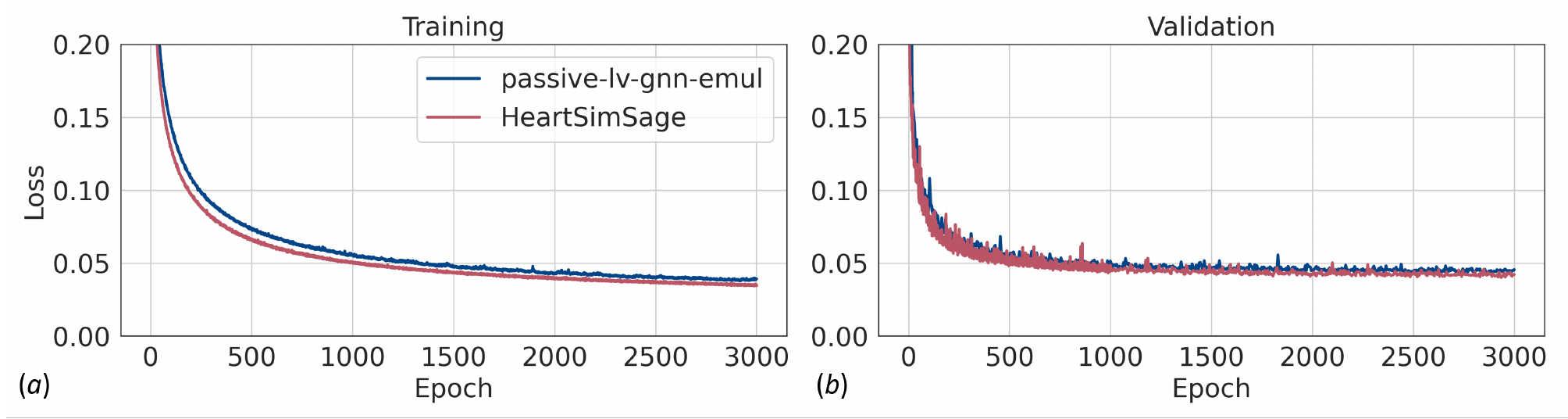}
 \end{center}
 \caption{Training (a) and validation (b) losses of the left ventricular dataset training between the case using \textit{passive-lv-gnn-emul} and \textit{HeartSimSage}.}
 \label{fig_cmp_lv}
\end{figure}

Our results show that both \textit{passive-lv-gnn-emul} and \textit{HeartSimSage} achieve comparable training and validation losses, with only marginal differences in accuracy (Table~\ref{table_cmp_lv}, Fig.~\ref{fig_cmp_lv}). However, \textit{HeartSimSage} exhibits faster training, likely due to its efficient message-passing scheme enabled by a rigorous partitioning strategy for neighboring node connections. It is worth noting that the original \textit{passive-lv-gnn-emul} implementation by Dalton et al. was developed in JAX, which can offer speed advantages for graph-based computation. Our comparison is based on a PyTorch reimplementation, which may not be fully optimized and could partially account for the observed slowdown. Additionally, we were unable to evaluate \textit{passive-lv-gnn-emul} on our biventricular dataset due to its inability to handle dynamic node topology. Overall, our findings highlight the strong generalization capability and versatility of \textit{HeartSimSage} for patient-specific cardiac mechanics modeling.

\begin{table}[ht]
\small
\centering
\begin{threeparttable}
\caption{Comparison between the dataset used in this work and Dalton et al.~\cite{dalton2022emulation}.}
\begin{tabular}{l l l}
\toprule
& Dalton et al.~\cite{dalton2022emulation} & This work \\
\arrayrulecolor{black!80}\midrule
Database & 2967 Left Ventricles (65 real) & 5000 Biventricles (150 real) \\
\arrayrulecolor{black!30}\midrule
\multirow{2}{*}{Pressure} & \multirow{2}{*}{$p_{\mathrm{lv}}=8\mathrm{mmHg}$} & $p_{\mathrm{lv}}=6-15\mathrm{mmHg}$ \\
& &$p_{\mathrm{rv}}=(0.3-0.8)p_{\mathrm{lv}}$ \\
\arrayrulecolor{black!30}\midrule
Surface B.C. & Fixed $\Gamma_{\mathrm{base}}$ only & Robin B.C. on $\Gamma_{\mathrm{base}}$ \& $\Gamma_{\mathrm{epi}}$ \\ \hline
Node Topology & Constant for all cases & Variable \\
\arrayrulecolor{black!30}\midrule
Loc. Rep. & One-hot boundary values & Laplace-Dirichlet solutions  \\
\arrayrulecolor{black!30}\midrule
Shape Parameters & PCA-derived latent values & Real geometric values \\
\arrayrulecolor{black!30}\midrule
Normalization & mean-std & max-min \\
\arrayrulecolor{black}\bottomrule
\end{tabular}
\begin{tablenotes}
    \footnotesize
    \item The pressure is assumed to be constant across all cases in Dalton et al.\cite{dalton2022emulation}. `B.C.' refers to boundary conditions. `Constant' node topology means that the node number, ordering, and connectivity remain the same for all cases. `Loc. Rep.' refers to location representation. In Dalton et al.\cite{dalton2022emulation}, the one-hot values are used to indicate whether a node is on or off a given surface. PCA stands for principal component analysis.
\end{tablenotes}
\label{table_lv_data_cmp}
\end{threeparttable}
\end{table}

\begin{table}[h]
    \centering
    \begin{threeparttable}
    \caption{\textit{HeartSimSage} performance on left ventricular mechanics dataset~\cite{dalton2022emulation}.}
    \begin{tabular}{c c c c c}
        \toprule
        Model & Train loss & Val loss & Speed (/epoch) & Comparison \\
        \arrayrulecolor{black!30}\midrule
        \color{red}{\textit{passive-lv-gnn-emul}} & 0.0390 & 0.0448 & 91s & / \\
        \textit{HeartSimSage} & 0.0350 & 0.0421 & 43s & +6.4\%  \\
        \arrayrulecolor{black}\bottomrule
    \end{tabular}
    \begin{tablenotes}
        \item See footnotes in Table~\ref{table_cmp_lr} for further details on the adopted convention.
    \end{tablenotes}
    \label{table_cmp_lv}
    \end{threeparttable}
\end{table}

\subsection{Limitations \& future work}

We will now highlight several limitations in our model. First, the model's accuracy is highly dependent on the quality of input data, particularly CT and MRI-based segmentations. Although data augmentation improves generalization, further validation on a larger and more diverse patient cohort is required. The current CT/MRI dataset may not fully capture the range of variations in human cardiac geometries. Future work will focus on expanding the dataset, especially by incorporating a broader spectrum of pathological cases, to improve the model’s generalizability and clinical relevance.

Second, our current approach utilizes basic geometric augmentations, including scaling, random shearing, and elastic deformations. To generate geometries that more closely resemble the dataset, we will explore advanced techniques such as variational autoencoders (VAE)~\cite{molnar2022representation, girin2020dynamical} and generative adversarial network autoencoders (GAN-AE)~\cite{zhang2023starnet} in a follow-up study.

Third, while we have extensively researched the design of the neighboring node connection strategy and compared their performance through rigorous sensitivity analysis to determine the optimal neighboring node connection strategy, the current version may not be optimal when considering pathological biventricular geometries, including hypertrophied ventricles, single ventricle physiology, or ventricles with structural details such as papillary muscles. Further studies are needed to assess the performance of the current approach in these cases and explore other effective node connection designs.

Fourth, although we incorporated fiber orientations as part of the node features, we applied the same fiber rules and parameters to generate the fiber directions for all cases. We did not vary the rule generation parameters for individual cases, which could have a meaningful impact on the displacement prediction~\cite{shi2024optimization}. However, as the node features contain the fiber orientations, the dataset still retains a degree of generality regarding these directions. Nevertheless, further evaluation of the sensitivity of fiber orientations is warranted.

Fifth, our method does not explicitly incorporate the governing physics equations, which may impact its training efficiency and prediction accuracy compared to traditional solvers that fully enforce these principles. However, recent advancements, such as the neural network finite element (NNFE) approach~\cite{motiwale2023high, motiwale2024neural, goodbrake2024neural} and physics-informed graph neural networks~\cite{dalton2023physics}, have demonstrated promising capabilities for accelerating cardiac mechanics simulations. Integrating similar concepts, such as virtual work principles, could further improve the predictive power and computational efficiency of our approach.

Additionally, we will expand the model's capabilities to address inverse problems, such as estimating material parameters from displacement fields obtained from image data, and perform uncertainty quantification (UQ) analysis to build confidence in AI-driven simulations for personalized diagnostics and treatment planning. We plan to extend this framework to other cardiac regions, such as the atria, and explore its application to other organs, including the stomach, esophagus, and uterus. We also aim to develop time-series neural network models to predict whole-heart motion throughout the cardiac cycle. This approach can potentially advance AI-driven, generalized solvers for a wide range of physics-based problems, further expanding the role of neural-network-based modeling in biomedical applications and other domains.

\section{Conclusion} 

We developed a novel attention-enhanced graph neural network (GNN), \textit{HeartSimSage}, as an emulator for finite element analysis (FEA), particularly designed and optimized for patient-specific cardiac mechanics modeling. Our model accurately and efficiently predicts the biventricular passive myocardial displacements from a patient-specific geometry, chamber pressures, and material parameters while accommodating diverse three-dimensional (3D) geometries, heterogeneous tissue properties, and physiological boundary conditions. A prominent feature of \textit{HeartSimSage} is that it supports a dynamic node topology with varying node counts, orderings, and connectivity. Our model incorporates GraphSAGE for optimized node interactions, Laplace-Dirichlet encoding for improved spatial representation, and subset-based training for enhanced efficiency. An attention-enhanced mechanism further refines information flow by adaptively weighting neighbor contributions and filtering irrelevant data, enabling the model to handle complex deformations effectively and efficiently. Achieving a computational speedup of approximately $1.3 \times 10^4$ times on GPU and $1.9 \times 10^2$ times on CPU compared to traditional FEA, \textit{HeartSimSage} maintains a reasonably low mean error of $0.13\% \pm 0.12\%$ in the predicted displacements. Differences in point-wise displacements were below $0.4~mm$ in most cases between FEA and \textit{HeartSimSage}-predictions.

We believe this work represents a significant advancement in neural-network-based FEA emulation of ventricular mechanics, offering a scalable and accurate alternative to traditional solvers. Integrating more biomechanically realistic tissue properties, uncertainty quantification, and expanding to other cardiac regions and organs will further enhance the model’s versatility and clinical applicability. It paves the way for AI-driven, generalized solvers that can address a broad range of physics-based problems, with potential applications in personalized diagnostics, treatment planning, and beyond.

\clearpage     

\section*{Acknowledgement}
The authors would like to acknowledge financial support from the American Heart Association’s Second Century Early Faculty Independence Award ($\#$24SCEFIA1260268) in performing this work. The authors would also like to acknowledge computing resources and services received through the Columbia Shared Research Computing Facility and the Ginsburg High-Performance Computing Cluster.

\bibliographystyle{model1-num-names}

\bibliography{Shi2020Paper} 

\begin{thebibliography}{119}
\expandafter\ifx\csname natexlab\endcsname\relax\def\natexlab#1{#1}\fi
\providecommand{\url}[1]{\texttt{#1}}
\providecommand{\href}[2]{#2}
\providecommand{\path}[1]{#1}
\providecommand{\DOIprefix}{doi:}
\providecommand{\ArXivprefix}{arXiv:}
\providecommand{\URLprefix}{URL: }
\providecommand{\Pubmedprefix}{pmid:}
\providecommand{\doi}[1]{\href{http://dx.doi.org/#1}{\path{#1}}}
\providecommand{\Pubmed}[1]{\href{pmid:#1}{\path{#1}}}
\providecommand{\bibinfo}[2]{#2}
\ifx\xfnm\relax \def\xfnm[#1]{\unskip,\space#1}\fi
\bibitem[{Trayanova et~al.(2011)Trayanova, Constantino, and Gurev}]{trayanova2011electromechanical}
\bibinfo{author}{N.~A. Trayanova}, \bibinfo{author}{J.~Constantino}, \bibinfo{author}{V.~Gurev},
\newblock \bibinfo{title}{Electromechanical models of the ventricles},
\newblock \bibinfo{journal}{American Journal of Physiology-Heart and Circulatory Physiology} \bibinfo{volume}{301} (\bibinfo{year}{2011}) \bibinfo{pages}{H279--H286}.
\bibitem[{Campbell and McCulloch(2011)}]{campbell2011multi}
\bibinfo{author}{S.~G. Campbell}, \bibinfo{author}{A.~D. McCulloch},
\newblock \bibinfo{title}{Multi-scale computational models of familial hypertrophic cardiomyopathy: genotype to phenotype},
\newblock \bibinfo{journal}{Journal of The Royal Society Interface} \bibinfo{volume}{8} (\bibinfo{year}{2011}) \bibinfo{pages}{1550--1561}.
\bibitem[{Marsden and Esmaily-Moghadam(2015)}]{marsden2015multiscale}
\bibinfo{author}{A.~L. Marsden}, \bibinfo{author}{M.~Esmaily-Moghadam},
\newblock \bibinfo{title}{Multiscale modeling of cardiovascular flows for clinical decision support},
\newblock \bibinfo{journal}{Applied Mechanics Reviews} \bibinfo{volume}{67} (\bibinfo{year}{2015}) \bibinfo{pages}{030804}.
\bibitem[{Palit et~al.(2015)Palit, Bhudia, Arvanitis, Turley, and Williams}]{palit2015computational}
\bibinfo{author}{A.~Palit}, \bibinfo{author}{S.~K. Bhudia}, \bibinfo{author}{T.~N. Arvanitis}, \bibinfo{author}{G.~A. Turley}, \bibinfo{author}{M.~A. Williams},
\newblock \bibinfo{title}{Computational modelling of left-ventricular diastolic mechanics: Effect of fibre orientation and right-ventricle topology},
\newblock \bibinfo{journal}{Journal of biomechanics} \bibinfo{volume}{48} (\bibinfo{year}{2015}) \bibinfo{pages}{604--612}.
\bibitem[{Mittal et~al.(2016)Mittal, Seo, Vedula, Choi, Liu, Huang, Jain, Younes, Abraham, and George}]{mittal2016computational}
\bibinfo{author}{R.~Mittal}, \bibinfo{author}{J.~H. Seo}, \bibinfo{author}{V.~Vedula}, \bibinfo{author}{Y.~J. Choi}, \bibinfo{author}{H.~Liu}, \bibinfo{author}{H.~H. Huang}, \bibinfo{author}{S.~Jain}, \bibinfo{author}{L.~Younes}, \bibinfo{author}{T.~Abraham}, \bibinfo{author}{R.~T. George},
\newblock \bibinfo{title}{Computational modeling of cardiac hemodynamics: current status and future outlook},
\newblock \bibinfo{journal}{Journal of Computational Physics} \bibinfo{volume}{305} (\bibinfo{year}{2016}) \bibinfo{pages}{1065--1082}.
\bibitem[{Sotiropoulos et~al.(2016)Sotiropoulos, Le, and Gilmanov}]{sotiropoulos2016fluid}
\bibinfo{author}{F.~Sotiropoulos}, \bibinfo{author}{T.~B. Le}, \bibinfo{author}{A.~Gilmanov},
\newblock \bibinfo{title}{Fluid mechanics of heart valves and their replacements},
\newblock \bibinfo{journal}{Annual Review of Fluid Mechanics} \bibinfo{volume}{48} (\bibinfo{year}{2016}) \bibinfo{pages}{259--283}.
\bibitem[{Vedula et~al.(2017)Vedula, Lee, Xu, Kuo, Hsiai, and Marsden}]{vedula2017method}
\bibinfo{author}{V.~Vedula}, \bibinfo{author}{J.~Lee}, \bibinfo{author}{H.~Xu}, \bibinfo{author}{C.-C.~J. Kuo}, \bibinfo{author}{T.~K. Hsiai}, \bibinfo{author}{A.~L. Marsden},
\newblock \bibinfo{title}{A method to quantify mechanobiologic forces during zebrafish cardiac development using 4-d light sheet imaging and computational modeling},
\newblock \bibinfo{journal}{PLoS computational biology} \bibinfo{volume}{13} (\bibinfo{year}{2017}) \bibinfo{pages}{e1005828}.
\bibitem[{Finsberg et~al.(2018)Finsberg, Xi, Tan, Zhong, Genet, Sundnes, Lee, and Wall}]{finsberg2018efficient}
\bibinfo{author}{H.~Finsberg}, \bibinfo{author}{C.~Xi}, \bibinfo{author}{J.~L. Tan}, \bibinfo{author}{L.~Zhong}, \bibinfo{author}{M.~Genet}, \bibinfo{author}{J.~Sundnes}, \bibinfo{author}{L.~C. Lee}, \bibinfo{author}{S.~T. Wall},
\newblock \bibinfo{title}{Efficient estimation of personalized biventricular mechanical function employing gradient-based optimization},
\newblock \bibinfo{journal}{International journal for numerical methods in biomedical engineering} \bibinfo{volume}{34} (\bibinfo{year}{2018}) \bibinfo{pages}{e2982}.
\bibitem[{Lee et~al.(2018)Lee, Vedula, Baek, Chen, Hsu, Ding, Chang, Kang, Small, Fei et~al.}]{lee2018spatial}
\bibinfo{author}{J.~Lee}, \bibinfo{author}{V.~Vedula}, \bibinfo{author}{K.~I. Baek}, \bibinfo{author}{J.~Chen}, \bibinfo{author}{J.~J. Hsu}, \bibinfo{author}{Y.~Ding}, \bibinfo{author}{C.-C. Chang}, \bibinfo{author}{H.~Kang}, \bibinfo{author}{A.~Small}, \bibinfo{author}{P.~Fei}, et~al.,
\newblock \bibinfo{title}{Spatial and temporal variations in hemodynamic forces initiate cardiac trabeculation},
\newblock \bibinfo{journal}{JCI insight} \bibinfo{volume}{3} (\bibinfo{year}{2018}).
\bibitem[{Pfaller et~al.(2019)Pfaller, H{\"o}rmann, Weigl, Nagler, Chabiniok, Bertoglio, and Wall}]{pfaller2019importance}
\bibinfo{author}{M.~R. Pfaller}, \bibinfo{author}{J.~M. H{\"o}rmann}, \bibinfo{author}{M.~Weigl}, \bibinfo{author}{A.~Nagler}, \bibinfo{author}{R.~Chabiniok}, \bibinfo{author}{C.~Bertoglio}, \bibinfo{author}{W.~A. Wall},
\newblock \bibinfo{title}{The importance of the pericardium for cardiac biomechanics: from physiology to computational modeling},
\newblock \bibinfo{journal}{Biomechanics and modeling in mechanobiology} \bibinfo{volume}{18} (\bibinfo{year}{2019}) \bibinfo{pages}{503--529}.
\bibitem[{Yoshida and Holmes(2021)}]{yoshida2021computational}
\bibinfo{author}{K.~Yoshida}, \bibinfo{author}{J.~W. Holmes},
\newblock \bibinfo{title}{Computational models of cardiac hypertrophy},
\newblock \bibinfo{journal}{Progress in biophysics and molecular biology} \bibinfo{volume}{159} (\bibinfo{year}{2021}) \bibinfo{pages}{75--85}.
\bibitem[{Schwarz et~al.(2023)Schwarz, Pegolotti, Pfaller, and Marsden}]{schwarz2023beyond}
\bibinfo{author}{E.~L. Schwarz}, \bibinfo{author}{L.~Pegolotti}, \bibinfo{author}{M.~R. Pfaller}, \bibinfo{author}{A.~L. Marsden},
\newblock \bibinfo{title}{Beyond cfd: Emerging methodologies for predictive simulation in cardiovascular health and disease},
\newblock \bibinfo{journal}{Biophysics Reviews} \bibinfo{volume}{4} (\bibinfo{year}{2023}).
\bibitem[{Green et~al.(2024)Green, Chan, Tulzer, Tulzer, and Yap}]{green2024myocardial}
\bibinfo{author}{L.~Green}, \bibinfo{author}{W.~X. Chan}, \bibinfo{author}{A.~Tulzer}, \bibinfo{author}{G.~Tulzer}, \bibinfo{author}{C.~H. Yap},
\newblock \bibinfo{title}{Myocardial biomechanical effects of fetal aortic valvuloplasty},
\newblock \bibinfo{journal}{Biomechanics and Modeling in Mechanobiology} \bibinfo{volume}{23} (\bibinfo{year}{2024}) \bibinfo{pages}{1433--1448}.
\bibitem[{Corral-Acero et~al.(2020)Corral-Acero, Margara, Marciniak, Rodero, Loncaric, Feng, Gilbert, Fernandes, Bukhari, Wajdan et~al.}]{corral2020digital}
\bibinfo{author}{J.~Corral-Acero}, \bibinfo{author}{F.~Margara}, \bibinfo{author}{M.~Marciniak}, \bibinfo{author}{C.~Rodero}, \bibinfo{author}{F.~Loncaric}, \bibinfo{author}{Y.~Feng}, \bibinfo{author}{A.~Gilbert}, \bibinfo{author}{J.~F. Fernandes}, \bibinfo{author}{H.~A. Bukhari}, \bibinfo{author}{A.~Wajdan}, et~al.,
\newblock \bibinfo{title}{The ‘digital twin’to enable the vision of precision cardiology},
\newblock \bibinfo{journal}{European heart journal} \bibinfo{volume}{41} (\bibinfo{year}{2020}) \bibinfo{pages}{4556--4564}.
\bibitem[{Qureshi et~al.(2023)Qureshi, Lip, Nordsletten, Williams, Aslanidi, and De~Vecchi}]{qureshi2023imaging}
\bibinfo{author}{A.~Qureshi}, \bibinfo{author}{G.~Y. Lip}, \bibinfo{author}{D.~A. Nordsletten}, \bibinfo{author}{S.~E. Williams}, \bibinfo{author}{O.~Aslanidi}, \bibinfo{author}{A.~De~Vecchi},
\newblock \bibinfo{title}{Imaging and biophysical modelling of thrombogenic mechanisms in atrial fibrillation and stroke},
\newblock \bibinfo{journal}{Frontiers in Cardiovascular Medicine} \bibinfo{volume}{9} (\bibinfo{year}{2023}) \bibinfo{pages}{1074562}.
\bibitem[{Qureshi et~al.(2022)Qureshi, Balmus, Lip, Williams, Nordsletten, Aslanidi, and De~Vecchi}]{qureshi2022mechanistic}
\bibinfo{author}{A.~Qureshi}, \bibinfo{author}{M.~Balmus}, \bibinfo{author}{G.~Lip}, \bibinfo{author}{S.~Williams}, \bibinfo{author}{D.~Nordsletten}, \bibinfo{author}{O.~Aslanidi}, \bibinfo{author}{A.~De~Vecchi},
\newblock \bibinfo{title}{Mechanistic modelling of virchows triad to assess thrombogenicity and stroke risk in atrial fibrillation patients},
\newblock \bibinfo{journal}{European Heart Journal-Digital Health} \bibinfo{volume}{3} (\bibinfo{year}{2022}) \bibinfo{pages}{ztac076--2788}.
\bibitem[{Bazzi et~al.(2022)Bazzi, Balouchzadeh, Pavey, Quirk, Yanagisawa, Vedula, Wagenseil, and Barocas}]{bazzi2022experimental}
\bibinfo{author}{M.~S. Bazzi}, \bibinfo{author}{R.~Balouchzadeh}, \bibinfo{author}{S.~N. Pavey}, \bibinfo{author}{J.~D. Quirk}, \bibinfo{author}{H.~Yanagisawa}, \bibinfo{author}{V.~Vedula}, \bibinfo{author}{J.~E. Wagenseil}, \bibinfo{author}{V.~H. Barocas},
\newblock \bibinfo{title}{Experimental and mouse-specific computational models of the fbln4 smko mouse to identify potential biomarkers for ascending thoracic aortic aneurysm},
\newblock \bibinfo{journal}{Cardiovascular engineering and technology} \bibinfo{volume}{13} (\bibinfo{year}{2022}) \bibinfo{pages}{558--572}.
\bibitem[{Crozier et~al.(2016)Crozier, Augustin, Neic, Prassl, Holler, Fastl, Hennemuth, Bredies, Kuehne, Bishop et~al.}]{crozier2016image}
\bibinfo{author}{A.~Crozier}, \bibinfo{author}{C.~Augustin}, \bibinfo{author}{A.~Neic}, \bibinfo{author}{A.~Prassl}, \bibinfo{author}{M.~Holler}, \bibinfo{author}{T.~Fastl}, \bibinfo{author}{A.~Hennemuth}, \bibinfo{author}{K.~Bredies}, \bibinfo{author}{T.~Kuehne}, \bibinfo{author}{M.~Bishop}, et~al.,
\newblock \bibinfo{title}{Image-based personalization of cardiac anatomy for coupled electromechanical modeling},
\newblock \bibinfo{journal}{Annals of biomedical engineering} \bibinfo{volume}{44} (\bibinfo{year}{2016}) \bibinfo{pages}{58--70}.
\bibitem[{Marx et~al.(2022)Marx, Niestrawska, Gsell, Caforio, Plank, and Augustin}]{marx2022robust}
\bibinfo{author}{L.~Marx}, \bibinfo{author}{J.~A. Niestrawska}, \bibinfo{author}{M.~A. Gsell}, \bibinfo{author}{F.~Caforio}, \bibinfo{author}{G.~Plank}, \bibinfo{author}{C.~M. Augustin},
\newblock \bibinfo{title}{Robust and efficient fixed-point algorithm for the inverse elastostatic problem to identify myocardial passive material parameters and the unloaded reference configuration},
\newblock \bibinfo{journal}{Journal of computational physics} \bibinfo{volume}{463} (\bibinfo{year}{2022}) \bibinfo{pages}{111266}.
\bibitem[{Balaban et~al.(2017)Balaban, Finsberg, Odland, Rognes, Ross, Sundnes, and Wall}]{balaban2017high}
\bibinfo{author}{G.~Balaban}, \bibinfo{author}{H.~Finsberg}, \bibinfo{author}{H.~H. Odland}, \bibinfo{author}{M.~E. Rognes}, \bibinfo{author}{S.~Ross}, \bibinfo{author}{J.~Sundnes}, \bibinfo{author}{S.~Wall},
\newblock \bibinfo{title}{High-resolution data assimilation of cardiac mechanics applied to a dyssynchronous ventricle},
\newblock \bibinfo{journal}{International journal for numerical methods in biomedical engineering} \bibinfo{volume}{33} (\bibinfo{year}{2017}) \bibinfo{pages}{e2863}.
\bibitem[{Gonzalo et~al.(2024)Gonzalo, Augustin, Bifulco, Telle, Chahine, Kassar, Guerrero-Hurtado, Dur{\'a}n, Mart{\'\i}nez-Legazpi, Flores et~al.}]{gonzalo2024multiphysics}
\bibinfo{author}{A.~Gonzalo}, \bibinfo{author}{C.~M. Augustin}, \bibinfo{author}{S.~F. Bifulco}, \bibinfo{author}{{\AA}.~Telle}, \bibinfo{author}{Y.~Chahine}, \bibinfo{author}{A.~Kassar}, \bibinfo{author}{M.~Guerrero-Hurtado}, \bibinfo{author}{E.~Dur{\'a}n}, \bibinfo{author}{P.~Mart{\'\i}nez-Legazpi}, \bibinfo{author}{O.~Flores}, et~al.,
\newblock \bibinfo{title}{Multiphysics simulations reveal haemodynamic impacts of patient-derived fibrosis-related changes in left atrial tissue mechanics},
\newblock \bibinfo{journal}{The Journal of Physiology} \bibinfo{volume}{602} (\bibinfo{year}{2024}) \bibinfo{pages}{6789--6812}.
\bibitem[{Mangion et~al.(2018)Mangion, Gao, Husmeier, Luo, and Berry}]{mangion2018advances}
\bibinfo{author}{K.~Mangion}, \bibinfo{author}{H.~Gao}, \bibinfo{author}{D.~Husmeier}, \bibinfo{author}{X.~Luo}, \bibinfo{author}{C.~Berry},
\newblock \bibinfo{title}{Advances in computational modelling for personalised medicine after myocardial infarction},
\newblock \bibinfo{journal}{Heart} \bibinfo{volume}{104} (\bibinfo{year}{2018}) \bibinfo{pages}{550--557}.
\bibitem[{Boyle et~al.(2019)Boyle, Zghaib, Zahid, Ali, Deng, Franceschi, Hakim, Murphy, Prakosa, Zimmerman et~al.}]{boyle2019computationally}
\bibinfo{author}{P.~M. Boyle}, \bibinfo{author}{T.~Zghaib}, \bibinfo{author}{S.~Zahid}, \bibinfo{author}{R.~L. Ali}, \bibinfo{author}{D.~Deng}, \bibinfo{author}{W.~H. Franceschi}, \bibinfo{author}{J.~B. Hakim}, \bibinfo{author}{M.~J. Murphy}, \bibinfo{author}{A.~Prakosa}, \bibinfo{author}{S.~L. Zimmerman}, et~al.,
\newblock \bibinfo{title}{Computationally guided personalized targeted ablation of persistent atrial fibrillation},
\newblock \bibinfo{journal}{Nature biomedical engineering} \bibinfo{volume}{3} (\bibinfo{year}{2019}) \bibinfo{pages}{870--879}.
\bibitem[{Niederer et~al.(2021)Niederer, Sacks, Girolami, and Willcox}]{niederer2021scaling}
\bibinfo{author}{S.~A. Niederer}, \bibinfo{author}{M.~S. Sacks}, \bibinfo{author}{M.~Girolami}, \bibinfo{author}{K.~Willcox},
\newblock \bibinfo{title}{Scaling digital twins from the artisanal to the industrial},
\newblock \bibinfo{journal}{Nature Computational Science} \bibinfo{volume}{1} (\bibinfo{year}{2021}) \bibinfo{pages}{313--320}.
\bibitem[{Stimm et~al.(2022)Stimm, Nordsletten, Jilberto, Miller, Berbero{\u{g}}lu, Kozerke, and Stoeck}]{stimm2022personalization}
\bibinfo{author}{J.~Stimm}, \bibinfo{author}{D.~A. Nordsletten}, \bibinfo{author}{J.~Jilberto}, \bibinfo{author}{R.~Miller}, \bibinfo{author}{E.~Berbero{\u{g}}lu}, \bibinfo{author}{S.~Kozerke}, \bibinfo{author}{C.~T. Stoeck},
\newblock \bibinfo{title}{Personalization of biomechanical simulations of the left ventricle by in-vivo cardiac dti data: Impact of fiber interpolation methods},
\newblock \bibinfo{journal}{Frontiers in Physiology} \bibinfo{volume}{13} (\bibinfo{year}{2022}) \bibinfo{pages}{2485}.
\bibitem[{Arevalo et~al.(2016)Arevalo, Vadakkumpadan, Guallar, Jebb, Malamas, Wu, and Trayanova}]{arevalo2016arrhythmia}
\bibinfo{author}{H.~J. Arevalo}, \bibinfo{author}{F.~Vadakkumpadan}, \bibinfo{author}{E.~Guallar}, \bibinfo{author}{A.~Jebb}, \bibinfo{author}{P.~Malamas}, \bibinfo{author}{K.~C. Wu}, \bibinfo{author}{N.~A. Trayanova},
\newblock \bibinfo{title}{Arrhythmia risk stratification of patients after myocardial infarction using personalized heart models},
\newblock \bibinfo{journal}{Nature communications} \bibinfo{volume}{7} (\bibinfo{year}{2016}) \bibinfo{pages}{11437}.
\bibitem[{Gillette et~al.(2022)Gillette, Gsell, Strocchi, Grandits, Neic, Manninger, Scherr, Roney, Prassl, Augustin et~al.}]{gillette2022personalized}
\bibinfo{author}{K.~Gillette}, \bibinfo{author}{M.~A. Gsell}, \bibinfo{author}{M.~Strocchi}, \bibinfo{author}{T.~Grandits}, \bibinfo{author}{A.~Neic}, \bibinfo{author}{M.~Manninger}, \bibinfo{author}{D.~Scherr}, \bibinfo{author}{C.~H. Roney}, \bibinfo{author}{A.~J. Prassl}, \bibinfo{author}{C.~M. Augustin}, et~al.,
\newblock \bibinfo{title}{A personalized real-time virtual model of whole heart electrophysiology},
\newblock \bibinfo{journal}{Frontiers in Physiology}  (\bibinfo{year}{2022}) \bibinfo{pages}{1860}.
\bibitem[{Brown et~al.(2024)Brown, Salvador, Shi, Pfaller, Hu, Harold, Hsiai, Vedula, and Marsden}]{brown2024modular}
\bibinfo{author}{A.~L. Brown}, \bibinfo{author}{M.~Salvador}, \bibinfo{author}{L.~Shi}, \bibinfo{author}{M.~R. Pfaller}, \bibinfo{author}{Z.~Hu}, \bibinfo{author}{K.~E. Harold}, \bibinfo{author}{T.~Hsiai}, \bibinfo{author}{V.~Vedula}, \bibinfo{author}{A.~L. Marsden},
\newblock \bibinfo{title}{A modular framework for implicit 3d--0d coupling in cardiac mechanics},
\newblock \bibinfo{journal}{Computer Methods in Applied Mechanics and Engineering} \bibinfo{volume}{421} (\bibinfo{year}{2024}) \bibinfo{pages}{116764}.
\bibitem[{Shi et~al.(2024)Shi, Chen, Takayama, and Vedula}]{shi2024optimization}
\bibinfo{author}{L.~Shi}, \bibinfo{author}{I.~Y. Chen}, \bibinfo{author}{H.~Takayama}, \bibinfo{author}{V.~Vedula},
\newblock \bibinfo{title}{An optimization framework to personalize passive cardiac mechanics},
\newblock \bibinfo{journal}{Computer Methods in Applied Mechanics and Engineering} \bibinfo{volume}{432} (\bibinfo{year}{2024}) \bibinfo{pages}{117401}.
\bibitem[{Jiang et~al.(2024)Jiang, Yan, Wang, Chen, and Cai}]{jiang2024highly}
\bibinfo{author}{Y.~Jiang}, \bibinfo{author}{Z.~Yan}, \bibinfo{author}{X.~Wang}, \bibinfo{author}{R.~Chen}, \bibinfo{author}{X.-C. Cai},
\newblock \bibinfo{title}{A highly parallel algorithm for simulating the elastodynamics of a patient-specific human heart with four chambers using a heterogeneous hyperelastic model},
\newblock \bibinfo{journal}{Journal of Computational Physics} \bibinfo{volume}{508} (\bibinfo{year}{2024}) \bibinfo{pages}{113027}.
\bibitem[{Feng et~al.(2024)Feng, Gao, and Luo}]{feng2024whole}
\bibinfo{author}{L.~Feng}, \bibinfo{author}{H.~Gao}, \bibinfo{author}{X.~Luo},
\newblock \bibinfo{title}{Whole-heart modelling with valves in a fluid--structure interaction framework},
\newblock \bibinfo{journal}{Computer Methods in Applied Mechanics and Engineering} \bibinfo{volume}{420} (\bibinfo{year}{2024}) \bibinfo{pages}{116724}.
\bibitem[{Augustin et~al.(2016)Augustin, Neic, Liebmann, Prassl, Niederer, Haase, and Plank}]{augustin2016anatomically}
\bibinfo{author}{C.~M. Augustin}, \bibinfo{author}{A.~Neic}, \bibinfo{author}{M.~Liebmann}, \bibinfo{author}{A.~J. Prassl}, \bibinfo{author}{S.~A. Niederer}, \bibinfo{author}{G.~Haase}, \bibinfo{author}{G.~Plank},
\newblock \bibinfo{title}{Anatomically accurate high resolution modeling of human whole heart electromechanics: a strongly scalable algebraic multigrid solver method for nonlinear deformation},
\newblock \bibinfo{journal}{Journal of computational physics} \bibinfo{volume}{305} (\bibinfo{year}{2016}) \bibinfo{pages}{622--646}.
\bibitem[{Bucelli et~al.(2023)Bucelli, Zingaro, Africa, Fumagalli, Dede', and Quarteroni}]{bucelli2023mathematical}
\bibinfo{author}{M.~Bucelli}, \bibinfo{author}{A.~Zingaro}, \bibinfo{author}{P.~C. Africa}, \bibinfo{author}{I.~Fumagalli}, \bibinfo{author}{L.~Dede'}, \bibinfo{author}{A.~Quarteroni},
\newblock \bibinfo{title}{A mathematical model that integrates cardiac electrophysiology, mechanics, and fluid dynamics: Application to the human left heart},
\newblock \bibinfo{journal}{International journal for numerical methods in biomedical engineering} \bibinfo{volume}{39} (\bibinfo{year}{2023}) \bibinfo{pages}{e3678}.
\bibitem[{Dalton et~al.(2022)Dalton, Gao, and Husmeier}]{dalton2022emulation}
\bibinfo{author}{D.~Dalton}, \bibinfo{author}{H.~Gao}, \bibinfo{author}{D.~Husmeier},
\newblock \bibinfo{title}{Emulation of cardiac mechanics using graph neural networks},
\newblock \bibinfo{journal}{Computer Methods in Applied Mechanics and Engineering} \bibinfo{volume}{401} (\bibinfo{year}{2022}) \bibinfo{pages}{115645}.
\bibitem[{Salvador et~al.(2023)Salvador, Regazzoni, Quarteroni et~al.}]{salvador2023fast}
\bibinfo{author}{M.~Salvador}, \bibinfo{author}{F.~Regazzoni}, \bibinfo{author}{A.~Quarteroni}, et~al.,
\newblock \bibinfo{title}{Fast and robust parameter estimation with uncertainty quantification for the cardiac function},
\newblock \bibinfo{journal}{Computer Methods and Programs in Biomedicine} \bibinfo{volume}{231} (\bibinfo{year}{2023}) \bibinfo{pages}{107402}.
\bibitem[{Salvador et~al.(2024)Salvador, Kong, Peirlinck, Parker, Chubb, Dubin, and Marsden}]{salvador2024digital}
\bibinfo{author}{M.~Salvador}, \bibinfo{author}{F.~Kong}, \bibinfo{author}{M.~Peirlinck}, \bibinfo{author}{D.~W. Parker}, \bibinfo{author}{H.~Chubb}, \bibinfo{author}{A.~M. Dubin}, \bibinfo{author}{A.~L. Marsden},
\newblock \bibinfo{title}{Digital twinning of cardiac electrophysiology for congenital heart disease},
\newblock \bibinfo{journal}{Journal of the Royal Society Interface} \bibinfo{volume}{21} (\bibinfo{year}{2024}) \bibinfo{pages}{20230729}.
\bibitem[{Yun et~al.(2014)Yun, Kim, Shin, Lee, Deshpande, and Kim}]{yun2014statistical}
\bibinfo{author}{Y.~Yun}, \bibinfo{author}{H.-C. Kim}, \bibinfo{author}{S.~Y. Shin}, \bibinfo{author}{J.~Lee}, \bibinfo{author}{A.~D. Deshpande}, \bibinfo{author}{C.~Kim},
\newblock \bibinfo{title}{Statistical method for prediction of gait kinematics with gaussian process regression},
\newblock \bibinfo{journal}{Journal of biomechanics} \bibinfo{volume}{47} (\bibinfo{year}{2014}) \bibinfo{pages}{186--192}.
\bibitem[{Tepole et~al.(2022)Tepole, Zhang, and Gomez}]{tepole2022data}
\bibinfo{author}{A.~B. Tepole}, \bibinfo{author}{J.~Zhang}, \bibinfo{author}{H.~Gomez},
\newblock \bibinfo{title}{Data-driven methods in biomechanics},
\newblock \bibinfo{journal}{Journal of biomechanical engineering} \bibinfo{volume}{144} (\bibinfo{year}{2022}) \bibinfo{pages}{120301}.
\bibitem[{Quarteroni et~al.(2025)Quarteroni, Gervasio, and Regazzoni}]{quarteroni2025combining}
\bibinfo{author}{A.~Quarteroni}, \bibinfo{author}{P.~Gervasio}, \bibinfo{author}{F.~Regazzoni},
\newblock \bibinfo{title}{Combining physics-based and data-driven models: advancing the frontiers of research with scientific machine learning},
\newblock \bibinfo{journal}{arXiv preprint arXiv:2501.18708}  (\bibinfo{year}{2025}).
\bibitem[{Peirlinck et~al.(2024)Peirlinck, Hurtado, Rausch, Tepole, and Kuhl}]{peirlinck2024universal}
\bibinfo{author}{M.~Peirlinck}, \bibinfo{author}{J.~A. Hurtado}, \bibinfo{author}{M.~K. Rausch}, \bibinfo{author}{A.~B. Tepole}, \bibinfo{author}{E.~Kuhl},
\newblock \bibinfo{title}{A universal material model subroutine for soft matter systems},
\newblock \bibinfo{journal}{Engineering with Computers}  (\bibinfo{year}{2024}) \bibinfo{pages}{1--23}.
\bibitem[{Sahli~Costabal et~al.(2020)Sahli~Costabal, Yang, Perdikaris, Hurtado, and Kuhl}]{sahli2020physics}
\bibinfo{author}{F.~Sahli~Costabal}, \bibinfo{author}{Y.~Yang}, \bibinfo{author}{P.~Perdikaris}, \bibinfo{author}{D.~E. Hurtado}, \bibinfo{author}{E.~Kuhl},
\newblock \bibinfo{title}{Physics-informed neural networks for cardiac activation mapping},
\newblock \bibinfo{journal}{Frontiers in Physics} \bibinfo{volume}{8} (\bibinfo{year}{2020}) \bibinfo{pages}{42}.
\bibitem[{Melis et~al.(2017)Melis, Clayton, and Marzo}]{melis2017bayesian}
\bibinfo{author}{A.~Melis}, \bibinfo{author}{R.~H. Clayton}, \bibinfo{author}{A.~Marzo},
\newblock \bibinfo{title}{Bayesian sensitivity analysis of a 1d vascular model with gaussian process emulators},
\newblock \bibinfo{journal}{International Journal for Numerical Methods in Biomedical Engineering} \bibinfo{volume}{33} (\bibinfo{year}{2017}) \bibinfo{pages}{e2882}.
\bibitem[{Ranftl et~al.(2023)Ranftl, M{\"u}ller, Windberger, Brenn, and von~der Linden}]{ranftl2023bayesian}
\bibinfo{author}{S.~Ranftl}, \bibinfo{author}{T.~S. M{\"u}ller}, \bibinfo{author}{U.~Windberger}, \bibinfo{author}{G.~Brenn}, \bibinfo{author}{W.~von~der Linden},
\newblock \bibinfo{title}{A bayesian approach to blood rheological uncertainties in aortic hemodynamics},
\newblock \bibinfo{journal}{International Journal for Numerical Methods in Biomedical Engineering} \bibinfo{volume}{39} (\bibinfo{year}{2023}) \bibinfo{pages}{e3576}.
\bibitem[{Khatibisepehr et~al.(2013)Khatibisepehr, Huang, and Khare}]{khatibisepehr2013design}
\bibinfo{author}{S.~Khatibisepehr}, \bibinfo{author}{B.~Huang}, \bibinfo{author}{S.~Khare},
\newblock \bibinfo{title}{Design of inferential sensors in the process industry: A review of bayesian methods},
\newblock \bibinfo{journal}{Journal of Process Control} \bibinfo{volume}{23} (\bibinfo{year}{2013}) \bibinfo{pages}{1575--1596}.
\bibitem[{Liu(2023)}]{liu2023bayesian}
\bibinfo{author}{Y.~Liu}, \bibinfo{title}{Bayesian Deep Learning With Random Feature-Based Gaussian Processes}, Ph.D. thesis, State University of New York at Stony Brook, \bibinfo{year}{2023}.
\bibitem[{Sena et~al.(2021)Sena, Erkilinc, Dippon, Shariati, Emmerich, Fischer, and Freund}]{sena2021bayesian}
\bibinfo{author}{M.~Sena}, \bibinfo{author}{M.~S. Erkilinc}, \bibinfo{author}{T.~Dippon}, \bibinfo{author}{B.~Shariati}, \bibinfo{author}{R.~Emmerich}, \bibinfo{author}{J.~K. Fischer}, \bibinfo{author}{R.~Freund},
\newblock \bibinfo{title}{Bayesian optimization for nonlinear system identification and pre-distortion in cognitive transmitters},
\newblock \bibinfo{journal}{Journal of Lightwave Technology} \bibinfo{volume}{39} (\bibinfo{year}{2021}) \bibinfo{pages}{5008--5020}.
\bibitem[{Dalton et~al.(2023)Dalton, Husmeier, and Gao}]{dalton2023physics}
\bibinfo{author}{D.~Dalton}, \bibinfo{author}{D.~Husmeier}, \bibinfo{author}{H.~Gao},
\newblock \bibinfo{title}{Physics-informed graph neural network emulation of soft-tissue mechanics},
\newblock \bibinfo{journal}{Computer Methods in Applied Mechanics and Engineering} \bibinfo{volume}{417} (\bibinfo{year}{2023}) \bibinfo{pages}{116351}.
\bibitem[{Regazzoni et~al.(2022)Regazzoni, Salvador, Ded{\`e}, and Quarteroni}]{regazzoni2022machine}
\bibinfo{author}{F.~Regazzoni}, \bibinfo{author}{M.~Salvador}, \bibinfo{author}{L.~Ded{\`e}}, \bibinfo{author}{A.~Quarteroni},
\newblock \bibinfo{title}{A machine learning method for real-time numerical simulations of cardiac electromechanics},
\newblock \bibinfo{journal}{Computer methods in applied mechanics and engineering} \bibinfo{volume}{393} (\bibinfo{year}{2022}) \bibinfo{pages}{114825}.
\bibitem[{Cicci et~al.(2024)Cicci, Fresca, Manzoni, and Quarteroni}]{cicci2024efficient}
\bibinfo{author}{L.~Cicci}, \bibinfo{author}{S.~Fresca}, \bibinfo{author}{A.~Manzoni}, \bibinfo{author}{A.~Quarteroni},
\newblock \bibinfo{title}{Efficient approximation of cardiac mechanics through reduced-order modeling with deep learning-based operator approximation},
\newblock \bibinfo{journal}{International Journal for Numerical Methods in Biomedical Engineering} \bibinfo{volume}{40} (\bibinfo{year}{2024}) \bibinfo{pages}{e3783}.
\bibitem[{Liang et~al.(2023)Liang, Liu, Elefteriades, and Sun}]{liang2023synergistic}
\bibinfo{author}{L.~Liang}, \bibinfo{author}{M.~Liu}, \bibinfo{author}{J.~Elefteriades}, \bibinfo{author}{W.~Sun},
\newblock \bibinfo{title}{Synergistic integration of deep neural networks and finite element method with applications of nonlinear large deformation biomechanics},
\newblock \bibinfo{journal}{Computer Methods in Applied Mechanics and Engineering} \bibinfo{volume}{416} (\bibinfo{year}{2023}) \bibinfo{pages}{116347}.
\bibitem[{Maher et~al.(2019)Maher, Wilson, and Marsden}]{maher2019accelerating}
\bibinfo{author}{G.~Maher}, \bibinfo{author}{N.~Wilson}, \bibinfo{author}{A.~Marsden},
\newblock \bibinfo{title}{Accelerating cardiovascular model building with convolutional neural networks},
\newblock \bibinfo{journal}{Medical \& biological engineering \& computing} \bibinfo{volume}{57} (\bibinfo{year}{2019}) \bibinfo{pages}{2319--2335}.
\bibitem[{Su et~al.(2020)Su, Zhang, Zou, Ghista, Le, and Chin}]{su2020generating}
\bibinfo{author}{B.~Su}, \bibinfo{author}{J.-M. Zhang}, \bibinfo{author}{H.~Zou}, \bibinfo{author}{D.~Ghista}, \bibinfo{author}{T.~T. Le}, \bibinfo{author}{C.~Chin},
\newblock \bibinfo{title}{Generating wall shear stress for coronary artery in real-time using neural networks: Feasibility and initial results based on idealized models},
\newblock \bibinfo{journal}{Computers in biology and medicine} \bibinfo{volume}{126} (\bibinfo{year}{2020}) \bibinfo{pages}{104038}.
\bibitem[{Ronneberger et~al.(2015)Ronneberger, Fischer, and Brox}]{ronneberger2015u}
\bibinfo{author}{O.~Ronneberger}, \bibinfo{author}{P.~Fischer}, \bibinfo{author}{T.~Brox},
\newblock \bibinfo{title}{U-net: Convolutional networks for biomedical image segmentation},
\newblock in: \bibinfo{booktitle}{Medical Image Computing and Computer-Assisted Intervention--MICCAI 2015: 18th International Conference, Munich, Germany, October 5-9, 2015, Proceedings, Part III 18}, \bibinfo{organization}{Springer}, \bibinfo{year}{2015}, pp. \bibinfo{pages}{234--241}.
\bibitem[{Liu(2018)}]{liu2018feature}
\bibinfo{author}{Y.~H. Liu},
\newblock \bibinfo{title}{Feature extraction and image recognition with convolutional neural networks},
\newblock in: \bibinfo{booktitle}{Journal of Physics: Conference Series}, volume \bibinfo{volume}{1087}, \bibinfo{organization}{IOP Publishing}, \bibinfo{year}{2018}, p. \bibinfo{pages}{062032}.
\bibitem[{Chen et~al.(2016)Chen, Jiang, Li, Jia, and Ghamisi}]{chen2016deep}
\bibinfo{author}{Y.~Chen}, \bibinfo{author}{H.~Jiang}, \bibinfo{author}{C.~Li}, \bibinfo{author}{X.~Jia}, \bibinfo{author}{P.~Ghamisi},
\newblock \bibinfo{title}{Deep feature extraction and classification of hyperspectral images based on convolutional neural networks},
\newblock \bibinfo{journal}{IEEE transactions on geoscience and remote sensing} \bibinfo{volume}{54} (\bibinfo{year}{2016}) \bibinfo{pages}{6232--6251}.
\bibitem[{Sveinsson~Cepero and Shadden(2024)}]{sveinsson2024seqseg}
\bibinfo{author}{N.~Sveinsson~Cepero}, \bibinfo{author}{S.~C. Shadden},
\newblock \bibinfo{title}{Seqseg: Learning local segments for automatic vascular model construction},
\newblock \bibinfo{journal}{Annals of Biomedical Engineering}  (\bibinfo{year}{2024}) \bibinfo{pages}{1--22}.
\bibitem[{Du et~al.(2024)Du, Parikh, Fan, Liu, and Wang}]{du2024conditional}
\bibinfo{author}{P.~Du}, \bibinfo{author}{M.~H. Parikh}, \bibinfo{author}{X.~Fan}, \bibinfo{author}{X.-Y. Liu}, \bibinfo{author}{J.-X. Wang},
\newblock \bibinfo{title}{Conditional neural field latent diffusion model for generating spatiotemporal turbulence},
\newblock \bibinfo{journal}{Nature Communications} \bibinfo{volume}{15} (\bibinfo{year}{2024}) \bibinfo{pages}{10416}.
\bibitem[{Kadry et~al.(2024)Kadry, Gupta, Sogbadji, Schaap, Petersen, Mizukami, Collet, Nezami, and Edelman}]{kadry2024diffusion}
\bibinfo{author}{K.~Kadry}, \bibinfo{author}{S.~Gupta}, \bibinfo{author}{J.~Sogbadji}, \bibinfo{author}{M.~Schaap}, \bibinfo{author}{K.~Petersen}, \bibinfo{author}{T.~Mizukami}, \bibinfo{author}{C.~Collet}, \bibinfo{author}{F.~R. Nezami}, \bibinfo{author}{E.~R. Edelman},
\newblock \bibinfo{title}{A diffusion model for simulation ready coronary anatomy with morpho-skeletal control},
\newblock in: \bibinfo{booktitle}{European Conference on Computer Vision}, \bibinfo{organization}{Springer}, \bibinfo{year}{2024}, pp. \bibinfo{pages}{396--412}.
\bibitem[{Guo et~al.(2024)Guo, Li, Chen, Sui, Chen, and Li}]{guo2024diffusion}
\bibinfo{author}{Y.~Guo}, \bibinfo{author}{Y.~Li}, \bibinfo{author}{L.~Chen}, \bibinfo{author}{D.~Sui}, \bibinfo{author}{Y.~Chen}, \bibinfo{author}{S.~Li},
\newblock \bibinfo{title}{A diffusion model approach for solving the inverse problem between cardiac electricalphysiology and electrocardiograph},
\newblock in: \bibinfo{booktitle}{2024 IEEE International Conference on Bioinformatics and Biomedicine (BIBM)}, \bibinfo{organization}{IEEE}, \bibinfo{year}{2024}, pp. \bibinfo{pages}{3244--3247}.
\bibitem[{Kissas et~al.(2020)Kissas, Yang, Hwuang, Witschey, Detre, and Perdikaris}]{kissas2020machine}
\bibinfo{author}{G.~Kissas}, \bibinfo{author}{Y.~Yang}, \bibinfo{author}{E.~Hwuang}, \bibinfo{author}{W.~R. Witschey}, \bibinfo{author}{J.~A. Detre}, \bibinfo{author}{P.~Perdikaris},
\newblock \bibinfo{title}{Machine learning in cardiovascular flows modeling: Predicting arterial blood pressure from non-invasive 4d flow mri data using physics-informed neural networks},
\newblock \bibinfo{journal}{Computer Methods in Applied Mechanics and Engineering} \bibinfo{volume}{358} (\bibinfo{year}{2020}) \bibinfo{pages}{112623}.
\bibitem[{Buoso et~al.(2021)Buoso, Joyce, and Kozerke}]{buoso2021personalising}
\bibinfo{author}{S.~Buoso}, \bibinfo{author}{T.~Joyce}, \bibinfo{author}{S.~Kozerke},
\newblock \bibinfo{title}{Personalising left-ventricular biophysical models of the heart using parametric physics-informed neural networks},
\newblock \bibinfo{journal}{Medical Image Analysis} \bibinfo{volume}{71} (\bibinfo{year}{2021}) \bibinfo{pages}{102066}.
\bibitem[{Caforio et~al.(2024)Caforio, Regazzoni, Pagani, Karabelas, Augustin, Haase, Plank, and Quarteroni}]{caforio2024physics}
\bibinfo{author}{F.~Caforio}, \bibinfo{author}{F.~Regazzoni}, \bibinfo{author}{S.~Pagani}, \bibinfo{author}{E.~Karabelas}, \bibinfo{author}{C.~Augustin}, \bibinfo{author}{G.~Haase}, \bibinfo{author}{G.~Plank}, \bibinfo{author}{A.~Quarteroni},
\newblock \bibinfo{title}{Physics-informed neural network estimation of material properties in soft tissue nonlinear biomechanical models},
\newblock \bibinfo{journal}{Computational Mechanics}  (\bibinfo{year}{2024}) \bibinfo{pages}{1--27}.
\bibitem[{Arzani et~al.(2021)Arzani, Wang, and D'Souza}]{arzani2021uncovering}
\bibinfo{author}{A.~Arzani}, \bibinfo{author}{J.-X. Wang}, \bibinfo{author}{R.~M. D'Souza},
\newblock \bibinfo{title}{Uncovering near-wall blood flow from sparse data with physics-informed neural networks},
\newblock \bibinfo{journal}{Physics of Fluids} \bibinfo{volume}{33} (\bibinfo{year}{2021}).
\bibitem[{Perdikaris(2024)}]{perdikaris2024probabilistic}
\bibinfo{author}{P.~Perdikaris}, \bibinfo{title}{Probabilistic data fusion and physics-informed machine learning: A new paradigm for modeling under uncertainty, and its application to accelerating the discovery of new materials.}, \bibinfo{type}{Technical Report}, Univ. of Pennsylvania, Philadelphia, PA (United States), \bibinfo{year}{2024}.
\bibitem[{Raissi et~al.(2024)Raissi, Perdikaris, Ahmadi, and Karniadakis}]{raissi2024physics}
\bibinfo{author}{M.~Raissi}, \bibinfo{author}{P.~Perdikaris}, \bibinfo{author}{N.~Ahmadi}, \bibinfo{author}{G.~E. Karniadakis},
\newblock \bibinfo{title}{Physics-informed neural networks and extensions},
\newblock \bibinfo{journal}{arXiv preprint arXiv:2408.16806}  (\bibinfo{year}{2024}).
\bibitem[{Garay et~al.(2024)Garay, Dunstan, Uribe, and Costabal}]{garay2024physics}
\bibinfo{author}{J.~Garay}, \bibinfo{author}{J.~Dunstan}, \bibinfo{author}{S.~Uribe}, \bibinfo{author}{F.~S. Costabal},
\newblock \bibinfo{title}{Physics-informed neural networks for parameter estimation in blood flow models},
\newblock \bibinfo{journal}{Computers in Biology and Medicine} \bibinfo{volume}{178} (\bibinfo{year}{2024}) \bibinfo{pages}{108706}.
\bibitem[{Hu et~al.(2024)Hu, Qi, and Chao}]{hu2024physics}
\bibinfo{author}{H.~Hu}, \bibinfo{author}{L.~Qi}, \bibinfo{author}{X.~Chao},
\newblock \bibinfo{title}{Physics-informed neural networks (pinn) for computational solid mechanics: Numerical frameworks and applications},
\newblock \bibinfo{journal}{Thin-Walled Structures}  (\bibinfo{year}{2024}) \bibinfo{pages}{112495}.
\bibitem[{Haghighat et~al.(2020)Haghighat, Raissi, Moure, Gomez, and Juanes}]{haghighat2020deep}
\bibinfo{author}{E.~Haghighat}, \bibinfo{author}{M.~Raissi}, \bibinfo{author}{A.~Moure}, \bibinfo{author}{H.~Gomez}, \bibinfo{author}{R.~Juanes},
\newblock \bibinfo{title}{A deep learning framework for solution and discovery in solid mechanics},
\newblock \bibinfo{journal}{arXiv preprint arXiv:2003.02751}  (\bibinfo{year}{2020}).
\bibitem[{Haghighat et~al.(2021)Haghighat, Raissi, Moure, Gomez, and Juanes}]{haghighat2021physics}
\bibinfo{author}{E.~Haghighat}, \bibinfo{author}{M.~Raissi}, \bibinfo{author}{A.~Moure}, \bibinfo{author}{H.~Gomez}, \bibinfo{author}{R.~Juanes},
\newblock \bibinfo{title}{A physics-informed deep learning framework for inversion and surrogate modeling in solid mechanics},
\newblock \bibinfo{journal}{Computer Methods in Applied Mechanics and Engineering} \bibinfo{volume}{379} (\bibinfo{year}{2021}) \bibinfo{pages}{113741}.
\bibitem[{Aghaee and Khan(2025)}]{aghaee2025pinning}
\bibinfo{author}{A.~Aghaee}, \bibinfo{author}{M.~O. Khan},
\newblock \bibinfo{title}{Pinning down the accuracy of physics-informed neural networks under laminar and turbulent-like aortic blood flow conditions},
\newblock \bibinfo{journal}{Computers in Biology and Medicine} \bibinfo{volume}{185} (\bibinfo{year}{2025}) \bibinfo{pages}{109528}.
\bibitem[{Pak et~al.(2023)Pak, Liu, Kim, Liang, Caballero, Onofrey, Ahn, Xu, McKay, Sun et~al.}]{pak2023patient}
\bibinfo{author}{D.~H. Pak}, \bibinfo{author}{M.~Liu}, \bibinfo{author}{T.~Kim}, \bibinfo{author}{L.~Liang}, \bibinfo{author}{A.~Caballero}, \bibinfo{author}{J.~Onofrey}, \bibinfo{author}{S.~S. Ahn}, \bibinfo{author}{Y.~Xu}, \bibinfo{author}{R.~McKay}, \bibinfo{author}{W.~Sun}, et~al.,
\newblock \bibinfo{title}{Patient-specific heart geometry modeling for solid biomechanics using deep learning},
\newblock \bibinfo{journal}{IEEE transactions on medical imaging}  (\bibinfo{year}{2023}).
\bibitem[{Saleh et~al.(2024)Saleh, Sommersperger, Navab, and Tombari}]{saleh2024physics}
\bibinfo{author}{M.~Saleh}, \bibinfo{author}{M.~Sommersperger}, \bibinfo{author}{N.~Navab}, \bibinfo{author}{F.~Tombari},
\newblock \bibinfo{title}{Physics-encoded graph neural networks for deformation prediction under contact},
\newblock \bibinfo{journal}{arXiv preprint arXiv:2402.03466}  (\bibinfo{year}{2024}).
\bibitem[{Salehi and Giannacopoulos(2022)}]{salehi2022physgnn}
\bibinfo{author}{Y.~Salehi}, \bibinfo{author}{D.~Giannacopoulos},
\newblock \bibinfo{title}{Physgnn: A physics--driven graph neural network based model for predicting soft tissue deformation in image--guided neurosurgery},
\newblock \bibinfo{journal}{Advances in Neural Information Processing Systems} \bibinfo{volume}{35} (\bibinfo{year}{2022}) \bibinfo{pages}{37282--37296}.
\bibitem[{Vaswani(2017)}]{vaswani2017attention}
\bibinfo{author}{A.~Vaswani},
\newblock \bibinfo{title}{Attention is all you need},
\newblock \bibinfo{journal}{Advances in Neural Information Processing Systems}  (\bibinfo{year}{2017}).
\bibitem[{Liu et~al.(2020)Liu, Ong, and Chen}]{liu2020graphsage}
\bibinfo{author}{J.~Liu}, \bibinfo{author}{G.~P. Ong}, \bibinfo{author}{X.~Chen},
\newblock \bibinfo{title}{Graphsage-based traffic speed forecasting for segment network with sparse data},
\newblock \bibinfo{journal}{IEEE Transactions on Intelligent Transportation Systems} \bibinfo{volume}{23} (\bibinfo{year}{2020}) \bibinfo{pages}{1755--1766}.
\bibitem[{Oh et~al.(2019)Oh, Cho, and Bruna}]{oh2019advancing}
\bibinfo{author}{J.~Oh}, \bibinfo{author}{K.~Cho}, \bibinfo{author}{J.~Bruna},
\newblock \bibinfo{title}{Advancing graphsage with a data-driven node sampling},
\newblock \bibinfo{journal}{arXiv preprint arXiv:1904.12935}  (\bibinfo{year}{2019}).
\bibitem[{Hamilton et~al.(2017)Hamilton, Ying, and Leskovec}]{hamilton2017inductive}
\bibinfo{author}{W.~Hamilton}, \bibinfo{author}{Z.~Ying}, \bibinfo{author}{J.~Leskovec},
\newblock \bibinfo{title}{Inductive representation learning on large graphs},
\newblock \bibinfo{journal}{Advances in neural information processing systems} \bibinfo{volume}{30} (\bibinfo{year}{2017}).
\bibitem[{Gao et~al.(2023)Gao, Zhou, Gao, and Zhuang}]{gao2023bayeseg}
\bibinfo{author}{S.~Gao}, \bibinfo{author}{H.~Zhou}, \bibinfo{author}{Y.~Gao}, \bibinfo{author}{X.~Zhuang},
\newblock \bibinfo{title}{Bayeseg: Bayesian modeling for medical image segmentation with interpretable generalizability},
\newblock \bibinfo{journal}{Medical Image Analysis} \bibinfo{volume}{89} (\bibinfo{year}{2023}) \bibinfo{pages}{102889}.
\bibitem[{Zhuang(2018)}]{zhuang2018multivariate}
\bibinfo{author}{X.~Zhuang},
\newblock \bibinfo{title}{Multivariate mixture model for myocardial segmentation combining multi-source images},
\newblock \bibinfo{journal}{IEEE transactions on pattern analysis and machine intelligence} \bibinfo{volume}{41} (\bibinfo{year}{2018}) \bibinfo{pages}{2933--2946}.
\bibitem[{Karim et~al.(2018)Karim, Blake, Inoue, Tao, Jia, Housden, Bhagirath, Duval, Varela, Behar et~al.}]{karim2018algorithms}
\bibinfo{author}{R.~Karim}, \bibinfo{author}{L.-E. Blake}, \bibinfo{author}{J.~Inoue}, \bibinfo{author}{Q.~Tao}, \bibinfo{author}{S.~Jia}, \bibinfo{author}{R.~J. Housden}, \bibinfo{author}{P.~Bhagirath}, \bibinfo{author}{J.-L. Duval}, \bibinfo{author}{M.~Varela}, \bibinfo{author}{J.~M. Behar}, et~al.,
\newblock \bibinfo{title}{Algorithms for left atrial wall segmentation and thickness--evaluation on an open-source ct and mri image database},
\newblock \bibinfo{journal}{Medical image analysis} \bibinfo{volume}{50} (\bibinfo{year}{2018}) \bibinfo{pages}{36--53}.
\bibitem[{Tobon-Gomez et~al.(2015)Tobon-Gomez, Geers, Peters, Weese, Pinto, Karim, Ammar, Daoudi, Margeta, Sandoval et~al.}]{tobon2015benchmark}
\bibinfo{author}{C.~Tobon-Gomez}, \bibinfo{author}{A.~J. Geers}, \bibinfo{author}{J.~Peters}, \bibinfo{author}{J.~Weese}, \bibinfo{author}{K.~Pinto}, \bibinfo{author}{R.~Karim}, \bibinfo{author}{M.~Ammar}, \bibinfo{author}{A.~Daoudi}, \bibinfo{author}{J.~Margeta}, \bibinfo{author}{Z.~Sandoval}, et~al.,
\newblock \bibinfo{title}{Benchmark for algorithms segmenting the left atrium from 3d ct and mri datasets},
\newblock \bibinfo{journal}{IEEE transactions on medical imaging} \bibinfo{volume}{34} (\bibinfo{year}{2015}) \bibinfo{pages}{1460--1473}.
\bibitem[{Wolterink et~al.(2016)Wolterink, Leiner, De~Vos, Coatrieux, Kelm, Kondo, Salgado, Shahzad, Shu, Snoeren et~al.}]{wolterink2016evaluation}
\bibinfo{author}{J.~M. Wolterink}, \bibinfo{author}{T.~Leiner}, \bibinfo{author}{B.~D. De~Vos}, \bibinfo{author}{J.-L. Coatrieux}, \bibinfo{author}{B.~M. Kelm}, \bibinfo{author}{S.~Kondo}, \bibinfo{author}{R.~A. Salgado}, \bibinfo{author}{R.~Shahzad}, \bibinfo{author}{H.~Shu}, \bibinfo{author}{M.~Snoeren}, et~al.,
\newblock \bibinfo{title}{An evaluation of automatic coronary artery calcium scoring methods with cardiac ct using the orcascore framework},
\newblock \bibinfo{journal}{Medical physics} \bibinfo{volume}{43} (\bibinfo{year}{2016}) \bibinfo{pages}{2361--2373}.
\bibitem[{Updegrove et~al.(2017)Updegrove, Wilson, Merkow, Lan, Marsden, and Shadden}]{updegrove2017simvascular}
\bibinfo{author}{A.~Updegrove}, \bibinfo{author}{N.~M. Wilson}, \bibinfo{author}{J.~Merkow}, \bibinfo{author}{H.~Lan}, \bibinfo{author}{A.~L. Marsden}, \bibinfo{author}{S.~C. Shadden},
\newblock \bibinfo{title}{Simvascular: an open source pipeline for cardiovascular simulation},
\newblock \bibinfo{journal}{Annals of biomedical engineering} \bibinfo{volume}{45} (\bibinfo{year}{2017}) \bibinfo{pages}{525--541}.
\bibitem[{Hang(2015)}]{hang2015tetgen}
\bibinfo{author}{S.~Hang},
\newblock \bibinfo{title}{Tetgen, a delaunay-based quality tetrahedral mesh generator},
\newblock \bibinfo{journal}{ACM Trans. Math. Softw} \bibinfo{volume}{41} (\bibinfo{year}{2015}) \bibinfo{pages}{11}.
\bibitem[{Ar{\'o}stica et~al.(2025)Ar{\'o}stica, Nolte, Brown, Gebauer, Karabelas, Jilberto, Salvador, Bucelli, Piersanti, Osouli et~al.}]{arostica2025software}
\bibinfo{author}{R.~Ar{\'o}stica}, \bibinfo{author}{D.~Nolte}, \bibinfo{author}{A.~Brown}, \bibinfo{author}{A.~Gebauer}, \bibinfo{author}{E.~Karabelas}, \bibinfo{author}{J.~Jilberto}, \bibinfo{author}{M.~Salvador}, \bibinfo{author}{M.~Bucelli}, \bibinfo{author}{R.~Piersanti}, \bibinfo{author}{K.~Osouli}, et~al.,
\newblock \bibinfo{title}{A software benchmark for cardiac elastodynamics},
\newblock \bibinfo{journal}{Computer Methods in Applied Mechanics and Engineering} \bibinfo{volume}{435} (\bibinfo{year}{2025}) \bibinfo{pages}{117485}.
\bibitem[{Shi et~al.(2025)Shi, Chen, and Vedula}]{shi2025leftatrium}
\bibinfo{author}{L.~Shi}, \bibinfo{author}{I.~Y. Chen}, \bibinfo{author}{V.~Vedula}, \bibinfo{title}{Personalized multiscale modeling of left atrial mechanics and blood flow}, \bibinfo{year}{2025}. \bibinfo{note}{\textit{in preparation}}.
\bibitem[{Liu and Marsden(2018)}]{liu2018unified}
\bibinfo{author}{J.~Liu}, \bibinfo{author}{A.~L. Marsden},
\newblock \bibinfo{title}{A unified continuum and variational multiscale formulation for fluids, solids, and fluid--structure interaction},
\newblock \bibinfo{journal}{Computer methods in applied mechanics and engineering} \bibinfo{volume}{337} (\bibinfo{year}{2018}) \bibinfo{pages}{549--597}.
\bibitem[{Holzapfel and Ogden(2009)}]{holzapfel2009constitutive}
\bibinfo{author}{G.~A. Holzapfel}, \bibinfo{author}{R.~W. Ogden},
\newblock \bibinfo{title}{Constitutive modelling of passive myocardium: a structurally based framework for material characterization},
\newblock \bibinfo{journal}{Philosophical Transactions of the Royal Society A: Mathematical, Physical and Engineering Sciences} \bibinfo{volume}{367} (\bibinfo{year}{2009}) \bibinfo{pages}{3445--3475}.
\bibitem[{Nolan et~al.(2014)Nolan, Gower, Destrade, Ogden, and McGarry}]{nolan2014robust}
\bibinfo{author}{D.~R. Nolan}, \bibinfo{author}{A.~L. Gower}, \bibinfo{author}{M.~Destrade}, \bibinfo{author}{R.~W. Ogden}, \bibinfo{author}{J.~McGarry},
\newblock \bibinfo{title}{A robust anisotropic hyperelastic formulation for the modelling of soft tissue},
\newblock \bibinfo{journal}{Journal of the mechanical behavior of biomedical materials} \bibinfo{volume}{39} (\bibinfo{year}{2014}) \bibinfo{pages}{48--60}.
\bibitem[{Bayer et~al.(2012)Bayer, Blake, Plank, and Trayanova}]{bayer2012novel}
\bibinfo{author}{J.~D. Bayer}, \bibinfo{author}{R.~C. Blake}, \bibinfo{author}{G.~Plank}, \bibinfo{author}{N.~A. Trayanova},
\newblock \bibinfo{title}{A novel rule-based algorithm for assigning myocardial fiber orientation to computational heart models},
\newblock \bibinfo{journal}{Annals of biomedical engineering} \bibinfo{volume}{40} (\bibinfo{year}{2012}) \bibinfo{pages}{2243--2254}.
\bibitem[{Simard et~al.(2003)Simard, Steinkraus, Platt et~al.}]{simard2003best}
\bibinfo{author}{P.~Y. Simard}, \bibinfo{author}{D.~Steinkraus}, \bibinfo{author}{J.~C. Platt}, et~al.,
\newblock \bibinfo{title}{Best practices for convolutional neural networks applied to visual document analysis.},
\newblock in: \bibinfo{booktitle}{Icdar}, volume~\bibinfo{volume}{3}, \bibinfo{organization}{Edinburgh}, \bibinfo{year}{2003}.
\bibitem[{Beauchamp and Olson(1973)}]{beauchamp1973corrections}
\bibinfo{author}{J.~J. Beauchamp}, \bibinfo{author}{J.~S. Olson},
\newblock \bibinfo{title}{Corrections for bias in regression estimates after logarithmic transformation},
\newblock \bibinfo{journal}{Ecology} \bibinfo{volume}{54} (\bibinfo{year}{1973}) \bibinfo{pages}{1403--1407}.
\bibitem[{{MSD Manual}(2024)}]{msd_heart_pressures}
\bibinfo{author}{{MSD Manual}}, \bibinfo{title}{Normal pressures in the heart and great vessels}, \bibinfo{year}{2024}. \URLprefix \url{https://www.msdmanuals.com/en-nz/professional/multimedia/table/normal-pressures-in-the-heart-and-great-vessels}, \bibinfo{note}{[Accessed September 4, 2024]}.
\bibitem[{Zhu et~al.(2022)Zhu, Vedula, Parker, Wilson, Shadden, and Marsden}]{zhu2022svfsi}
\bibinfo{author}{C.~Zhu}, \bibinfo{author}{V.~Vedula}, \bibinfo{author}{D.~Parker}, \bibinfo{author}{N.~Wilson}, \bibinfo{author}{S.~Shadden}, \bibinfo{author}{A.~Marsden},
\newblock \bibinfo{title}{svfsi: a multiphysics package for integrated cardiac modeling},
\newblock \bibinfo{journal}{Journal of Open Source Software} \bibinfo{volume}{7} (\bibinfo{year}{2022}) \bibinfo{pages}{4118}.
\bibitem[{Deshai et~al.(2024)Deshai, Ambikapathy, and Kumar}]{deshai2024deep}
\bibinfo{author}{N.~Deshai}, \bibinfo{author}{A.~Ambikapathy}, \bibinfo{author}{R.~Kumar}, \bibinfo{title}{Deep Learning And Its Applications}, \bibinfo{publisher}{Cipher Publisher}, \bibinfo{year}{2024}.
\bibitem[{Huang et~al.(2023)Huang, Qin, Zhou, Zhu, Liu, and Shao}]{huang2023normalization}
\bibinfo{author}{L.~Huang}, \bibinfo{author}{J.~Qin}, \bibinfo{author}{Y.~Zhou}, \bibinfo{author}{F.~Zhu}, \bibinfo{author}{L.~Liu}, \bibinfo{author}{L.~Shao},
\newblock \bibinfo{title}{Normalization techniques in training dnns: Methodology, analysis and application},
\newblock \bibinfo{journal}{IEEE transactions on pattern analysis and machine intelligence} \bibinfo{volume}{45} (\bibinfo{year}{2023}) \bibinfo{pages}{10173--10196}.
\bibitem[{Huang(2022)}]{huang2022normalization}
\bibinfo{author}{L.~Huang}, \bibinfo{title}{Normalization Techniques in Deep Learning}, \bibinfo{publisher}{Springer}, \bibinfo{year}{2022}.
\bibitem[{Hinton(2007)}]{hinton2007learning}
\bibinfo{author}{G.~E. Hinton},
\newblock \bibinfo{title}{Learning multiple layers of representation},
\newblock \bibinfo{journal}{Trends in cognitive sciences} \bibinfo{volume}{11} (\bibinfo{year}{2007}) \bibinfo{pages}{428--434}.
\bibitem[{LeCun et~al.(2015)LeCun, Bengio, and Hinton}]{lecun2015deep}
\bibinfo{author}{Y.~LeCun}, \bibinfo{author}{Y.~Bengio}, \bibinfo{author}{G.~Hinton},
\newblock \bibinfo{title}{Deep learning},
\newblock \bibinfo{journal}{nature} \bibinfo{volume}{521} (\bibinfo{year}{2015}) \bibinfo{pages}{436--444}.
\bibitem[{Ali et~al.(2014)Ali, Faraj, Koya, Ali, and Faraj}]{ali2014data}
\bibinfo{author}{P.~J.~M. Ali}, \bibinfo{author}{R.~H. Faraj}, \bibinfo{author}{E.~Koya}, \bibinfo{author}{P.~J.~M. Ali}, \bibinfo{author}{R.~H. Faraj},
\newblock \bibinfo{title}{Data normalization and standardization: a technical report},
\newblock \bibinfo{journal}{Mach Learn Tech Rep} \bibinfo{volume}{1} (\bibinfo{year}{2014}) \bibinfo{pages}{1--6}.
\bibitem[{Goodfellow et~al.(2016)Goodfellow, Bengio, Courville, and Bengio}]{goodfellow2016deep}
\bibinfo{author}{I.~Goodfellow}, \bibinfo{author}{Y.~Bengio}, \bibinfo{author}{A.~Courville}, \bibinfo{author}{Y.~Bengio}, \bibinfo{title}{Deep learning}, volume~\bibinfo{volume}{1}, \bibinfo{publisher}{MIT press Cambridge}, \bibinfo{year}{2016}.
\bibitem[{Hinton et~al.(2011)Hinton, Krizhevsky, and Wang}]{hinton2011transforming}
\bibinfo{author}{G.~E. Hinton}, \bibinfo{author}{A.~Krizhevsky}, \bibinfo{author}{S.~D. Wang},
\newblock \bibinfo{title}{Transforming auto-encoders},
\newblock in: \bibinfo{booktitle}{Artificial Neural Networks and Machine Learning--ICANN 2011: 21st International Conference on Artificial Neural Networks, Espoo, Finland, June 14-17, 2011, Proceedings, Part I 21}, \bibinfo{organization}{Springer}, \bibinfo{year}{2011}, pp. \bibinfo{pages}{44--51}.
\bibitem[{He et~al.(2016)He, Zhang, Ren, and Sun}]{he2016deep}
\bibinfo{author}{K.~He}, \bibinfo{author}{X.~Zhang}, \bibinfo{author}{S.~Ren}, \bibinfo{author}{J.~Sun},
\newblock \bibinfo{title}{Deep residual learning for image recognition},
\newblock in: \bibinfo{booktitle}{Proceedings of the IEEE conference on computer vision and pattern recognition}, \bibinfo{year}{2016}, pp. \bibinfo{pages}{770--778}.
\bibitem[{Kou et~al.(2014)Kou, Caballero, Dulgheru, Voilliot, De~Sousa, Kacharava, Athanassopoulos, Barone, Baroni, Cardim et~al.}]{kou2014echocardiographic}
\bibinfo{author}{S.~Kou}, \bibinfo{author}{L.~Caballero}, \bibinfo{author}{R.~Dulgheru}, \bibinfo{author}{D.~Voilliot}, \bibinfo{author}{C.~De~Sousa}, \bibinfo{author}{G.~Kacharava}, \bibinfo{author}{G.~D. Athanassopoulos}, \bibinfo{author}{D.~Barone}, \bibinfo{author}{M.~Baroni}, \bibinfo{author}{N.~Cardim}, et~al.,
\newblock \bibinfo{title}{Echocardiographic reference ranges for normal cardiac chamber size: results from the norre study},
\newblock \bibinfo{journal}{European Heart Journal--Cardiovascular Imaging} \bibinfo{volume}{15} (\bibinfo{year}{2014}) \bibinfo{pages}{680--690}.
\bibitem[{Kawel-Boehm et~al.(2025)Kawel-Boehm, Hetzel, Ambale-Venkatesh, Captur, Chin, Fran{\c{c}}ois, Jerosch-Herold, Luu, Raisi-Estabragh, Starekova et~al.}]{kawel2025reference}
\bibinfo{author}{N.~Kawel-Boehm}, \bibinfo{author}{S.~J. Hetzel}, \bibinfo{author}{B.~Ambale-Venkatesh}, \bibinfo{author}{G.~Captur}, \bibinfo{author}{C.~W. Chin}, \bibinfo{author}{C.~J. Fran{\c{c}}ois}, \bibinfo{author}{M.~Jerosch-Herold}, \bibinfo{author}{J.~M. Luu}, \bibinfo{author}{Z.~Raisi-Estabragh}, \bibinfo{author}{J.~Starekova}, et~al.,
\newblock \bibinfo{title}{Reference ranges (“normal values”) for cardiovascular magnetic resonance (cmr) in adults and children: 2025 update},
\newblock \bibinfo{journal}{Journal of cardiovascular magnetic resonance}  (\bibinfo{year}{2025}) \bibinfo{pages}{101853}.
\bibitem[{CHEEMA et~al.(1996)CHEEMA, Jalal, FEROZE, and Khan}]{cheema1996dimensions}
\bibinfo{author}{S.~A. CHEEMA}, \bibinfo{author}{A.~Jalal}, \bibinfo{author}{N.~FEROZE}, \bibinfo{author}{J.~S. Khan},
\newblock \bibinfo{title}{Dimensions of mitral valve of normal human hearts in pakistani subjects},
\newblock \bibinfo{journal}{Pakistan Heart Journal} \bibinfo{volume}{29} (\bibinfo{year}{1996}).
\bibitem[{Kitzman et~al.(1988)Kitzman, Scholz, Hagen, Ilstrup, and Edwards}]{kitzman1988age}
\bibinfo{author}{D.~W. Kitzman}, \bibinfo{author}{D.~G. Scholz}, \bibinfo{author}{P.~T. Hagen}, \bibinfo{author}{D.~M. Ilstrup}, \bibinfo{author}{W.~D. Edwards},
\newblock \bibinfo{title}{Age-related changes in normal human hearts during the first 10 decades of life. part ii (maturity): a quantitative anatomic study of 765 specimens from subjects 20 to 99 years old},
\newblock in: \bibinfo{booktitle}{Mayo Clinic Proceedings}, volume~\bibinfo{volume}{63}, \bibinfo{organization}{Elsevier}, \bibinfo{year}{1988}, pp. \bibinfo{pages}{137--146}.
\bibitem[{Wold et~al.(1987)Wold, Esbensen, and Geladi}]{wold1987principal}
\bibinfo{author}{S.~Wold}, \bibinfo{author}{K.~Esbensen}, \bibinfo{author}{P.~Geladi},
\newblock \bibinfo{title}{Principal component analysis},
\newblock \bibinfo{journal}{Chemometrics and intelligent laboratory systems} \bibinfo{volume}{2} (\bibinfo{year}{1987}) \bibinfo{pages}{37--52}.
\bibitem[{Kherif and Latypova(2020)}]{kherif2020principal}
\bibinfo{author}{F.~Kherif}, \bibinfo{author}{A.~Latypova},
\newblock \bibinfo{title}{Principal component analysis},
\newblock in: \bibinfo{booktitle}{Machine learning}, \bibinfo{publisher}{Elsevier}, \bibinfo{year}{2020}, pp. \bibinfo{pages}{209--225}.
\bibitem[{Greenacre et~al.(2022)Greenacre, Groenen, Hastie, d’Enza, Markos, and Tuzhilina}]{greenacre2022principal}
\bibinfo{author}{M.~Greenacre}, \bibinfo{author}{P.~J. Groenen}, \bibinfo{author}{T.~Hastie}, \bibinfo{author}{A.~I. d’Enza}, \bibinfo{author}{A.~Markos}, \bibinfo{author}{E.~Tuzhilina},
\newblock \bibinfo{title}{Principal component analysis},
\newblock \bibinfo{journal}{Nature Reviews Methods Primers} \bibinfo{volume}{2} (\bibinfo{year}{2022}) \bibinfo{pages}{100}.
\bibitem[{Mrowca et~al.(2018)Mrowca, Zhuang, Wang, Haber, Fei-Fei, Tenenbaum, and Yamins}]{mrowca2018flexible}
\bibinfo{author}{D.~Mrowca}, \bibinfo{author}{C.~Zhuang}, \bibinfo{author}{E.~Wang}, \bibinfo{author}{N.~Haber}, \bibinfo{author}{L.~F. Fei-Fei}, \bibinfo{author}{J.~Tenenbaum}, \bibinfo{author}{D.~L. Yamins},
\newblock \bibinfo{title}{Flexible neural representation for physics prediction},
\newblock \bibinfo{journal}{Advances in neural information processing systems} \bibinfo{volume}{31} (\bibinfo{year}{2018}).
\bibitem[{Li et~al.(2018)Li, Wu, Tedrake, Tenenbaum, and Torralba}]{li2018learning}
\bibinfo{author}{Y.~Li}, \bibinfo{author}{J.~Wu}, \bibinfo{author}{R.~Tedrake}, \bibinfo{author}{J.~B. Tenenbaum}, \bibinfo{author}{A.~Torralba},
\newblock \bibinfo{title}{Learning particle dynamics for manipulating rigid bodies, deformable objects, and fluids},
\newblock \bibinfo{journal}{arXiv preprint arXiv:1810.01566}  (\bibinfo{year}{2018}).
\bibitem[{Moln{\'a}r and Tam{\'a}s(2022)}]{molnar2022representation}
\bibinfo{author}{S.~Moln{\'a}r}, \bibinfo{author}{L.~Tam{\'a}s},
\newblock \bibinfo{title}{Representation learning for point clouds with variational autoencoders},
\newblock in: \bibinfo{booktitle}{European conference on computer vision}, \bibinfo{organization}{Springer}, \bibinfo{year}{2022}, pp. \bibinfo{pages}{727--737}.
\bibitem[{Girin et~al.(2020)Girin, Leglaive, Bie, Diard, Hueber, and Alameda-Pineda}]{girin2020dynamical}
\bibinfo{author}{L.~Girin}, \bibinfo{author}{S.~Leglaive}, \bibinfo{author}{X.~Bie}, \bibinfo{author}{J.~Diard}, \bibinfo{author}{T.~Hueber}, \bibinfo{author}{X.~Alameda-Pineda},
\newblock \bibinfo{title}{Dynamical variational autoencoders: A comprehensive review},
\newblock \bibinfo{journal}{arXiv preprint arXiv:2008.12595}  (\bibinfo{year}{2020}).
\bibitem[{Zhang et~al.(2023)Zhang, Wang, Lin, Nicholas, Shen, and Miao}]{zhang2023starnet}
\bibinfo{author}{Y.~Zhang}, \bibinfo{author}{H.~Wang}, \bibinfo{author}{G.~Lin}, \bibinfo{author}{V.~C.~H. Nicholas}, \bibinfo{author}{Z.~Shen}, \bibinfo{author}{C.~Miao},
\newblock \bibinfo{title}{Starnet: Style-aware 3d point cloud generation},
\newblock \bibinfo{journal}{arXiv preprint arXiv:2303.15805}  (\bibinfo{year}{2023}).
\bibitem[{Motiwale et~al.(2023)Motiwale, Zhang, and Sacks}]{motiwale2023high}
\bibinfo{author}{S.~Motiwale}, \bibinfo{author}{W.~Zhang}, \bibinfo{author}{M.~S. Sacks},
\newblock \bibinfo{title}{High-speed high-fidelity cardiac simulations using a neural network finite element approach},
\newblock in: \bibinfo{booktitle}{International Conference on Functional Imaging and Modeling of the Heart}, \bibinfo{organization}{Springer}, \bibinfo{year}{2023}, pp. \bibinfo{pages}{537--544}.
\bibitem[{Motiwale et~al.(2024)Motiwale, Zhang, Feldmeier, and Sacks}]{motiwale2024neural}
\bibinfo{author}{S.~Motiwale}, \bibinfo{author}{W.~Zhang}, \bibinfo{author}{R.~Feldmeier}, \bibinfo{author}{M.~S. Sacks},
\newblock \bibinfo{title}{A neural network finite element approach for high speed cardiac mechanics simulations},
\newblock \bibinfo{journal}{Computer Methods in Applied Mechanics and Engineering} \bibinfo{volume}{427} (\bibinfo{year}{2024}) \bibinfo{pages}{117060}.
\bibitem[{Goodbrake et~al.(2024)Goodbrake, Motiwale, and Sacks}]{goodbrake2024neural}
\bibinfo{author}{C.~Goodbrake}, \bibinfo{author}{S.~Motiwale}, \bibinfo{author}{M.~S. Sacks},
\newblock \bibinfo{title}{A neural network finite element method for contact mechanics},
\newblock \bibinfo{journal}{Computer Methods in Applied Mechanics and Engineering} \bibinfo{volume}{419} (\bibinfo{year}{2024}) \bibinfo{pages}{116671}.
\bibitem[{Simo and Taylor(1991)}]{simo1991quasi}
\bibinfo{author}{J.~C. Simo}, \bibinfo{author}{R.~L. Taylor},
\newblock \bibinfo{title}{Quasi-incompressible finite elasticity in principal stretches. continuum basis and numerical algorithms},
\newblock \bibinfo{journal}{Computer methods in applied mechanics and engineering} \bibinfo{volume}{85} (\bibinfo{year}{1991}) \bibinfo{pages}{273--310}.

\end{thebibliography}

\appendix

\section{Passive mechanics formulation}\label{sec:passive-mechanics-description} 

Let $\pmb{\Omega_X}$ and $\pmb{\Omega_x}$ be bounded open sets in $\mathcal{R}^3$ with Lipschitz boundaries and represent the reference and current configurations, respectively. In this study, $\pmb{\Omega_X}$ denotes the stress-free reference configuration of the myocardium while $\pmb{\Omega_x}$ represents the deformed myocardium at any instant during passive expansion. We define the boundary, $\Gamma_\mathbf{x} := \Gamma_{\mathrm{base}}\cup\Gamma_{\mathrm{epi}}\cup\Gamma_{\mathrm{endo-lv}}\cup\Gamma_{\mathrm{endo-rv}}$, as the union of the basal plane ($\Gamma_{\mathrm{base}}$), epicardium ($\Gamma_{\mathrm{epi}}$), and endocardium of the left and the right ventricles ($\Gamma_{\mathrm{endo-lv}}$, $\Gamma_{\mathrm{endo-rv}}$). The motion of the body is then characterized by a deformation map, $\varphi : \pmb{\Omega_X} \rightarrow \pmb{\Omega_x}$ such that $\mathbf{x} = \varphi\left({\mathbf{X}}, t\right)$, where $\mathbf{x}$ is the current position of a material point at a time $t$ that was originally at $\mathbf{X}$ in the reference configuration. The displacement, $\mathbf{u}$ and velocity, $\mathbf{v}$ of a material particle are defined as,

\begin{equation}
\begin{split}
\mathbf{u} := & \varphi(\mathbf{X},t) - \mathbf{X} = \mathbf{x}(\mathbf{X},t) - \mathbf{X} \\
\mathbf{v} := & \frac{\partial \varphi}{\partial \mathbf{t}} \bigg|_\mathbf{X} = \frac{d \mathbf{u}}{dt}
\end{split}
\end{equation}

\noindent where, $d(\bullet)/dt$ is the total time derivative. The deformation gradient ($\mathbf{F}$), its Jacobian determinant ($J$), the right Cauchy-Green deformation tensor ($\mathbf{C}$), and the Green-Lagrange strain tensor ($\mathbf{E}$) are defined as,

\begin{equation}
\begin{split}
\mathbf{F} := \frac{\partial \varphi}{\partial\mathbf{X}}, \quad J := \det \big(\mathbf{F}\big), \quad \mathbf{C}=\mathbf{F}^T\mathbf{F}, \quad \mathbf{E}=\frac{1}{2}\big(\mathbf{C}- \mathbf{I}\big)
\end{split}
\end{equation}

\noindent We perform a multiplicative decomposition of the deformation gradient tensor to define $\overline{\mathbf{F}}$ and $\overline{\mathbf{C}}$ as,

\begin{equation}
\overline{\mathbf{F}} := J^{-1/3}\mathbf{F}, \quad \overline{\mathbf{C}} := J^{-2/3}\mathbf{C}
\end{equation}

\noindent that represent the splitting of local deformation gradient ($\mathbf{F}$) into volume-preserving (isochoric, $\overline{\mathbf{F}}$) and volume-changing (dilatational, $J^{1/3}\mathbf{I}$) components. The mechanical behavior of the hyperelastic material is characterized by a Gibbs free energy per unit mass, $G(\overline{\mathbf{C}}, p)$, where $p$ is the thermodynamic pressure, and can be decoupled into isochoric ($G_\mathrm{iso}$) and volumetric components ($G_\mathrm{vol}$) as~\cite{liu2018unified},

\begin{equation}
G(\overline{\mathbf{C}},p) = G_\mathrm{iso} (\overline{\mathbf{C}}) + G_\mathrm{vol} (p)
\end{equation}

\noindent In the absence of body forces, the equations of the motion are then given as,

\begin{subequations}
\begin{align}
  \frac{d\mathbf{u}}{dt} - \mathbf{v} = & \ \mathbf{0} \quad \mathrm{in} \ \Omega_\mathbf{x} \label{eq_kin2}\\
  \beta_{\theta}(p)\frac{dp}{dt} + \nabla_{\mathbf{x}}\cdot\mathbf{v} = & \ 0 \quad \mathrm{in} \ \Omega_\mathbf{x} \label{eq_mass2} \\
  \rho(p)\frac{d\mathbf{v}}{dt} + \nabla_\mathbf{x} p - \nabla_\mathbf{x} \cdot \pmb{\sigma}_{dev} = & \ \mathbf{0} \quad \mathrm{in} \ \Omega_\mathbf{x} \label{eq_linMom2}
\end{align}
\label{eq_govEqs1}
\end{subequations}

\noindent where, $\beta_{\theta}$ is the isothermal compressibility coefficient, $\rho$ is the density of the material, $\pmb{\sigma}_{dev}$ represents the deviatoric part of the Cauchy stress tensor, and $\nabla_\mathbf{x}$ is the gradient operator defined in the spatial coordinates. In the above formulation, $p$ is an independent variable while $\rho$ and $\beta_{\theta}$ are dependent on $p$. Further, the first equation (Eq. (\ref{eq_kin2})) represents the kinematic relation between the displacement and the velocity of the body, and the following two equations (Eqs. (\ref{eq_mass2}, \ref{eq_linMom2})) represent the conservation of mass and linear momentum, respectively. 

If $\rho_0$ is the density of the material in the reference configuration, the constitutive relations of the hyperelastic material are represented in terms of the specific Gibbs free energy components (per unit mass) as~\cite{liu2018unified},

\begin{subequations}
\begin{align}
    \rho(p) := \Bigg(\frac{\partial G_{vol}(p)}{\partial p}\Bigg)^{-1} \ , \quad \beta(p) := \frac{1}{\rho} \frac{d\rho}{dp} = & -\rho(p) \frac{\partial^2G_{vol}(p)}{\partial p^2 }, \label{eq_rho_beta} \\ 
    \pmb{\sigma}_{dev} := J^{-1}\overline{\mathbf{F}} \big(\mathbb{P}:\overline{\mathbf{S}}\big) \overline{\mathbf{F}}^{T} + 2\mu_v \mathrm{dev}[\mathbf{d}] \ , &\quad \overline{\mathbf{S}} = 2 \frac{\partial (\rho_0 G_{iso})}{\partial \overline{\mathbf{C}}} \label{eq_dev_stress} \\ 
    \pmb{\sigma} := \pmb{\sigma}_{dev} - p\mathbf{I} 
\end{align}
\label{eq_constRel}
\end{subequations} 

\noindent where, $\mu_v$ is the dynamic shear viscosity, $\mathrm{dev}[\mathbf{d}]$ is the deviatoric part of the rate of deformation tensor, $\mathbf{d} := \frac{1}{2}\big(\nabla_\mathbf{x} \mathbf{v} + (\nabla_\mathbf{x} \mathbf{v})^T\big)$, and $\mathbb{P} = \mathbb{I} - \frac{1}{3}\big(\mathbf{C}^{-1} \otimes \mathbf{C}\big) $ is the projection tensor. The first term in Eq. (\ref{eq_dev_stress}) for $\pmb{\sigma}_{dev}$ represents the isochoric elastic stress, while the second term is the viscous shear stress.

To model the near-incompressibility of the myocardium, we employ the ST91 volumetric strain energy model developed by Simo and Taylor in 1991~\cite{simo1991quasi} given in the form of specific Gibbs free energy as,

\begin{equation}
    G_{\mathrm{vol}}(p)\bigg\vert^{ST91} = \frac{-p^2 + p\sqrt{p^2 + \kappa^2}}{2 \kappa \rho_0} - \frac{\kappa}{2\rho_0} \ln \bigg( \frac{\sqrt{p^2 + \kappa^2} - p}{\kappa} \bigg)
\label{eq_ST91_Gvol}
\end{equation}

\noindent where $\kappa$ is the bulk modulus and $\rho_0$ is the density in the reference configuration. The corresponding $\rho(p)$ and $\beta_{\theta}(p)$ (Eq. \ref{eq_rho_beta}) are given by,

\begin{equation}
    \rho (p)\bigg\vert^\mathrm{ST91} = \frac{\rho_0}{\kappa} \Big( \sqrt{p^2 + \kappa^2} + p \Big) \ , \ \ \beta_{\theta}(p) \bigg\vert^\mathrm{ST91} = \frac{1}{\sqrt{p^2 + \kappa^2}}
\label{eq_ST91_rho}
\end{equation}

\section{Shape parameters description}\label{sec:shape}

To describe the shape of the 3D biventricle model in a general way, we defined the shape parameters in the way as shown in Fig.~\ref{table_shape-cooeficients}. 

Specifically, there are 7 geometric points: $\mathbf{x}^{\mathrm{lv}}$, and $\mathbf{x}^{\mathrm{rv}}$ the centers of the left and right ventricular chambers, $\mathbf{x}_{\mathrm{base}}^{\mathrm{lv}}$, and $\mathbf{x}_{\mathrm{base}}^{\mathrm{rv}}$ the centers of the left and right ventricular chambers at the base plane, $\mathbf{x}_{\mathrm{apex}}^{\mathrm{lv}}$, and $\mathbf{x}_{\mathrm{apex}}^{\mathrm{rv}}$ the apex points of the left and right ventricular chambers, and $\mathbf{x}_{\mathrm{center}}$ the center of the whole geometry.

There are 17 dimensions: the dimensions of the whole structure $\dim_{\mathrm{x}}$, $\dim_{\mathrm{y}}$, $\dim_{\mathrm{z}}$, $\dim^{\mathrm{lv}}_x$, the 
dimensions of the LV chamber $\dim^{\mathrm{lv}}_y$, $\dim^{\mathrm{lv}}_z$, the 
dimensions of the RV chamber $\dim^{\mathrm{rv}}_x$, $\dim^{\mathrm{rv}}_y$, $\dim^{\mathrm{rv}}_z$. $\dim_{\mathrm{base}}^{\mathrm{lv,x}}$, the length of the two intersection points of the line parallel to $x$-axis passing through $\mathbf{x}_{\mathrm{base}}^{\mathrm{lv}}$; $\dim_{\mathrm{base}}^{\mathrm{lv,y}}$, the length of the two intersection points of the line parallel to $y$-axis passing through $\mathbf{x}_{\mathrm{base}}^{\mathrm{lv}}$; $\dim_{\mathrm{base}}^{\mathrm{rv,x}}$, the length of the two intersection points of the line parallel to $x$-axis passing through $\mathbf{x}_{\mathrm{base}}^{\mathrm{rv}}$; $\dim_{\mathrm{base}}^{\mathrm{rv}}$, the length of the two intersection points of the line parallel to $y$-axis passing through $\mathbf{x}_{\mathrm{base}}^{\mathrm{rv}}$. $\dim_{\mathrm{mid}}^{\mathrm{lv,x}}$, the length of the two intersection points of the line parallel to $x$-axis passing through $\mathbf{x}^{\mathrm{lv}}$; $\dim_{\mathrm{mid}}^{\mathrm{lv,y}}$, the length of the two intersection points of the line parallel to $y$-axis passing through $\mathbf{x}^{\mathrm{lv}}$; $\dim_{\mathrm{mid}}^{\mathrm{rv,x}}$, the length of the two intersection points of the line parallel to $x$-axis passing through $\mathbf{x}^{\mathrm{rv}}$; $\dim_{\mathrm{mid}}^{\mathrm{rv,y}}$, the length of the two intersection points of the line parallel to $y$-axis passing through $\mathbf{x}^{\mathrm{rv}}$.

There are 7 thicknesses for the left ventricle: $t_{\mathrm{base}}^{\mathrm{lv,x}}$, the shortest distance between the higher-$x$-value point of $\dim_{\mathrm{base}}^{\mathrm{lv,x}}$ and the epicardium surface; $t_{\mathrm{base}}^{\mathrm{lv,y+}}$, the shortest distance between the higher-$y$-value of $\dim_{\mathrm{base}}^{\mathrm{lv,y}}$ and the epicardium surface; $t_{\mathrm{base}}^{\mathrm{lv,y-}}$, the shortest distance between the lower-$y$-value of $\dim_{\mathrm{base}}^{\mathrm{lv,y}}$ and the epicardium surface; $t_{\mathrm{mid}}^{\mathrm{lv,x}}$, the shortest distance between the higher-$x$-value point of $\dim_{\mathrm{mid}}^{\mathrm{lv,x}}$ and the epicardium surface; $t_{\mathrm{mid}}^{\mathrm{lv,y+}}$, the shortest distance between the higher-$y$-value of $\dim_{\mathrm{mid}}^{\mathrm{lv,y}}$ and the epicardium surface; $t_{\mathrm{mid}}^{\mathrm{lv,y-}}$, the shortest distance between the lower-$y$-value of $\dim_{\mathrm{mid}}^{\mathrm{lv,y}}$ and the epicardium surface; $t_{\mathrm{apex}}^{\mathrm{lv}}$, the shortest distance between $\mathbf{x}_{\mathrm{apex}}^{\mathrm{lv}}$ and the epicardium surface. Similar definitions apply to the 7 right ventricle thicknesses $t_{\mathrm{base}}^{\mathrm{rv,x}}$, $t_{\mathrm{base}}^{\mathrm{rv,y+}}$, $t_{\mathrm{base}}^{\mathrm{rv,y-}}$, $t_{\mathrm{mid}}^{\mathrm{rv,x}}$, $t_{\mathrm{mid}}^{\mathrm{rv,y+}}$, $t_{\mathrm{mid}}^{\mathrm{rv,y-}}$, and $t_{\mathrm{apex}}^{\mathrm{rv}}$. There are 2 septum thicknesses $t_{\mathrm{base}}^{\mathrm{sep}}$  defined as the shortest distance between the lower-$x$-value point of $\dim_{\mathrm{base}}^{\mathrm{lv,x}}$ and the endo-rv surface; and $t_{\mathrm{mid}}^{\mathrm{sep}}$ defined as shortest distance between the lower-$x$-value point of $\dim_{\mathrm{mid}}^{\mathrm{lv,x}}$ and the endo-rv surface.

\begin{table}[H]
\begin{center}
\footnotesize
\caption{Shape parameters for the biventricle geometry.}
\label{table_shape-cooeficients}
\vspace{4pt}
\begin{tabular}{c c c c c c c c}
\toprule
$\mathbf{x}_{\mathrm{base}}^{\mathrm{lv}}$ & $\mathbf{x}_{\mathrm{base}}^{\mathrm{rv}}$ & $\mathbf{x}^{\mathrm{lv}}$ & $\mathbf{x}^{\mathrm{rv}}$ & $\mathbf{x}_{\mathrm{apex}}^{\mathrm{lv}}$ & $\mathbf{x}_{\mathrm{apex}}^{\mathrm{rv}}$ & $\mathbf{x}_{\mathrm{center}}$ \\
\arrayrulecolor{black!30}\midrule
$t_{\mathrm{base}}^{\mathrm{lv,x}}$ & $t_{\mathrm{base}}^{\mathrm{lv,y+}}$ & $t_{\mathrm{base}}^{\mathrm{lv,y-}}$ & $t_{\mathrm{mid}}^{\mathrm{lv,x}}$ & $t_{\mathrm{mid}}^{\mathrm{lv,y+}}$ & $t_{\mathrm{mid}}^{\mathrm{lv,y-}}$ & $t_{\mathrm{apex}}^{\mathrm{lv}}$ \\
\arrayrulecolor{black!30}\midrule
$t_{\mathrm{base}}^{\mathrm{rv,x}}$ & $t_{\mathrm{base}}^{\mathrm{rv,y+}}$ & $t_{\mathrm{base}}^{\mathrm{rv,y-}}$ & $t_{\mathrm{mid}}^{\mathrm{rv,x}}$ & $t_{\mathrm{mid}}^{\mathrm{rv,y+}}$ & $t_{\mathrm{mid}}^{\mathrm{rv,y-}}$ & $t_{\mathrm{apex}}^{\mathrm{rv}}$ \\
\arrayrulecolor{black!30}\midrule
$t_{\mathrm{base}}^{\mathrm{sep}}$ & $t_{\mathrm{mid}}^{\mathrm{sep}}$ & $\dim_{\mathrm{x}}$ & $\dim_{\mathrm{y}}$ & $\dim_{\mathrm{z}}$ & $\dim_{\mathrm{base}}^{\mathrm{lv,x}}$ & $\dim_{\mathrm{base}}^{\mathrm{lv,y}}$ \\
\arrayrulecolor{black!30}\midrule
$\dim^{\mathrm{lv}}_x$ & $\dim^{\mathrm{lv}}_y$ & $\dim^{\mathrm{lv}}_z$ & $\dim_{\mathrm{mid}}^{\mathrm{lv,x}}$ & $\dim_{\mathrm{mid}}^{\mathrm{lv,y}}$ & $\dim_{\mathrm{base}}^{\mathrm{rv,x}}$ & $\dim_{\mathrm{base}}^{\mathrm{rv,y}}$ \\
\arrayrulecolor{black!30}\midrule
$\dim^{\mathrm{rv}}_x$ & $\dim^{\mathrm{rv}}_y$ & $\dim^{\mathrm{rv}}_z$ & $\dim_{\mathrm{mid}}^{\mathrm{rv,x}}$ & $\dim_{\mathrm{mid,y}}^{\mathrm{rv}}$ &  &  \\
\arrayrulecolor{black}\bottomrule
\end{tabular}
\end{center}
\end{table}

\begin{figure}[h] 
 \begin{center}
 \includegraphics[width=1\textwidth]{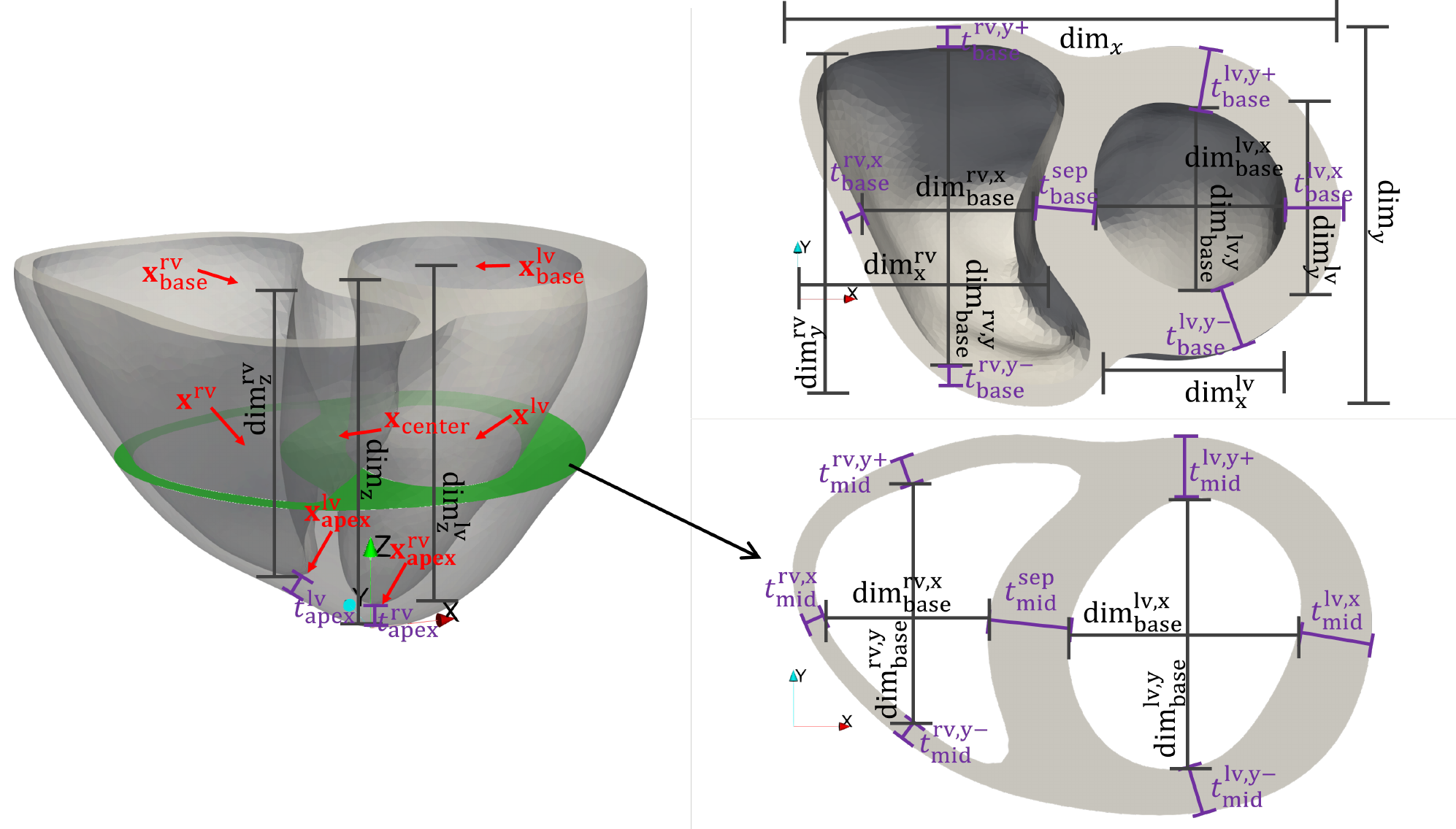}
 \end{center}
 \caption{Illustration of the shape parameters definition on the geometry.}
 \label{fig_shape}
\end{figure}

\section{Material parameters and pressures for the cases shown in Fig.~\ref{fig_data_results}}

\begin{table}[ht]
\begin{center}
\footnotesize
\caption{Material parameters and pressures for the cases in Fig.~\ref{fig_data_results}.}
\label{table_mat_params_fig}
\vspace{4pt}
\begin{tabular}{c c c c c c c c c }
\toprule
Case & $a$  & $b$ & $a_f$ & $b_f$ & $a_s$ & $b_s$ & $p_{\mathrm{lv}}$    &    $p_{\mathrm{rv}}$   \\
\arrayrulecolor{black!30}\midrule
1 & $3.5\times 10^3$ & $16.3$  & $1.3\times 10^5$  & $8.5$ & $3.5 \times 10^3$ & $19.3$ & $6.3$  & $4.5$   \\ 
2 & $2.5\times 10^2$ & $3.9$  & $1.3\times 10^5$  & $10.7$ & $8.0 \times 10^3$ & $6.7$ & $7.6$  & $5.6$   \\
3 & $4.3\times 10^3$ & $4.3$  & $6.1\times 10^4$  & $26.2$ & $2.7 \times 10^3$ & $27.3$ & $7.5$  & $4.0$   \\
4 & $1.4\times 10^3$ & $26.5$  & $2.4\times 10^5$  & $22.9$ & $1.9 \times 10^3$ & $20.7$ & $8.5$  & $5.5$   \\
5 & $6.0\times 10^2$ & $20.1$  & $6.5\times 10^5$  & $18.8$ & $9.4 \times 10^4$ & $7.6$ & $14.6$  & $11.2$   \\
6 & $8.1\times 10^3$ & $15.5$  & $3.1\times 10^4$  & $15.7$ & $5.4 \times 10^4$ & $6.8$ & $10.6$  & $7.8$   \\
7 & $5.1\times 10^2$ & $24.2$  & $5.6\times 10^5$  & $14.3$ & $3.2 \times 10^4$ & $10.8$ & $7.8$  & $5.8$   \\
8 & $3.7\times 10^3$ & $17.7$  & $3.7\times 10^5$  & $26.3$ & $3.7 \times 10^3$ & $2.4$ & $8.1$  & $5.5$   \\               
\arrayrulecolor{black}\bottomrule
\end{tabular}

\begin{tablenotes}
    \centering
    \footnotesize
    \item $a_{(\cdot)}$: $\text{dyn}/\text{cm}^2$; $b_{(\cdot)}$ is dimensionless; $p_{(\cdot)}$: $\text{mmHg}$.
\end{tablenotes}

\end{center}
\end{table}

\end{document}